\documentclass{aa}  

\usepackage{graphicx}
\usepackage{xcolor}
 \usepackage{lscape}
\usepackage{txfonts}
\usepackage{newtxtext,newtxmath}
\usepackage{longtable}
\usepackage{adjustbox}
\newcommand{\nicer}{{\rm NICER}}
\def\gtsima{$\; \buildrel > \over \sim \;$}
\def\ltsima{$\; \buildrel < \over \sim \;$}
\def\gsim{\lower.5ex\hbox{\gtsima}}
\def\lsim{\lower.5ex\hbox{\ltsima}}
\begin{document}

   \title{The inner flow geometry in MAXI J1820+070 during hard and hard-intermediate states}
\titlerunning{The inner flow geometry in MAXI J1820+070 }
\authorrunning{B. De Marco et al.}

   \author{B. De Marco,
          \inst{1,2}
          A. A. Zdziarski,
          \inst{2}
          G. Ponti,
          \inst{3,4}
          G. Migliori,
          \inst{5,6}
          T. M. Belloni,
          \inst{3}
          A. Segovia Otero,
           \inst{7}
           M. A. Dzie{\l}ak,
           \inst{2}\newline
           \and
           E. V. Lai
           \inst{2}
          }

   \institute{Departament de F\'isica, EEBE, Universitat Polit\`ecnica de Catalunya, Av. Eduard Maristany 16, E-08019 Barcelona, Spain
               \thanks{ \email{barbara.de.marco@upc.edu}}
              \and
              Nicolaus Copernicus Astronomical Center, Polish Academy of Sciences, Bartycka 18, PL-00-716 Warszawa, Poland
               \and
              INAF-Osservatorio Astronomico di Brera, Via E. Bianchi 46, I-23807 Merate (LC), Italy
               \and
              Max-Planck-Institut f\"ur Extraterrestrische Physik, Giessenbachstrasse, D-85748, Garching, Germany
               \and
              Istituto di Radioastronomia - INAF, Via P. Gobetti 101, I-40129 Bologna, Italy
               \and
              Dipartimento di Fisica e Astronomia, Universit\`a di Bologna, Via Gobetti 93/2, I-40129 Bologna, Italy
               \and
              Department of Astronomy and Theoretical Physics, Lund Observatory, Box 43, SE-221 00 Lund, Sweden
             }

    \date{Received ...; accepted ...}

% \abstract{}{}{}{}{} 
% 5 {} token are mandatory
 
  \abstract
  % context heading (optional)
  %{} leave it empty if necessary  
   {We present a systematic X-ray spectral-timing study of the recently discovered, exceptionally bright black hole X-ray binary system MAXI J1820+070. Our analysis focuses on the first part of the 2018 outburst, covering the rise throughout the hard state, the bright hard and hard-intermediate states, and the transition to the soft-intermediate state.} 
   %
  % aims heading (mandatory)
   {We address the issue of constraining the geometry of the innermost accretion flow and its evolution throughout an outburst.} 
  % methods heading (mandatory)
   {We employed two independent X-ray spectral-timing methods applied to archival \nicer\ data of MAXI J1820+070. We first identified and tracked the evolution of a characteristic frequency of soft X-ray thermal reverberation lags (lags of the thermally reprocessed disc emission after the irradiation of variable hard X-ray photons). This frequency is sensitive to intrinsic changes in the relative distance between the X-ray source and the disc. Then, we studied the spectral evolution of the quasi-thermal component responsible for the observed thermal reverberation lags. We did so by analysing high-frequency covariance spectra, which single out spectral components that vary in a linearly correlated way on the shortest sampled timescales and are thus produced in the innermost regions of the accretion flow.}
  % results heading (mandatory)
   {The frequency of thermal reverberation lags steadily increases throughout most of the outburst, implying that the relative distance between the X-ray source and the disc decreases as the source softens. However, near transition this evolution breaks, showing a sudden increase (decrease) in lag amplitude (frequency). 
  On the other hand, the temperature of the quasi-thermal component in covariance spectra, due to disc irradiation and responsible for the observed soft reverberation lags, consistently increases throughout all the analysed observations.}
  % conclusions heading (optional), leave it empty if necessary 
   {This study proposes an alternative interpretation to the recently proposed contracting corona scenario. Assuming a constant height for the X-ray source, the steady increase in the reverberation lag frequency and in the irradiated disc temperature in high-frequency covariance spectra can be explained in terms of a decrease in the disc inner radius as the source softens. The behaviour of thermal reverberation lags near transition might be related to the relativistic plasma ejections detected at radio wavelengths, suggesting a causal connection between the two phenomena.
Throughout most of the hard and hard-intermediate state, the disc is consistent with being truncated (with an inner radius $R_{\rm in}\gsim 10 R_{\rm g}$), reaching close to the innermost stable circular orbit only near transition. }


   \maketitle

\section{Introduction}
\label{sec:intro}
Black hole X-ray binaries (BHXRBs) are ideal laboratories for studying in detail the evolution of the accretion flow as a function of the accretion state during an outburst \citep[e.g.][]{Done2007}. 
During the soft state, thermal emission with a peak temperature of  $\sim$ 1 keV is observed. The low or undetected rapid X-ray variability \citep[on timescales between milliseconds and hundreds of seconds; e.g.][]{Gierlinski2005,Belloni2005} suggests the accretion flow is stable against fast changes in the mass accretion rate in this state \citep[][]{Lyubarskii1997,Churazov2001}. A geometrically thin, optically thick accretion disc \citep[][]{Shakura1973,Novikov1973} extending close to the innermost stable circular orbit (ISCO) of the black hole (BH) is the accepted paradigm to explain the thermally dominated soft state \citep[e.g.][]{Gilfanov2010}.
However, this standard model is unable to explain the hard spectral states, which are instead dominated by rapidly variable \citep[][]{Gierlinski2005,MunozDarias2011,Heil2012} emission from a hot plasma with peak temperatures of a few tens to hundreds of keV \citep[e.g.][]{Joinet2008,Motta2009,Buisson2019,Zdziarski2021b}. In the hard state the primary broadband X-ray continuum more likely results from Compton up-scattering processes of less energetic photons (thermal and/or synchrotron) by the electrons in the hot plasma \citep[][]{Poutanen1998,Gierlinski1999,Poutanen2009}. On the other hand, the standard disc is observed to be cooler ($kT\lsim$0.3 keV; e.g. \citealt{Kolehmainen2014,ZdziarskiBDM2020}) and more variable \citep[][]{Wilkinson2009, BDM2015a} than in the soft state.
Transitions between hard and soft states are characterised by abrupt changes in the X-ray spectral and timing properties of the source \citep[e.g.][]{Belloni2010,Bogensberger2020}. The intermediate states during which the transition occurs have been further classified into hard-intermediate and soft-intermediate states, based on their significantly different X-ray timing properties \citep{Homan2005,Belloni2005}. The observed changes characterising the intermediate states are likely a consequence of major, but still not well-understood, variations in the physical properties of the accretion flow.

All the models proposed to explain the hard state postulate that the hot phase of the accretion flow is located at relatively small distances from the BH; however, its exact geometry is unknown, and a number of viable possibilities currently exist. 
For example, the hot flow might fill the inner regions of a truncated optically thick and geometrically thin accretion disc, with the transition region characterised by an overlap between the two phases \citep[i.e. a corona sandwiching the inner disc and/or disrupted disc clumps embedded in the hot flow; e.g.][]{Done1999,Yuan2014,Poutanen2018}. 
Alternatively, the standard disc might reach close to the BH and the hard X-ray source may be a compact region, possibly located on the rotation axis of the BH (the so-called lamppost; e.g. \citealt{Garcia2014,Niedzwiecki2016}) and linked to the base of a jet \citep[][]{Miller2006}. As a matter of fact, radio observations indicate the presence of a bipolar relativistically expanding and outflowing plasma associated with this state (\citealt{Fender2004}; \citeyear{Fender2006}; \citealt{Corbel2001}; \citeyear{Corbel2004}; \citealt{Brocksopp2002}; \citealt{Miller-Jones2011}). However, the exact relation (if any) between the X-ray source and the jet is still unclear \citep[e.g.][]{Yuan2019}.

The main challenge for these models is explaining the high level of X-ray spectral and timing complexity that characterises the hard state, including: complex broadband spectral curvature \citep[][]{Makishima1986,Nowak2011}; evolution of characteristic variability timescales \citep[][]{Done2007}; detection of hard X-ray lags (hard band flux variations lagging behind soft band variations) in Comptonised emission-dominated regions of the spectrum \citep[][]{Nowak1999,Grinberg2014}; and the recurrent appearance of low-frequency quasi-periodic oscillations (QPOs) during high luminosity hard and hard-intermediate states \citep[e.g.][]{Rodriguez2002,Casella2004,Sobolewska2006,Ingram2019a}.  
Truncated disc models can (at least qualitatively) explain most of the observed X-ray spectral and timing properties of BHXRBs and their evolution during an outburst, provided some spectral stratification of the hot flow is postulated for the hard state \citep[][]{Done2007,Axelsson2018,Mahmoud2018a}. Within this scenario the key ingredient is an evolving inner disc radius, moving closer to the BH as the source evolves from the hard to the soft state, and receding again as the source goes back to quiescence \citep[][]{Esin1997,Yuan2014}.
On the other hand, a compact X-ray source with a disc inner radius at $\sim R_{{\rm ISCO}}$ already at moderate Eddington ratios is usually invoked to explain the broad profile of Fe K emission lines in the disc reflection component, which are observed particularly in bright hard states \citep[][]{Miller2006,Garcia2015,Furst2015}. Within this model, variations in the vertical extent of the primary hard X-ray source have been proposed to explain the evolution of some of the X-ray spectral and timing properties throughout the hard state \citep[][]{Kara2019,Buisson2019}.

Theoretical arguments and supporting observational evidence exist for the optically thick and geometrically thin disc being highly truncated during quiescence and in the early stages of an outburst ($R_{\rm{in}}\sim 10^{3-4}\ R_{\rm{g}}$ for $M_{\rm{BH}}\sim 10 M_{\odot}$, where $R_{\rm{g}}=GM_{\rm{BH}}/c^{2}$ is the gravitational radius; e.g. \citealt{Lasota1996,Narayan1997,Dubus2001,McClintock2001,Esin2001,Bernardini2016}). Nonetheless, the question of when the disc reaches the ISCO (whether early in the hard state, in the bright hard state, or during the intermediate states characterising the hard-to-soft state transition) remains unanswered. Attempts to settle this issue have so far yielded controversial results (see \citealt{Bambi2021} for a review). Among the main problems is the difficulty in obtaining a long, almost uninterrupted monitoring of an outburst with high throughput X-ray detectors covering energy ranges with significant contribution from all the main spectral components. In particular, while the rise in luminosity throughout the hard state can be quite slow (of the order of a few weeks to months), the hard-intermediate states preceding the transition to the soft-intermediate states are spanned in a very short amount of time (a few days), and thus they are very difficult to catch \citep[][]{Belloni2005,Dunn2010}. Yet, these phases are of the utmost importance for understanding how the accretion flow evolves. 

The \emph{Neutron star Interior Composition Explorer} (\nicer; \citealt{Gendrau2016}) is making significant inroads in this context by performing high cadence, high quality X-ray observations of Galactic transient sources. 
Among the sources observed by \nicer\ is the very bright BHXRB MAXI J1820+070 (ASASSN-18ey),  which was discovered in March 2018 \citep[][]{Tucker2018}. 
During the 2018 outburst the source rapidly brightened, reaching extremely high X-ray luminosities, with a peak X-ray flux of about 4 Crab. Such a high brightness and  low absorbing column density to the source ($\rm{N_{H}\sim 10^{21} cm^{-2}}$, \citealt{Uttley2018}) make MAXI J1820+070 an excellent target for detailed studies of BHXRB outbursts and state transitions. 
Intensive monitoring at radio wavelengths, followed by high spatial resolution \emph{Chandra} observations, revealed the ejection of long-lived discrete relativistic jets \citep[][]{Bright2019,Espinasse2020}. Major changes in the X-ray variability properties of the source preceded these ejections by a few hours \citep[][]{Homan2020}, suggesting that these events marked the transition from the hard to the soft state.

Since its discovery, \nicer\ has been observing the source regularly, providing an incredibly rich archive of data covering all the phases of the outburst.
Here we focus on the first part of the 2018 outburst, covering the hard and hard-intermediate states before transition to the soft-intermediate states. We employ two independent methods to constrain the inner accretion flow geometry and its evolution. Both methods rely on the application of X-ray spectral-timing analysis techniques \citep[][]{Uttley2014} to study the quasi-thermal soft component produced by the hard X-ray irradiation of the inner disc \citep[e.g.][]{ZdziarskiBDM2020}. Since the hard X-ray flux impinging on the cold disc medium is variable, it will induce a temporal response in the disc quasi-thermal emission, which is the focus of this work.

The first method (Sect. \ref{sec:softlags}) is based on studying the evolution of high-frequency ($\nu\gsim$1 Hz) soft X-ray lags, namely soft band flux responding to rapid hard band flux variations with a time delay. These lags are thought to be the signature of X-ray thermal reverberation, namely the time-delayed response of the re-emitted quasi-thermal disc emission to variations in the hard X-ray irradiating flux \citep[e.g.][]{Uttley2011}. Being strongly dependent on the light path between the X-ray source and the reprocessing region, X-ray reverberation lags can be used to map the geometry of the inner accretion flow and its evolution. To date, such lags have been detected in some of the best monitored BHXRBs \citep[][]{Uttley2011,BDM2015b,BDM2016,BDM2017,WangJ2020}, including MAXI J1820+070 \citep[][]{Kara2019}.

The second method (Sect. \ref{sec:covar}) is based on studying the evolution of covariance spectra \citep[][]{Wilkinson2009,Uttley2014}, which are used to single out the spectral components responsible for the observed thermal reverberation lags. The peak temperature of the quasi-thermal component responding to the hard X-ray irradiation of the inner disc is linked to the disc inner radius, thus providing an alternative way of constraining the geometry of the inner accretion flow \citep[][]{ZdziarskiBDM2020,Zdziarski2021a}.

For our computations we make use of the latest results from the optical spectroscopy of MAXI J1820+070, which yield a BH mass of $M_{\rm{BH}} = (5.95 \pm 0.22) {\rm sin}^{-3}i\ M_{\odot}$ \citep[][]{Torres2020}. Hereafter we assume a fiducial value of $M_{\rm{BH}} =7.6\pm1.7\ M_{\odot}$. This encompasses the range of values inferred from current constraints on the orbital ($66^\circ<i<81^\circ$; \citealt{Torres2020}), and jet inclination ($63\pm 3^\circ$; \citealt{Atri2020}; $64\pm 5^\circ$, \citealt{Wood2021}). We use the distance to the source of $d = 2.96\pm0.33\ \rm{kpc}$ \citep[where the uncertainty is for 68 per cent confidence;][]{Atri2020}.

\section{Observations and data reduction}
\label{sec:data}

We analysed data from \nicer\ X-ray Timing Instrument (XTI) observations of MAXI J1820+070 between MJD 58189 and MJD 58306. These observations cover the first part of the 2018 outburst, namely the hard and hard-intermediate accretion states, up to the transition to the soft-intermediate states (the observations analysed are listed in Table \ref{tab1}, the observation identification number, ObsID, increases progressively with the time of the observation; we note that our sample also includes the observations previously analysed in \citealt{Kara2019}). For brevity, the different observations will be hereafter referred to as O\emph{xxx}, with \emph{xxx} being the last three digits of their ObsID.

The data were reprocessed with the NICERDAS tools in HEASOFT v6.28 and \nicer\ calibration files as of July 27, 2020. Calibrated, unfiltered, all Measurement/Power Unit (MPU) merged files ({\tt ufa}) were created using {\tt NICERL2} task. 
We applied standard filtering criteria \citep[e.g.][]{Stevens2018}, using {\tt NIMAKETIME} and {\tt NICERCLEAN} routines. 
We extracted light curves in the energy range $13-15\ \rm{keV}$ (corresponding to the range where \nicer\ XTI effective area quickly drops) to check for periods of high particles background. Time intervals showing background events with rate $>2 \ \rm{counts\ s^{-1}}$ \citep[e.g.][]{Ludlam2018} were discarded.

Of the 56 focal plane modules (FPMs) of \nicer\ XTI, FPMs 11, 20, 22, and 60 are not operational. 
Following `\nicer\ Analysis Tips \& Caveats\footnote{\url{https://heasarc.gsfc.nasa.gov/docs/nicer/data_analysis/nicer_analysis_tips.html}},' FPMs 14 and 34 were removed from the analysis as they can exhibit increased detector noise. 
The single observations were screened to identify additional FPMs showing anomalous behaviour, which, for example, can manifest as the total number of registered counts significantly exceeding the average counts from all detectors. However, none was found during the analysed observations. Occasionally (particularly during the brightest phases of the outburst before the transition to the soft-intermediate states), the number of active FPMs was reduced to avoid telemetry saturation and an increase in dead time. Once accounting for this reduction and the outlined filtering criteria, the net number of FPMs used for this analysis ranges between $50$ and $27$, as reported in Table \ref{tab1}.

The resulting event lists were used to extract light curves in different energy bands with a time bin of $0.4\ \rm{ms}$. Custom ancillary response files (ARFs) and redistribution matrix files (RMFs) were computed for the specific subset of FPMs of each observation. This was done by combining the corresponding per-module, publicly distributed ARFs and RMFs.
Fits were performed using Xspec v.12.11.1 and errors are hereafter reported at the 90 per cent confidence level.

In order to speed-up the reduction process, short good time intervals (GTIs) with length $<10$ s were removed. 
This led us to the exclusion of observations O128, O129, O154, O181, O191, O192, O193, because of their resulting very short net exposure ($\lsim200$ s).
The effective, on-source, exposure times of analysed observations after screening and filtering of short GTIs are listed in Table \ref{tab1}.

\section{Hardness-intensity diagram}
\label{sec:HID} 

\begin{figure}
        \includegraphics[width=\columnwidth]{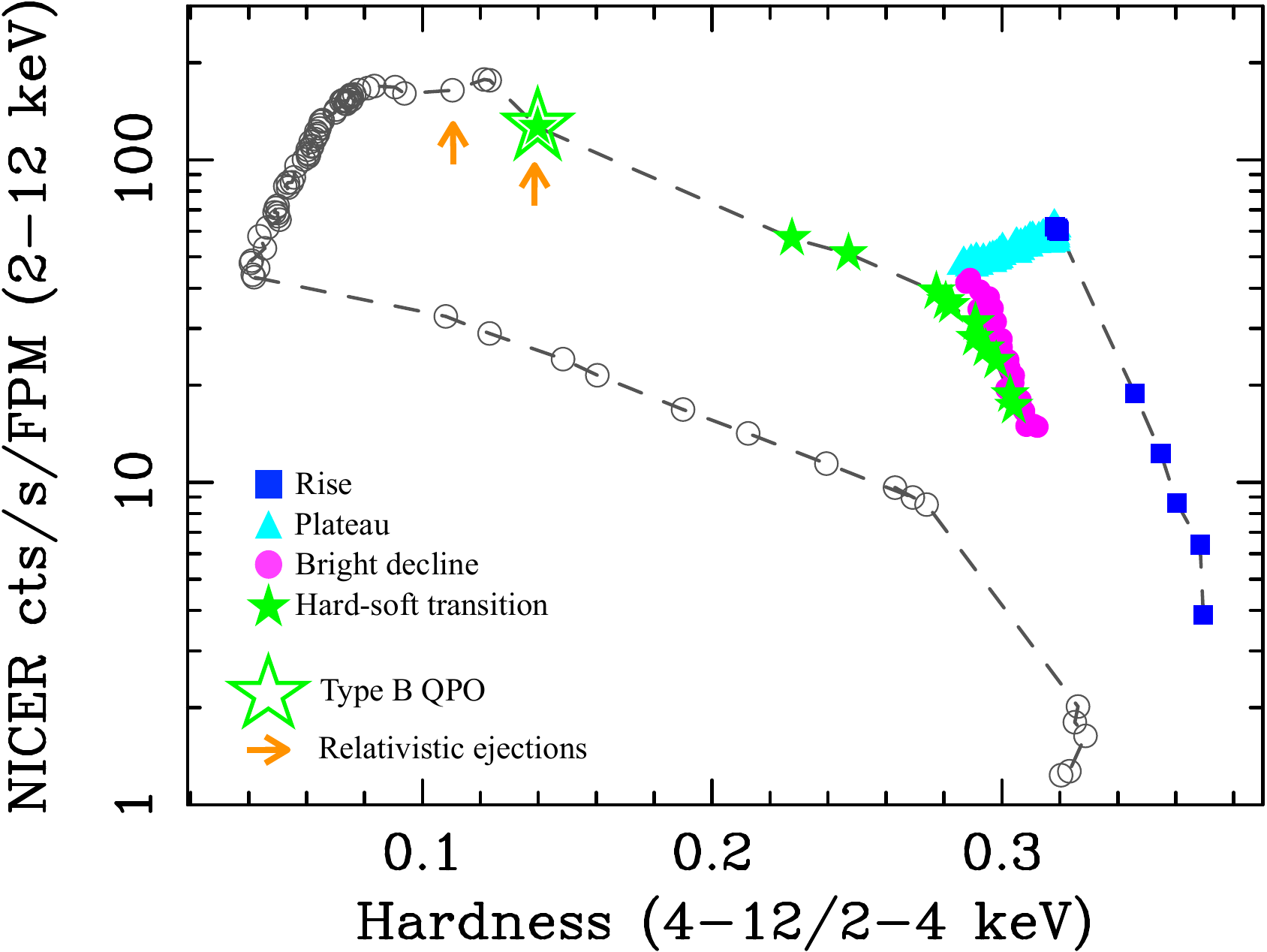}
    \caption{HID of MAXI J1820+070 during its 2018 outburst (from MJD 58189 to MJD 58305). The hardness is computed as the ratio between the 4--12 keV and 2--4 keV \nicer\ count rates. On the y axis we report the total 2--12 keV \nicer\ count rate re-normalised for the number of used FPMs. Highlighted in different colours and symbols are the phases of the outburst that are the focus of our work. The grey empty circles show the subsequent spectral evolution of the source up until MJD 58404. The epochs of the appearance of a type-B QPO (MJD$\sim$58305.7; \citealt{Homan2020}) and the launch of relativistic discrete radio ejecta (MJD 58305.6$\pm0.04$; \citealt{Wood2021}, and $\sim$MJD 58306, \citealt{Bright2019,Espinasse2020}) are marked, respectively, by the bigger star and orange arrows.}
    \label{fig:hid}
\end{figure}

We built the hardness-intensity diagram (HID; Fig. \ref{fig:hid}) of the source in order to select observations corresponding to the accretion states of interest. To this aim we extracted \nicer\ count rates in the energy ranges 2--4 keV and 4--12 keV. The total 2--12 keV count rate was normalised for the number of FPMs used for the analysis ({\tt N}$_{\rm{FPM}}$, Table \ref{tab1}). In some cases {\tt N}$_{\rm{FPM}}$ changed during a single observation. In these cases the observation was split so as to have datasets characterised by a constant number of FPMs (each point in the HID corresponds to either a single \nicer\ observation or the part of an observation characterised by constant {\tt N}$_{\rm{FPM}}$). We note that the HID of Fig. \ref{fig:hid} reports also observations excluded from the analysis because too short (Sect. \ref{sec:data}).

During the first part of the outburst, we identified four main phases: an initial brightening at almost constant hardness (which lasted for about 12 days, `rise' in Fig. \ref{fig:hid}), a persistent (lasting about 2 months) bright hard state (`plateau' in Fig. \ref{fig:hid}), a sharp dimming phase (lasting about one month) at slightly lower, and almost constant hardness (`bright decline' in Fig. \ref{fig:hid}), and the quick (lasting about two weeks) transition to the soft-intermediate state (`hard-soft transition' in Fig. \ref{fig:hid}). These phases are the focus of our analysis.
We note that the last analysed dataset includes only the first part ($\sim$64 ks) of O197, corresponding to an on-source effective exposure of $\sim 12$ ks. This spans the phases immediately before the fast transition from the hard-intermediate to the soft-intermediate state (i.e. the $\sim47$ ks preceding the transition, of which $\sim$8 ks of effective time) and the transition itself. The excluded part of O197 corresponds to the soft-intermediate state. In this phase X-ray spectral-timing measurements are noisier due to a drop in the intrinsic variability of the source.
As shown by \citet[][]{Homan2020}, apart from the drop of broadband X-ray variability strength, the transition from the hard-intermediate to the soft-intermediate state was also marked by an intense X-ray flare and the switch from type-C to type-B QPO at MJD$\sim$58305.7. This epoch is highlighted in the HID of Fig. \ref{fig:hid}. Also shown in Fig. \ref{fig:hid} is the estimated epoch of the launch of discrete relativistic ejecta, detected first at radio wavebands \citep[][]{Bright2019} and later on, at larger scales, in the X-rays \citep[][]{Espinasse2020}. An earlier ejection has been recently reported \citep[][]{Wood2021}, coinciding with O197, and preceding the switch to a type-B QPO. The empty circles in Fig. \ref{fig:hid} show the spectral changes of the source during the remaining part of the 2018 outburst, till MJD 58404. 

\section{X-ray spectral-timing analysis}
\label{sec:spec-tim}

We measured the Fourier cross-spectrum to extract frequency-dependent time lags and covariance spectra \citep[e.g.][]{Nowak1999,Uttley2014}. Given that the on-source exposure time of each \nicer\ observation is usually quite short (a few hundred to a few thousand seconds; see Table \ref{tab1}), we adopted the approach of combining consecutive observations in order to increase the signal-to-noise ratio (S/N) of spectral-timing products. Before combining consecutive observations we verified that they do not show deviations from stationarity \citep[][]{Vaughan2003}.
The main reason for this is that non-stationary behaviour is likely associated with substantial changes in the physical properties of the accretion flow, possibly including geometry, whose evolution we aim to test here. Details on this selection are reported in Appendix \ref{sec:PSD}.
In addition, we combined only datasets with the same number of FPMs. This is relevant for the fit of covariance spectra (Sect. \ref{sec:covar}), since the response of the instrument depends on the specific FPMs used for the analysis (Sect. \ref{sec:data}). 
The final groups of combined observations can be inferred from Table \ref{tab1}.

\
At the beginning of the outburst, the power spectral density (PSD) changed relatively slowly. In this phase of the outburst we split large groups of observations having compatible PSDs into smaller groups each characterised by sufficiently long net exposure time ($\sim$ 10--15 ks).
On the other hand, before the hard-soft transition the PSD changed quickly, thus preventing us from combining consecutive observations and obtaining longer exposures (e.g. O191, O192, and O193 were discarded because of the short exposures and the inability to combine them Sect. \ref{sec:data}).

\begin{figure}
        \includegraphics[width=\columnwidth]{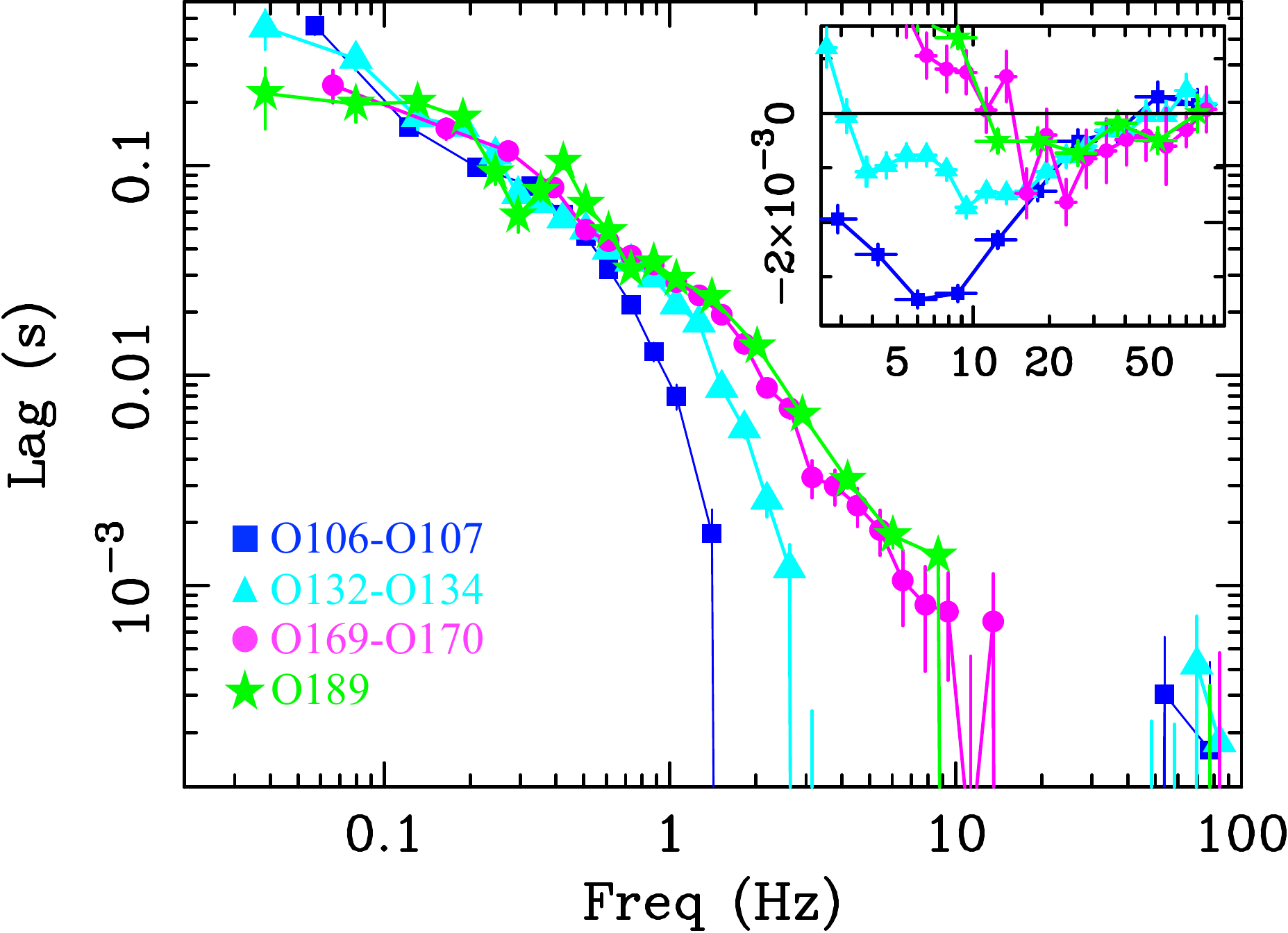}\\

    \caption{ 0.5--1 keV vs. 2--5 keV lag-frequency spectrum of some of the analysed observations of MAXI J1820+070 during the rise, plateau, bright decline, and hard-soft transition phases (with colour codes and symbols as in Fig. \ref{fig:hid}). The positive lags dominating at low frequencies ($\lsim$ 2 Hz) are hard X-ray lags. The plot illustrates the hump-like structures characterising these low-frequency hard lags and their complex evolution throughout the first part of the outburst. At high frequencies ($\gsim$ 2 Hz) the (negative) soft X-ray lag becomes dominant (see inset, with the y axis in linear scale).}
    \label{fig:hardsoftlags}
\end{figure}

\begin{figure*}
        \includegraphics[width=0.68\columnwidth]{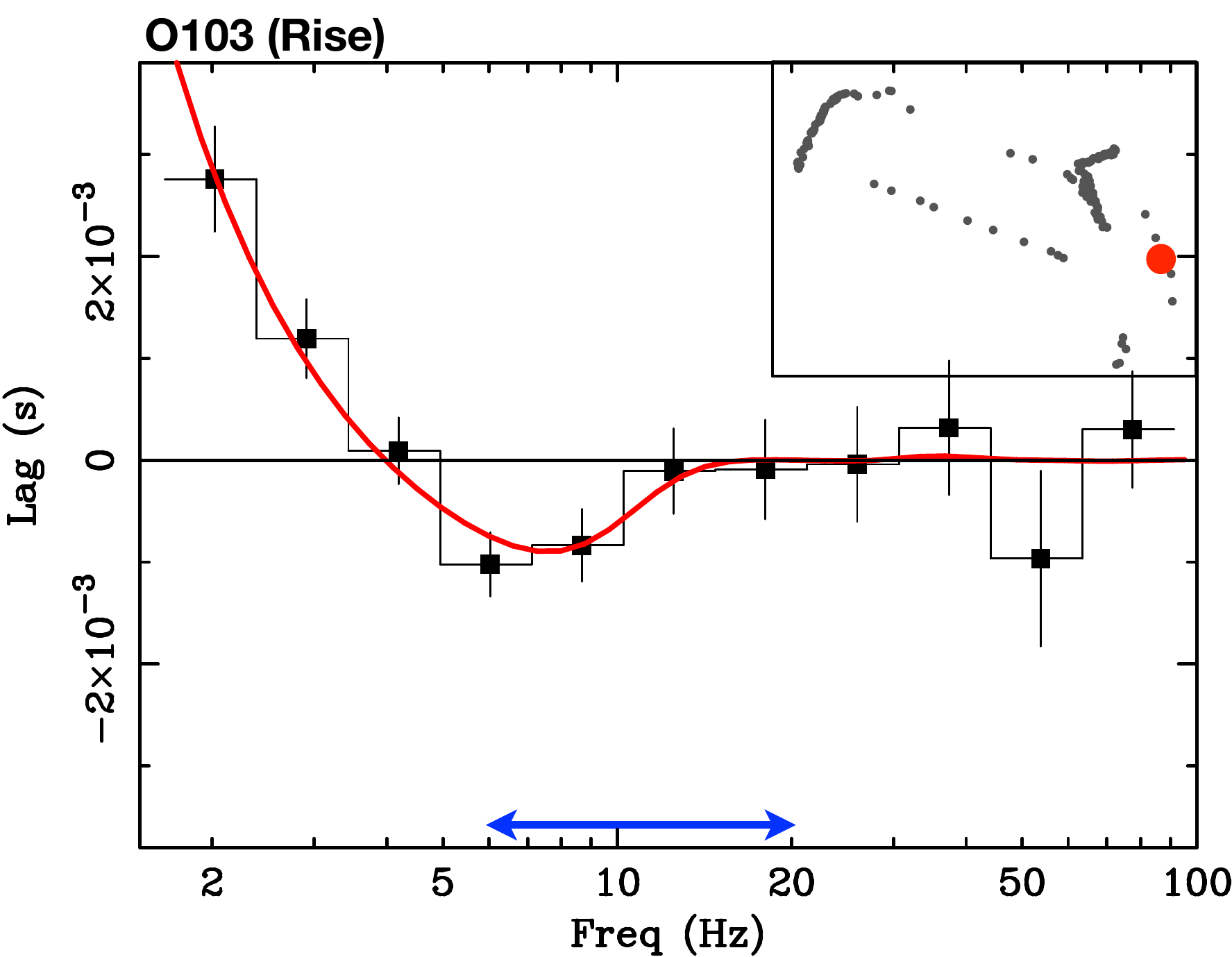}
        \includegraphics[width=0.68\columnwidth]{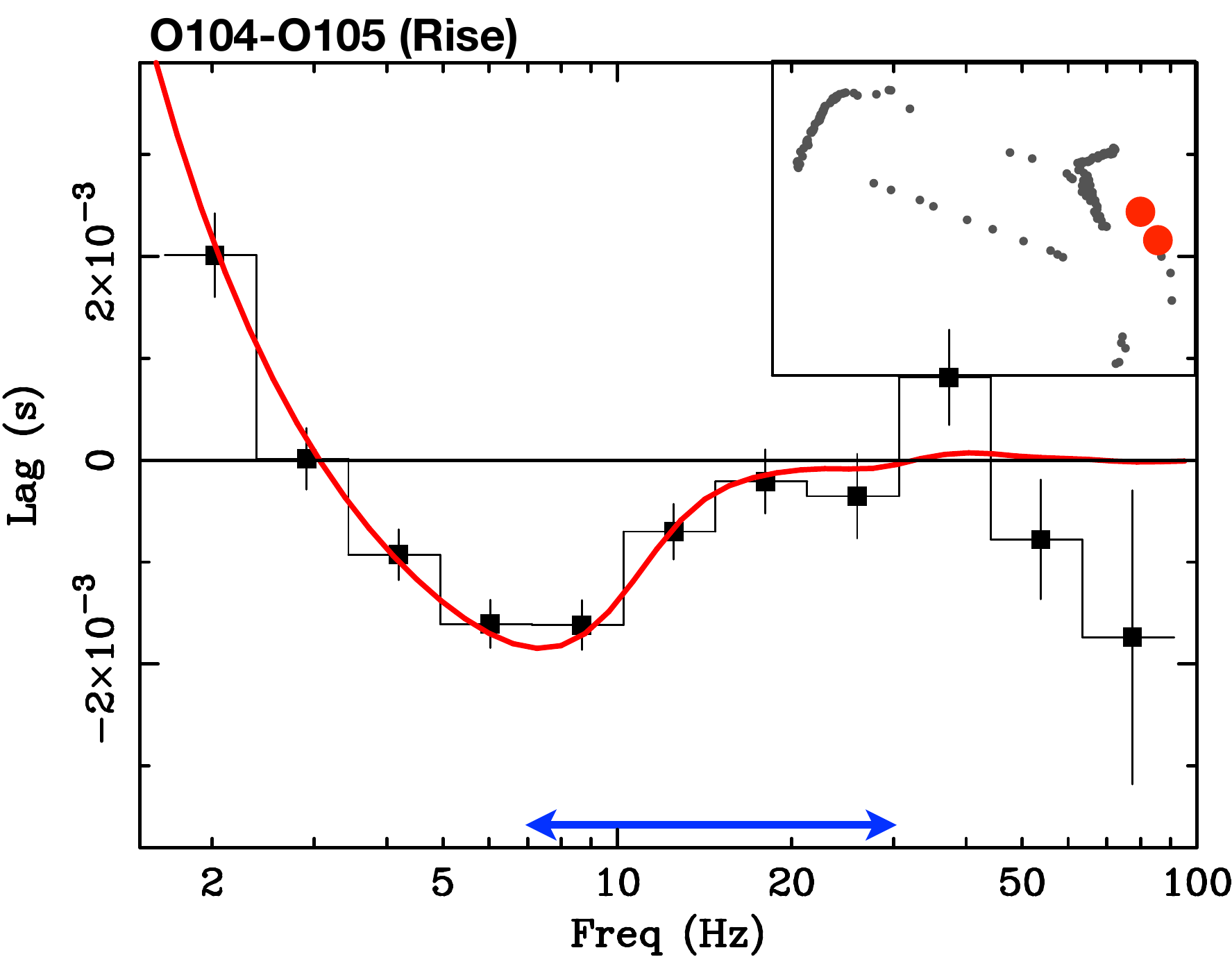}
        \includegraphics[width=0.68\columnwidth]{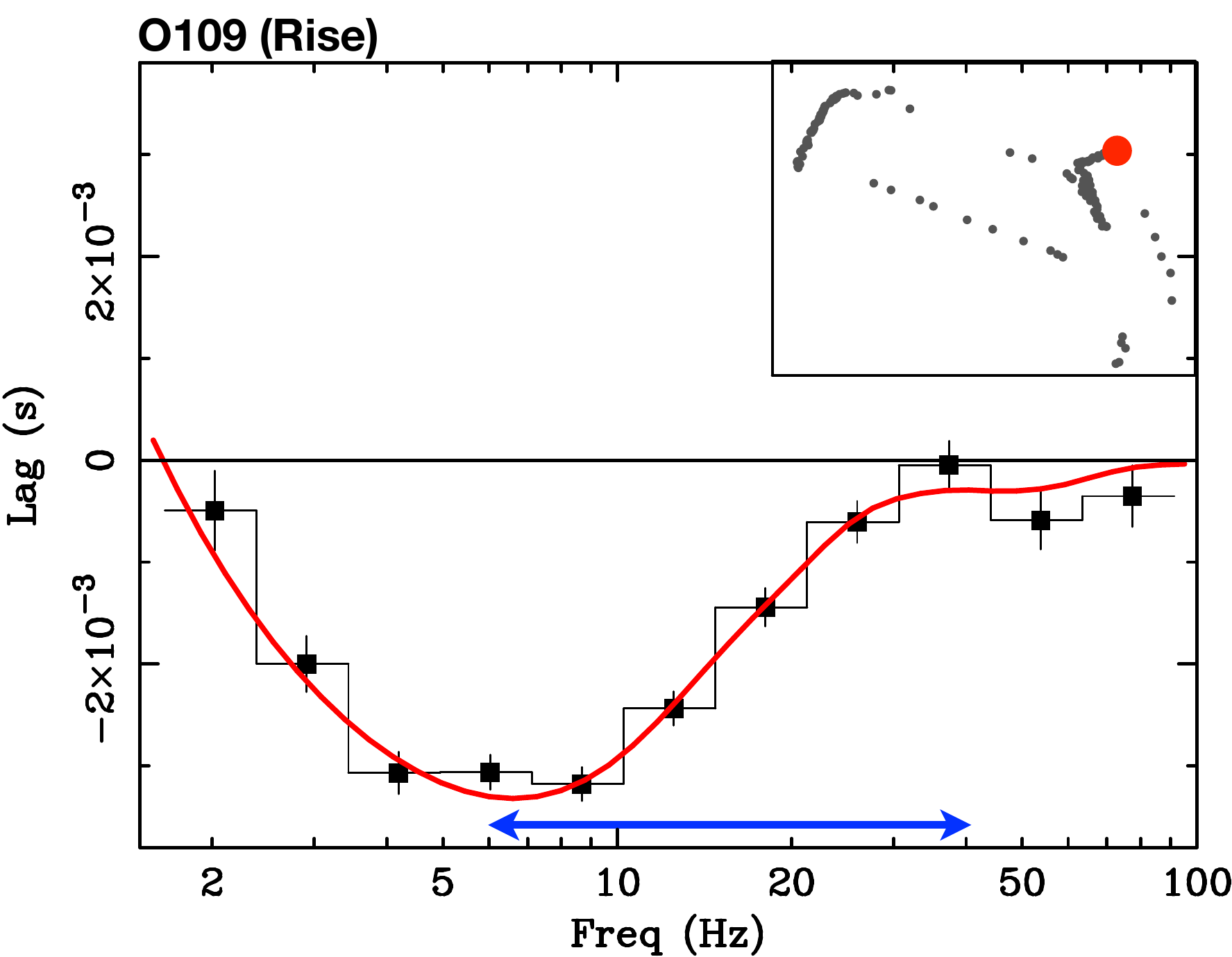}\\
        \includegraphics[width=0.68\columnwidth]{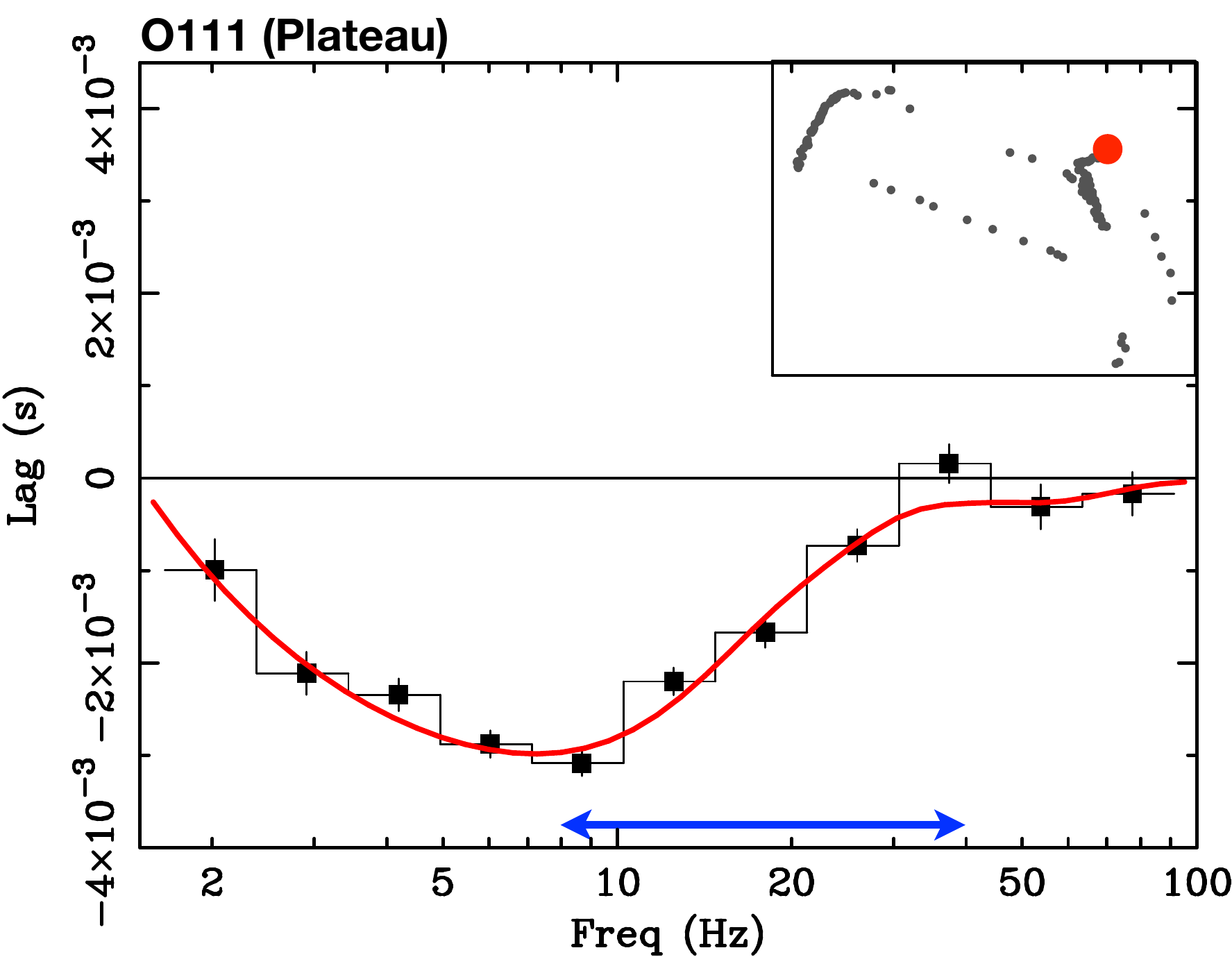}
        \includegraphics[width=0.68\columnwidth]{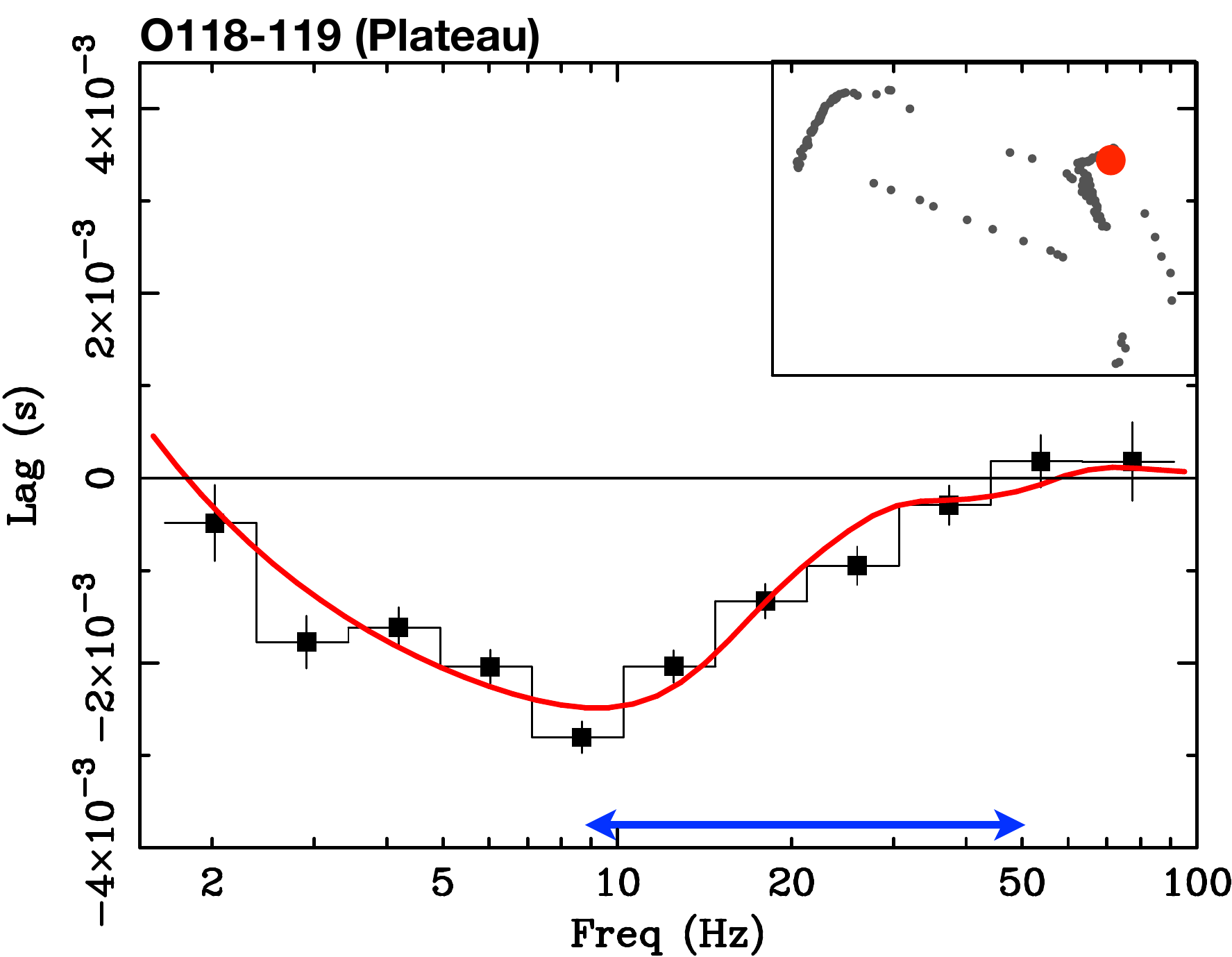}
        \includegraphics[width=0.68\columnwidth]{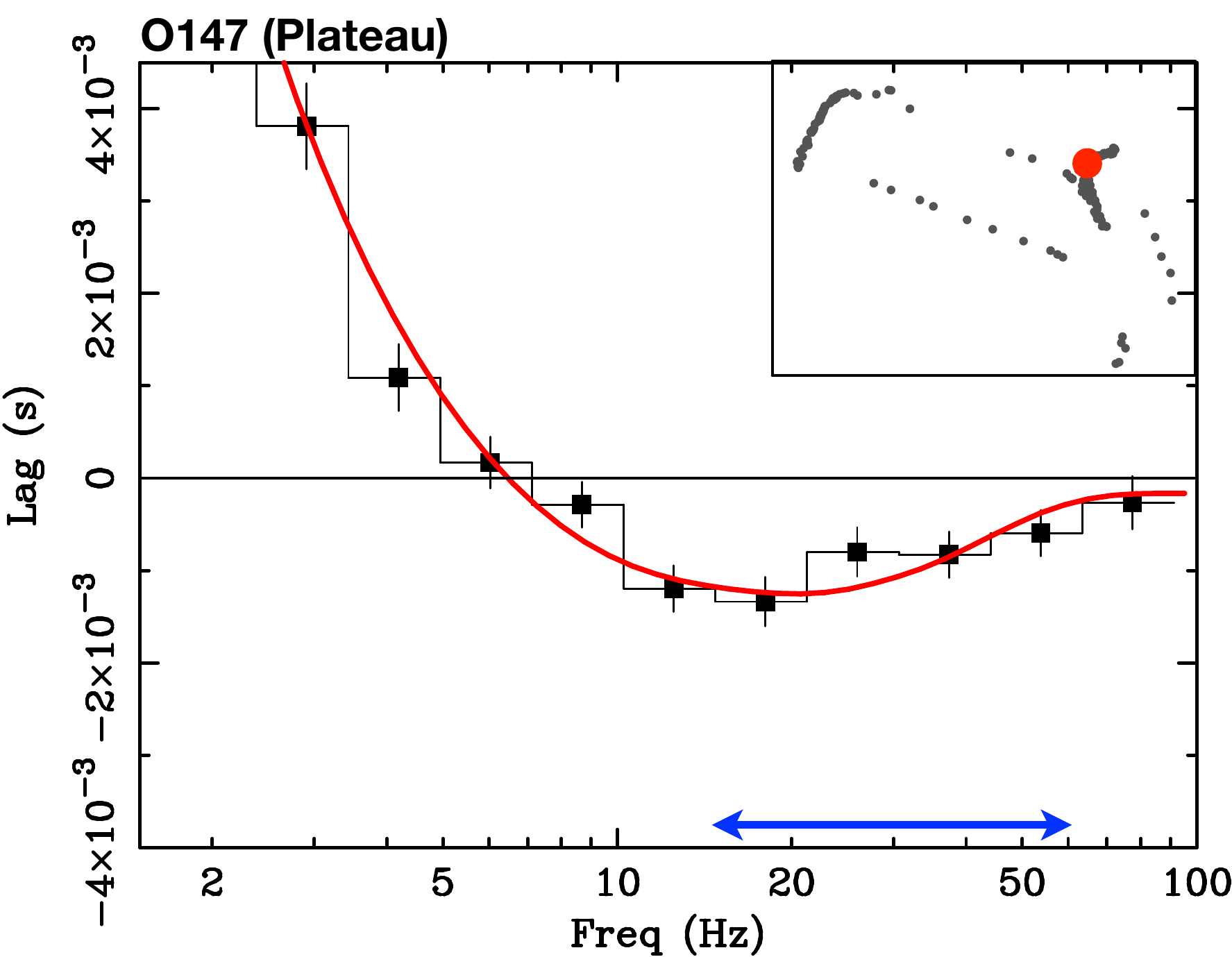}\\        
        \includegraphics[width=0.68\columnwidth]{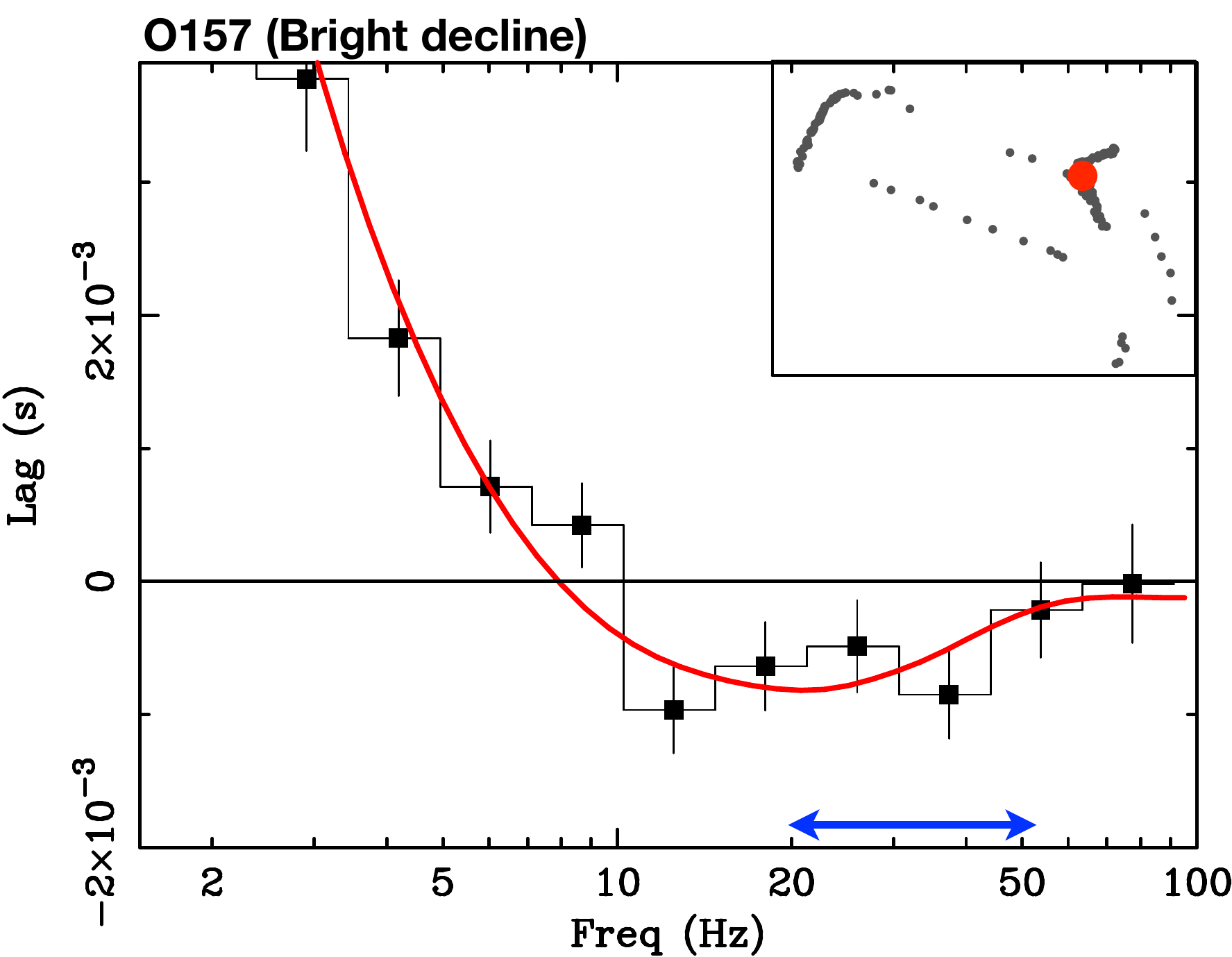}
        \includegraphics[width=0.68\columnwidth]{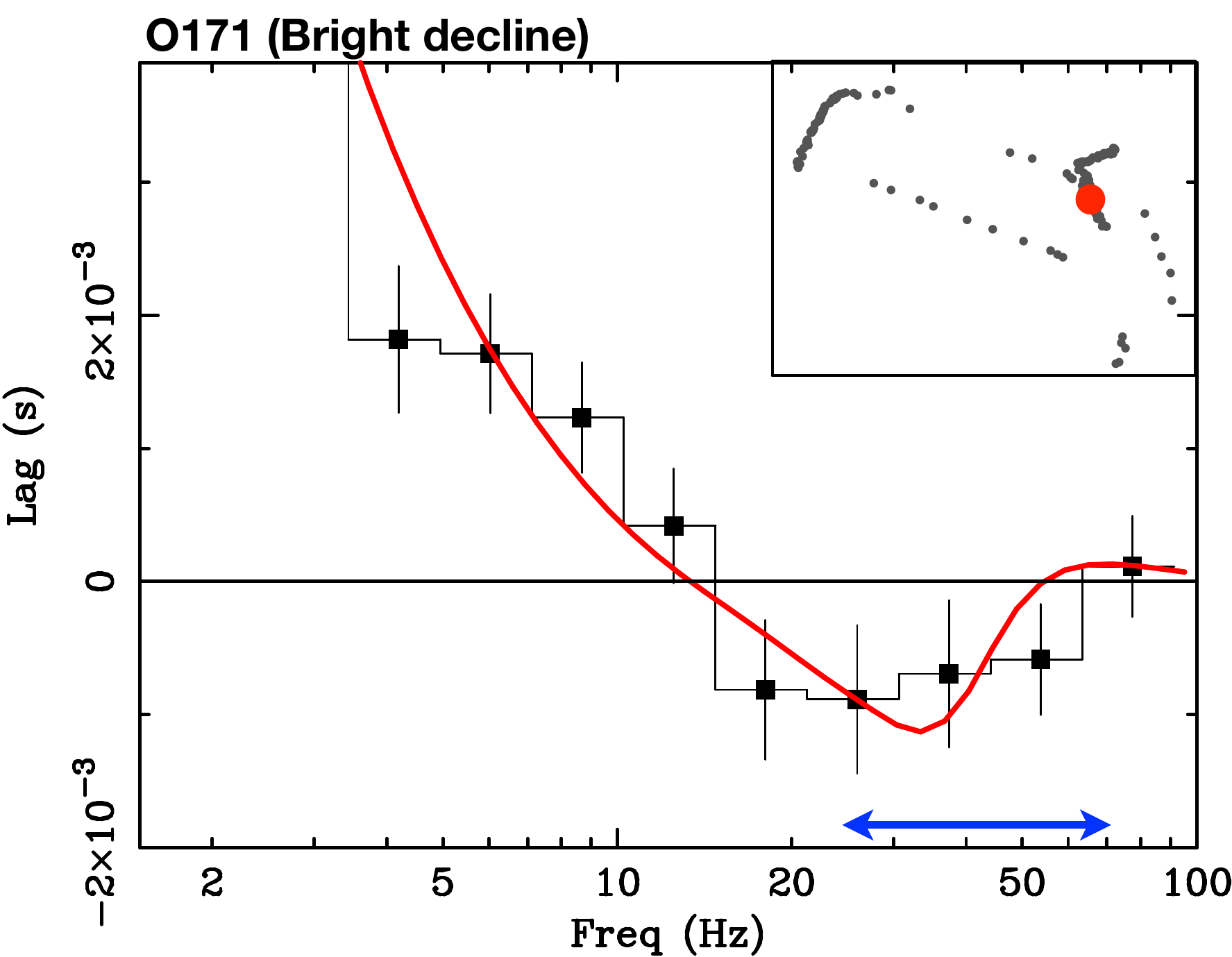}
        \includegraphics[width=0.68\columnwidth]{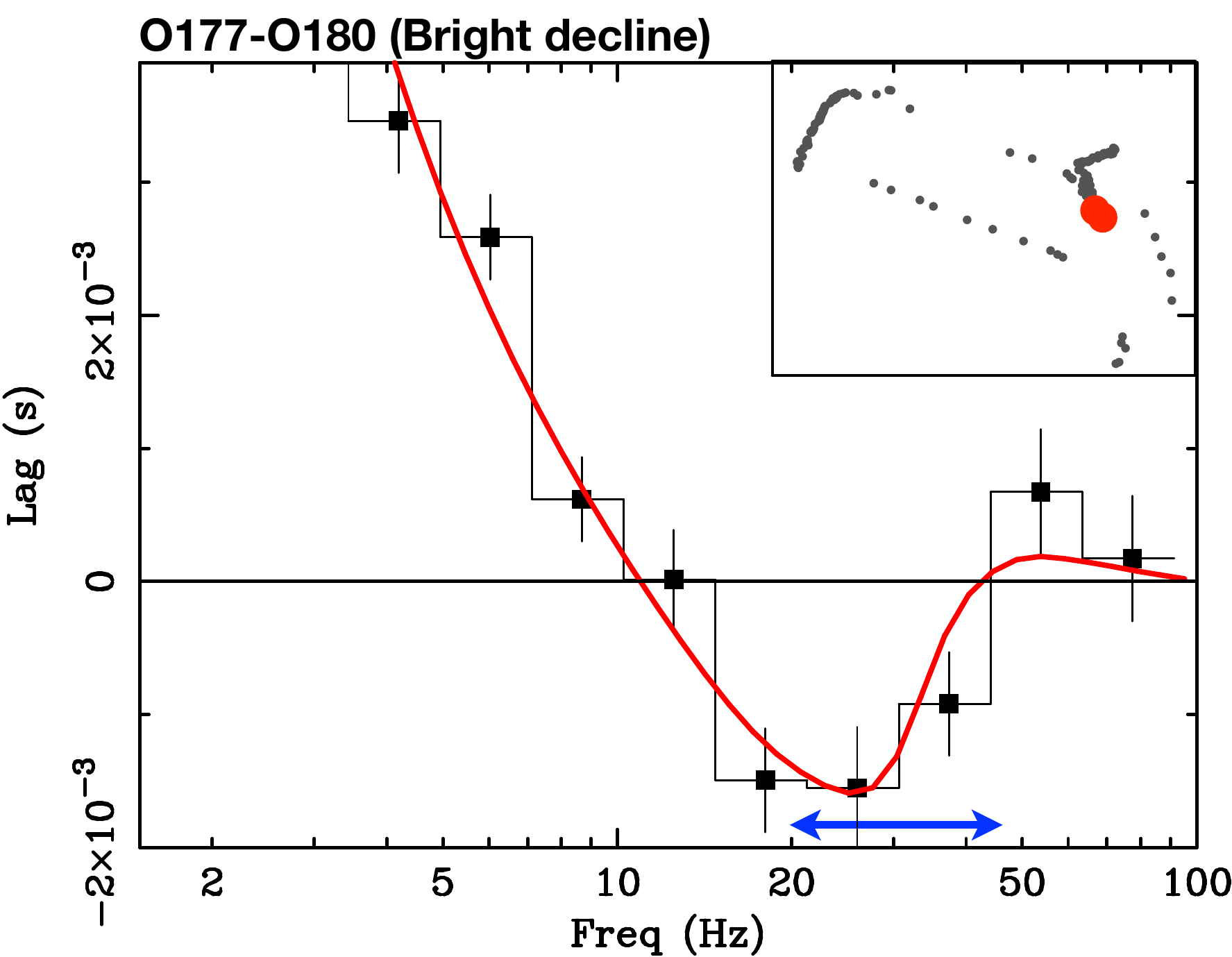}\\
        \includegraphics[width=0.68\columnwidth]{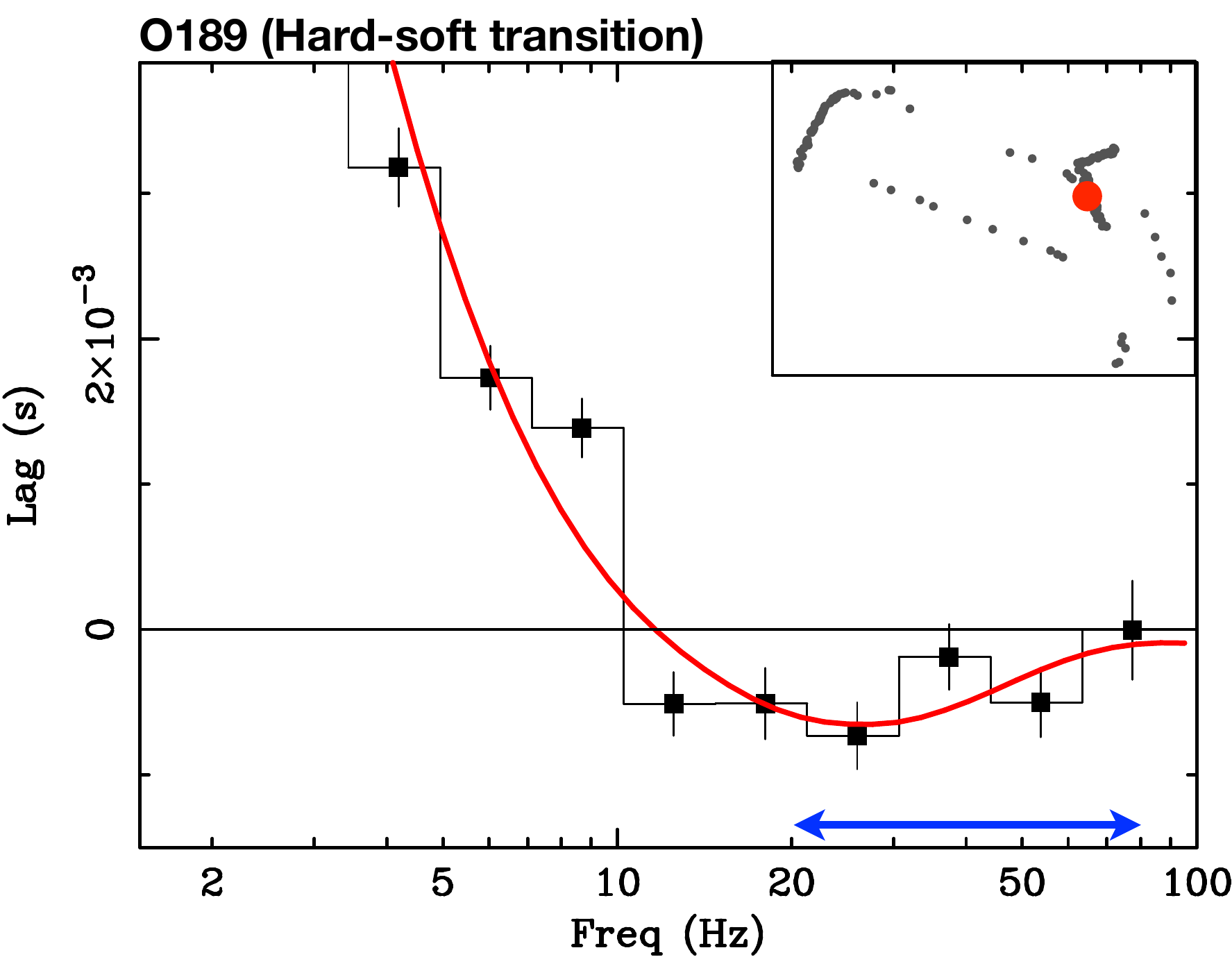}
        \includegraphics[width=0.68\columnwidth]{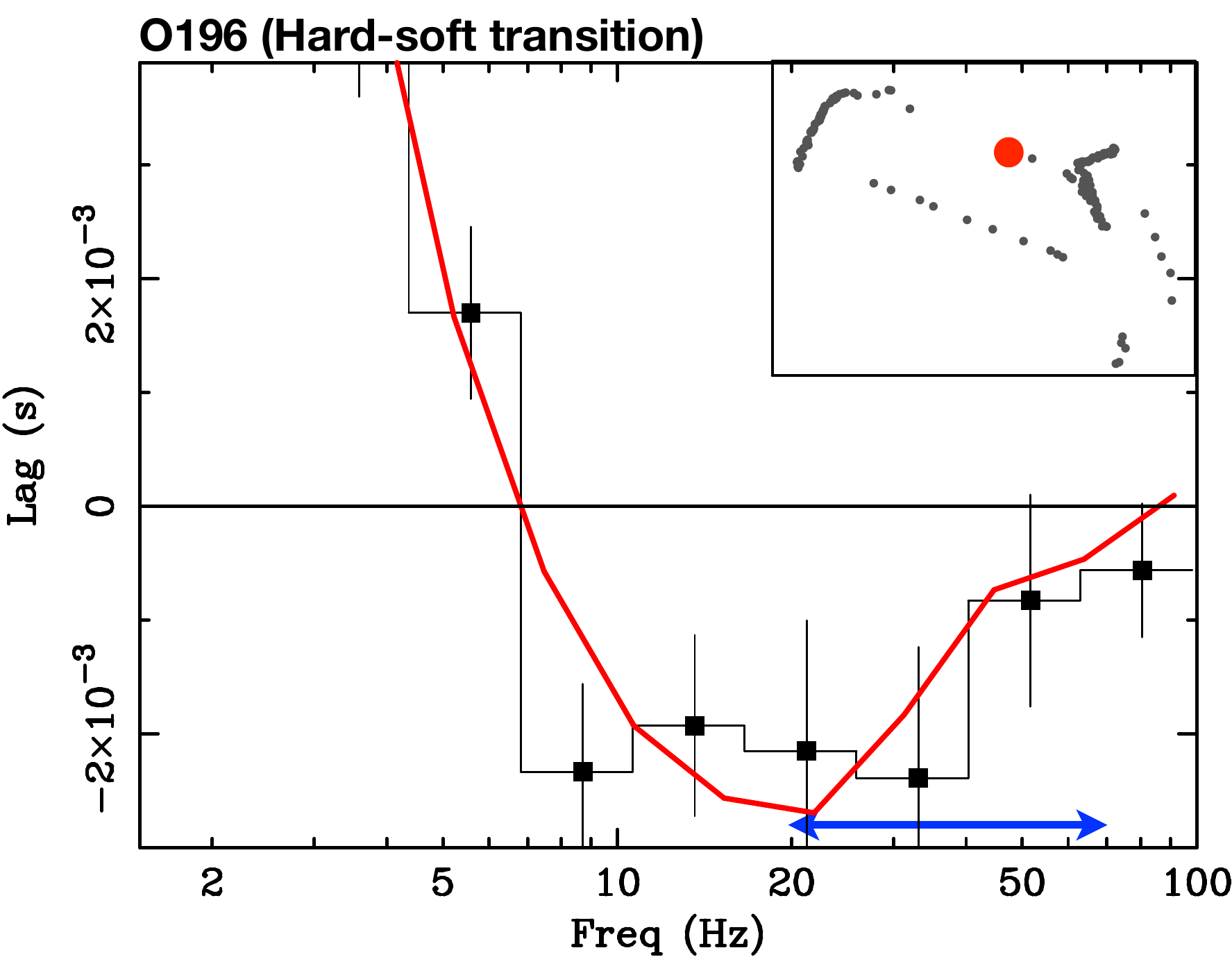}
        \includegraphics[width=0.68\columnwidth]{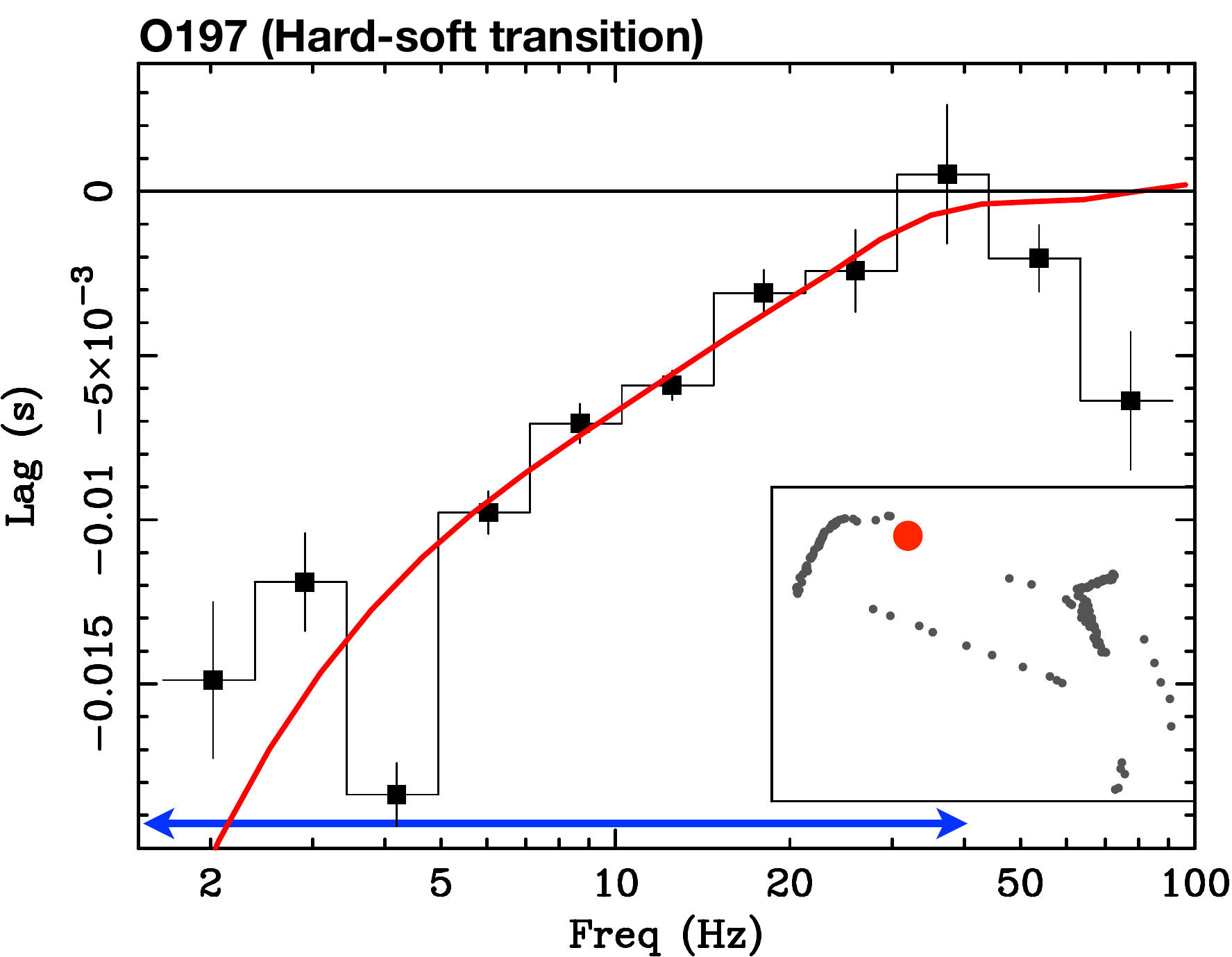}
    \caption{ 0.5--1 keV vs. 2--5 keV lag-frequency spectra of selected observations of MAXI J1820+070. The observations illustrate the diverse behaviour of soft (negative) lags during different phases of the outburst (as also reported in the insets). The solid red curve is the best-fit {\tt RELTRANS} model. The blue arrow marks the frequency range used for the extraction of covariance spectra. }
    \label{fig:softlags}
\end{figure*}

\subsection{The evolution of high-frequency soft reverberation lags}
\label{sec:softlags}

Soft X-ray lags (soft variations lagging behind hard variations) in BHXRBs are interpreted as a signature of thermal reprocessing in the disc. Light travel time delays due to the extra path between the X-ray source and the disc are thought to contribute significantly to the observed soft lags (thermal reverberation; e.g. \citealt{Uttley2011,BDM2015b}; \citeyear{BDM2017}). As a consequence, these lags can be used as a diagnostic of the geometry of the innermost accretion flow. 
Soft X-ray lags ascribable to thermal reverberation were consistently observed in the Plateau phase of MAXI J1820+070, with clear indications of an evolution of lag amplitude and  frequency \citep{Kara2019}.

We measured soft X-ray lags throughout the first part of the outburst of MAXI J1820+070. 
Following results reported in \cite{Zdziarski2021b} we selected a soft (0.5--1 keV) and a hard (2--5 keV) energy band. The selected soft band shows a soft excess that includes some contribution from disc emission \citep[figure 6 in ][]{Zdziarski2021b}, the hard band is instead dominated by a hard X-ray Comptonisation component. Although the analysis of \cite{Zdziarski2021b} does not cover all the phases considered here, we verified that these choices remain appropriate even in the latest phases.
We did not include photons with energies $>5$ keV to avoid strong contribution from Fe K line photons. 
We computed frequency-dependent time lags between the selected bands as $\tau(\nu)=\phi(\nu)/2\pi\nu$, where $\phi(\nu)$ is the phase of the cross spectrum as a function of frequency \citep[e.g.][]{Nowak1999}. We corrected the lags for dead-time-induced cross-channel talk \citep[][]{vanderKlis1987,Vaughan1999}, which introduces instrumental $\pm\pi$-phase lags. However, we verified that up to very high frequencies ($\sim100$ Hz) the contribution from the source dominates, thus making cross-channel talk negligible.

A number of examples among the obtained lag-frequency spectra are shown in Fig. \ref{fig:hardsoftlags} (which reports the broad-frequency band $\sim0.03-100$ Hz lag spectrum) and Fig. \ref{fig:softlags} (which focuses on the high-frequency $\sim 2-100$ Hz range behaviour). The lags are re-binned using a multiplicative re-binning factor of 1.2, with each frequency bin containing at least 500 points \citep[][]{Ingram2019b}.
At low frequencies ($\lsim$ 1--2 Hz) the lag-frequency spectra are dominated by (positive) hard X-ray lags (variations in hard X-ray photons follow variations in soft X-ray photons; see Fig. \ref{fig:hardsoftlags}). These lags are commonly observed in BHXRBs at frequencies $\lsim 1$ Hz and are thought to be intrinsic to the primary hard X-ray continuum \citep[e.g.][]{Nowak1999,Pottschmidt2000,Grinberg2014,Rapisarda2016}.\\
\begin{figure*}
        \includegraphics[width=\columnwidth]{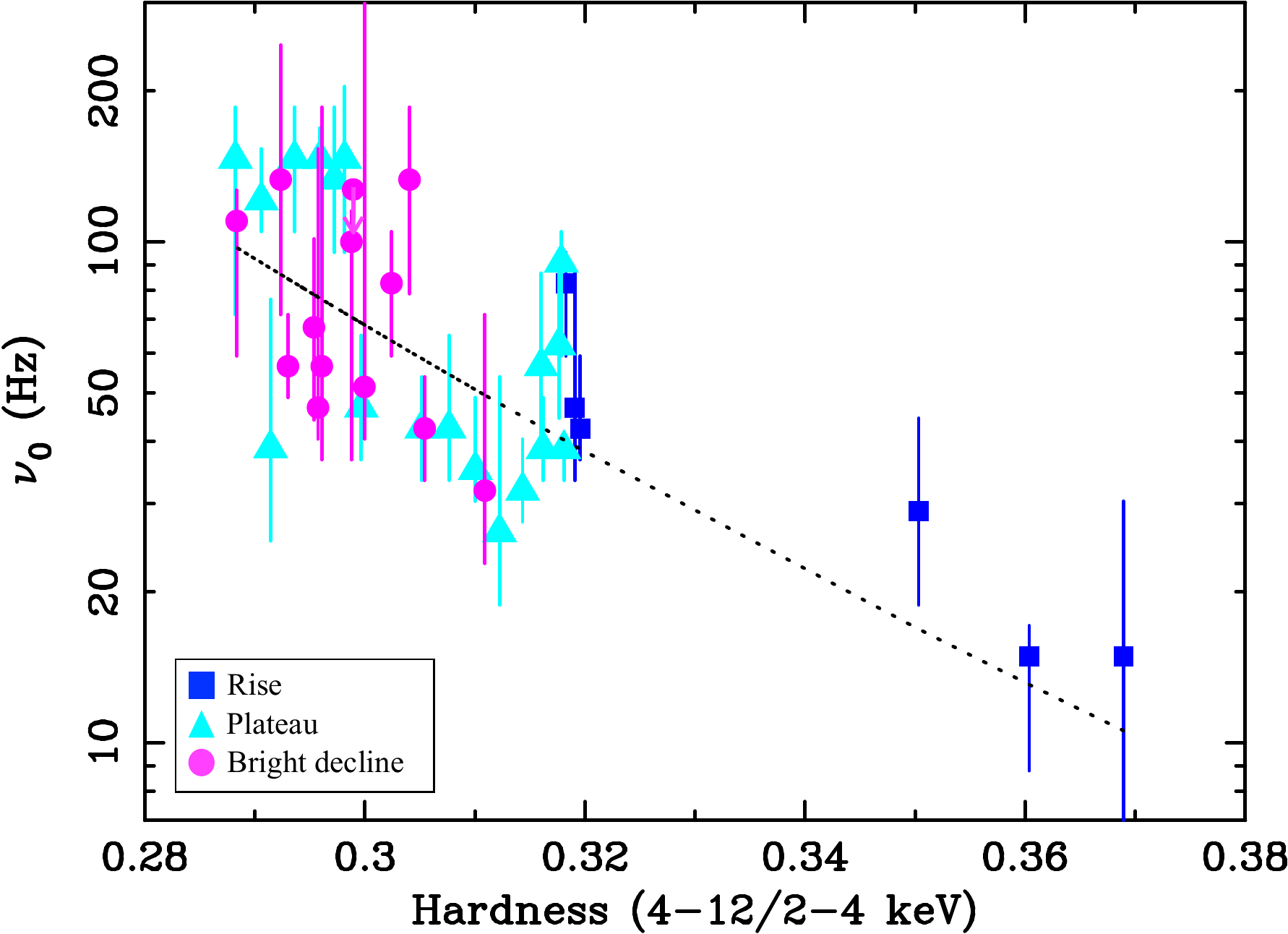}               
        \hspace{0.2cm}  
        \includegraphics[width=\columnwidth]{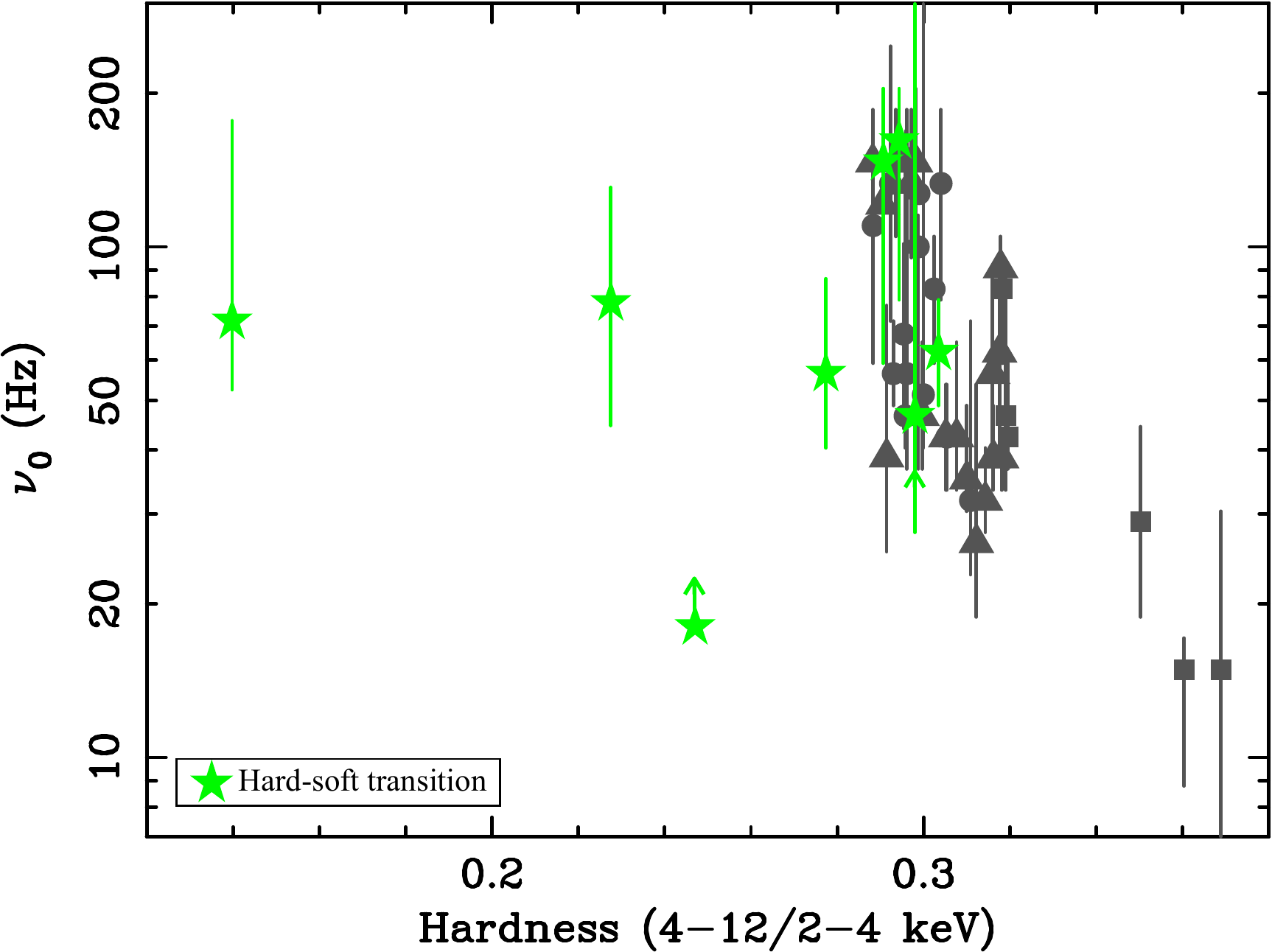}
    \caption{Evolution of $\nu_{0}$ in lag-frequency spectra as a function of spectral hardness. The \emph{left panel} reports results from the rise, plateau, and bright decline phases, which are observed to follow the same trend. The dotted black line is the best-fit model. The \emph{right panel} reports results from the hard-soft transition phase as compared to the other phases (grey symbols), which is observed to break from the observed trend at higher spectral hardness.The arrows are 90 per cent lower and upper limits.}
    \label{fig:nu0_vs_HR}
\end{figure*}
At higher frequencies a (negative) soft X-ray lag is clearly observed in almost all the analysed observations. Only during observations O101 and O102 the detection is marginal\footnote{For O101 and O102 we measured an average soft lag of $\tau=1.0\pm0.5$ ms and $\tau=1.5\pm0.4$ ms over the frequency interval $\sim$2.5--10 Hz and $\sim$5--20 Hz, respectively}.  
The soft X-ray lag typically starts to be observed at frequencies $\gsim 2$ Hz (inset in Fig. \ref{fig:hardsoftlags} and Fig. \ref{fig:softlags}), after getting to its maximum amplitude (in absolute value) it starts decreasing, reaching zero-amplitude at a given very high frequency (hereafter $\nu_{0}$).

As clear from Fig. \ref{fig:softlags}, soft X-ray lags appear to vary both in amplitude and in characteristic frequencies throughout the analysed observations. The maximum measured amplitudes range between $\sim 3$ ms and $\sim 0.2$ ms\footnote{We note that the measured values of lag amplitude are a factor of $\sim3$ longer than those reported in \cite{Kara2019}. This discrepancy has been verified by considering the same selected observations reported in \cite{Kara2019} and using the same energy bands. Details are reported in Appendix \ref{sec:lagampl}.}. 
However, the most compelling observed phenomenon is the net increase in lag amplitude in the last analysed dataset (O197), as compared to previous ones (Fig. \ref{fig:softlags}, bottom-right panel). Here the soft X-ray lag amplitude suddenly increases to $\sim$15--20 ms and dominates the entire sampled frequency range (see bottom panel of Fig. \ref{fig:softlags}), with no clear evidence of hard X-ray lags (down to $\sim$0.05 Hz). 
This dataset corresponds to the phases immediately before and during the transition as identified by the appearance of a type-B QPO \citep[][]{Homan2020}. However, the disappearance of the hard lags/appearance of a long $\sim$15--20 ms lag precedes the appearance of the type-B QPO. Indeed, we verified that the long, dominant soft X-ray lag is already present when considering only the $\sim50$ ks preceding the transition itself (i.e. the first $\sim 8$ ks of effective on-source time during O197, Sect. \ref{sec:HID}).
Such a long reverberation lag is observed also right after the transition from the soft to the hard state (from O262 to O265). This will be studied in more detail in a follow-up paper. It is worth noting that a much longer lag ($\sim$ 20 s) has been reported by \cite{Buisson2021} in a \emph{NuSTAR} observation performed on MJD 58306.35 (quasi-simultaneous to \nicer\ observation O198), that is, right after the transition to the soft-intermediate state.

The evolution of soft X-ray reverberation lags informs us about the evolution of the geometry of the system. From a visual inspection of the lag-frequency spectra, the observed maximum amplitude of the reverberation lag clearly shows a complex evolutionary pattern (the trend of observed maximum reverberation lag amplitude as a function of spectral hardness is shown and discussed in more detail in Appendix \ref {sec:lagampl}).
However, it is well known that the presence of both primary continuum and reprocessed photons in the energy bands used for the computation of cross-spectra has the effect of diluting the amplitude of soft lags \citep[e.g.][]{Uttley2014}. In particular, it can be easily shown that the amount of dilution also depends on the amplitude of the lags intrinsic to the primary X-ray continuum, being minimum when the continuum has no intrinsic interband lags. Nonetheless, the primary continuum in BHXRBs is characterised by (positive) hard X-ray lags, which show a complex evolution during the outburst \citep[e.g.][]{Pottschmidt2000,Grinberg2014,Altamirano2015,Reig2018}. As clear from Fig. \ref{fig:hardsoftlags}, this complexity characterises also MAXI J1820+070 (see also \citealt{WangY2020}). As a consequence, the hard X-ray lags in the continuum can bias the maximum observed amplitude of the reverberation lag, and their behaviour throughout the outburst can potentially distort the observed evolution of soft reverberation lags.

Nonetheless, the frequency at which the soft lag goes to zero, $\nu_{0}$, has been shown to be unaffected by dilution of continuum photons \citep[e.g. Fig.  8 of ][]{Wilkins2013,Uttley2014,Mizumoto2018}. This frequency is also unaffected by the presence of hard X-ray lags associated with the primary continuum, because these lags are commonly observed to steadily decrease with frequency, eventually becoming negligible \citep[e.g.][]{Nowak1999,BDM2015b}. This behaviour is also seen in MAXI J1820+070, as shown in Fig. \ref{fig:hardsoftlags}. Thanks to the suppression of hard X-ray lags towards high frequencies, $\nu_{0}$ directly depends on the reverberation lag intrinsic (and unbiased) amplitude as $\nu_{0}\propto \tau_{\rm rev}^{-1}$. We note that $\nu_{0}$ is unbiased by the presence of hard lags even when these lags are not totally suppressed at high frequencies (see Appendix \ref{sec:nuzero}). Therefore, $\nu_{0}$ can be used to study, in a model-independent way, the evolution of soft reverberation lags.

In order to constrain $\nu_{0}$ in an objective way we modelled the frequency-dependent lag profile. Because of the marginal detection of soft X-ray lags during O101 and O102, and given that no significant differences in the lag are observed, only for this part of the analysis these two observations were combined. 
For the fits we used the {\tt RELTRANS} spectral-timing model (\citealt{Ingram2019c}) in the version presented in \citeauthor{Mastroserio2019} (\citeyear{Mastroserio2019,Mastroserio2020}). 
This model assumes a lamppost geometry for the source of Comptonised hard X-ray photons and describes the hard X-ray lags as due to spectral pivoting of the inverse-Compton spectrum \citep[][]{Mastroserio2018,Mastroserio2019}. Hard X-ray lags generally show hump-like features at specific frequencies which might be associated with a more structured Comptonising region (e.g. \citealt{Mahmoud2018a}; \citeyear{Mahmoud2018b}). 
These complexities are clearly observed in the lag-frequency spectra of MAXI J1820+070 (see Fig. \ref{fig:hardsoftlags}) and did not allow us to always obtain good fits of both the hard and soft lags on a broad range of frequencies.
Therefore, we excluded most of the hard lags complex profile from the fits, restricting the modelling of the lag-frequency spectra to frequencies ranging between $\sim1-2$ Hz and $100$ Hz. For O197 we used the frequency range $6-100$ Hz because of the anomalous behaviour observed at lower frequencies (lack of visible hard lags) that did not allow us to obtain good fits over a broader frequency range. Since this observation catches the source right before and during the transition, it is likely to present additional spectral-timing complexities not included in current models.

We note that, for a given value of the BH mass, the parameters of the model influencing the position of $\nu_{0}$ are the X-ray source height, $h$, and the disc inner radius, $R_{\rm in}$. However, these parameters are degenerate when fitting lag-frequency spectra alone. For this reason, and because the geometry of the Comptonising region might be more complex than assumed in the model, we do not discuss the inferred best-fit values for these parameters, and use the best-fit models only to obtain a phenomenological description of the soft X-ray lag profile. Additional technical details of these fits are reported in Appendix \ref{sec:lagfits}.

The derived best-fit models provide a satisfactory description of the high-frequency lag spectrum at each epoch (see the examples in Fig. \ref{fig:softlags}, where the red solid curve represents the best-fit model).     
We estimated $\nu_{0}$ as the frequency at which the best-fit model approaches zero at high frequencies\footnote{We noted that, given the low S/N of some observations at high frequencies, the errors on $\nu_0$ were not well constrained because the model that defines the 90\% confidence upper limit tends asymptotically to zero. Therefore, we set an arbitrarily small threshold of $\lsim0.05$ ms for the absolute amplitude of the negative lag, below which we consider the model to be consistent with zero lag. We verified that assuming even smaller values for this threshold returns consistent results.}. 
The errors on $\nu_{0}$ are obtained by accounting for the uncertainty in the parameters of the model (as described in Appendix \ref{sec:lagfits}).
The derived values of $\nu_{0}$ are listed in Table \ref{tab2}. These values are plotted as a function of spectral hardness in Fig. \ref{fig:nu0_vs_HR}.

For most of the analysed observations (more specifically those corresponding to the rise, plateau, and bright decline phases), the data show a steady trend of increasing $\nu_{0}$ with decreasing spectral hardness (Fig. \ref{fig:nu0_vs_HR} left panel). A linear model in log-log space to describe this trend in the rise, plateau, and bright decline is preferred to a constant model at $>99.99$ per cent confidence level (i.e. corresponding to a $\Delta\chi^2\sim154$ for a difference of 1 degree of freedom; Fig. \ref{fig:nu0_vs_HR} left panel).

Notably this trend appears to break during the hard-intermediate states preceding the transition (hard-soft transition phase in Fig. \ref{fig:nu0_vs_HR}, right panel). Indeed, $\nu_{0}$ is observed to decrease by a factor of $\sim$2--3 at spectral hardness $\lsim 0.28$, remaining systematically lower than observed in most of the bright hard state. During observation O197 (right before and during the transition) we estimate $\nu_{0}=71^{+105}_{-19}$ Hz (Table \ref{tab2}).

An increase in $\nu_{0}$ (as characteristic of most of the analysed part of the outburst, Fig. \ref{fig:nu0_vs_HR}, left panel) indicates a decrease in the intrinsic lag amplitude, and thus a decrease in the relative distances between the Comptonising plasma and the reprocessing region in the disc. 
This might be due to the inner disc truncation radius moving inwards or the Comptonising region decreasing in its height or vertical extent. The behaviour close to transition (Fig. \ref{fig:nu0_vs_HR}, right panel) is instead more puzzling and might be associated with the presence of a jet \citep[][see Sect. \ref{sec:discussion1}]{Bright2019,Homan2020,Espinasse2020}.

The intrinsic amplitudes that can be inferred from $\nu_{0}$ correspond to Euclidean distances of the order of several tens-to-hundreds of $R_{\rm{g}}$ (see also discussion in Sect. \ref{sec:discussion2}). However, these distances represent the weighted mean over all the light travel paths of the Comptonised photons to the disc \citep[e.g.][]{Wilkins2013}, and can be significantly larger than the minimum travelled distance set by the X-ray source height and the inner disc radius (\citealt{Mahmoud2019}; see also Sect. \ref{sec:discussion2}). The distribution of light travel paths depends on the geometrical parameters of the disc and the Comptonising plasma, and is encoded in the impulse-response function \citep[e.g.][]{Gilfanov2000,Poutanen2002,Wilkins2013}. However, simultaneous fits of cross-spectra and time-averaged spectra, possibly with models that account for more  complicated geometries for the X-ray source, are needed in order to constrain the impulse-response function.
In the following section we employ an alternative technique to obtain tighter constraints on the geometry of the inner accretion flow responsible for the observed soft reverberation lags in MAXI J1820+070.

\subsection{Constraining the inner radius of the disc from covariance spectra}
\label{sec:covar}

The hard X-ray flux that irradiates the inner parts of the accretion disc is partly reflected (i.e. producing absorption features and emission lines plus a Compton scattering continuum) and partly reprocessed as a quasi-thermal continuum in the soft X-ray band. The latter adds up to the disc black body component due to internal dissipation \citep[e.g.][]{Gierlinski2008,ZdziarskiBDM2020} and is expected to give a significant contribution in accretion states energetically dominated by the hot Comptonising plasma. Therefore, the observed disc black body temperature at the inner radius of the disc is $T_{\rm in}=f_{\rm col} T_{{\rm eff}}$, where $T_{{\rm eff}}$ is the effective temperature due to the internal dissipation and irradiation processes (such as $T^4_{\rm{eff}}= T^4_{\rm{eff, dis}}+T^4_{\rm{eff, irr}}$), and $f_{\rm col}>1$ is the colour-correction factor accounting for spectral hardening \citep[e.g.][]{Ebisawa1993,Shimura1995,Davis2005,Salvesen2021}.

As discussed in  \cite{ZdziarskiBDM2020} and \cite{Zdziarski2021a}, $T_{\rm in}$ can be estimated from the Stefan-Boltzmann law, as:
\begin{eqnarray}\label{eq:S-B_law}
T_{\rm in} \gsim f_{\rm col} T_{\rm{eff, irr}} = f_{\rm col} \left[\frac{(1-a)F_{\rm{\rm irr}}(R_{\rm in})}{\sigma}\right]^{1/4}
,\end{eqnarray}
where $a$ is the disc albedo, $(1-a)$ is the fraction of irradiating flux reprocessed as a quasi-thermal component, $F_{\rm{irr}}$ is the total irradiating hard X-ray flux impinging on the inner radius of the disc $R_{\rm{in}}$, and $\sigma$ is the Stefan–Boltzmann constant.

Assuming a standard emissivity profile for the X-ray source (i.e. $\propto R^{-q}$, with the emissivity index $q=3$) and a small scale height compared to the disc radius (such as in a coronal geometry or a lamppost relatively close to the BH horizon, this assumption is discussed in Sect. \ref{sec:discussion1}), the flux irradiating the inner edge of the disc is $F_{\rm irr}(R_{\rm in})\propto R_{\rm in}^{-2}$ \citep[][]{ZdziarskiBDM2020}. Therefore, following \cite{ZdziarskiBDM2020}, Eq. \ref{eq:S-B_law} translates into a lower limit on $R_{\rm in}$:
\begin{eqnarray}\label{eq:S-B_law2}
\frac{R_{\rm in}}{R_{\rm g}}  \gsim  10 \frac{\mathcal{R}^{1/2}(1-a)^{1/2} l_{\rm irr}^{1/2}}{(kT_{\rm eff}/1\ {\rm keV})^2 (M/10M_{\odot})^{1/2}}
,\end{eqnarray}
where $l_{\rm irr}=4\pi d^2 F_{\rm irr}/L_{\rm Edd}$ (assuming isotropy) is the total (i.e. including contribution from both variable and constant components) Eddington-scaled (with $L_{\rm Edd}$ for pure hydrogen) luminosity of the irradiating hard X-ray source, and $k$ is the Boltzmann constant. The factor $\mathcal{R}$ is the reflection fraction and was introduced by \cite{ZdziarskiBDM2020} to account for the possible reduction of hard X-ray flux impinging on the disc (e.g. as a consequence of anisotropic emission; see discussion in Sect. \ref{sec:discussion2}). 
The approximate equality in Eq. \ref{eq:S-B_law2} holds for the case of null disc internal dissipation ($T^4_{\rm{eff}}\sim T^4_{\rm{eff, irr}}$). Under this condition an estimate of the disc truncation radius can be obtained. We note that this additional quasi-thermal component is incorporated in state-of-the-art reflection spectra \citep[e.g.][]{Tomsick2018}, but for the remainder of the paper we fit it as a separate component in order to constrain its temperature.

The quasi-thermal emission due to irradiation responds to the fast variability of the hard X-ray flux (thermal reverberation), as inferred from the detection of high-frequency soft X-ray lags (Sect. \ref{sec:softlags}). Previous analyses showed that the thermal response of the disc to variable irradiation also produces a soft, disc black body-like component on top of the hard X-ray Comptonisation continuum in high-frequency covariance spectra \citep[][]{Wilkinson2009,Uttley2011}.
These spectra single out components varying in a linearly correlated way (coherently) and produced in the innermost regions of the accretion flow, thus can be used to study the quasi-thermal emission produced via irradiation of the inner disc.

We estimated the temperature of the quasi-thermal component responding to hard X-ray irradiation of the inner disc, from the fits of high-frequency covariance spectra ($kT_{\rm in,covar}$). We emphasise that these are the frequencies where the soft X-ray lags ascribed to thermal reverberation are observed, thus allowing us to constrain the geometry of the region of the disc producing these delays. Indeed, the temperature inferred for this region was used together with Eq. \ref{eq:S-B_law2} in order to obtain constraints on the inner radius of the irradiated disc.
In order to extract the covariance spectra we chose a broad energy band as reference (0.5--10 keV) and computed cross-spectra between the reference band and a series of adjacent energy bins. We followed the standard approach of removing the contribution of each energy bin from the reference band before computing each cross spectrum (see \citealt{Uttley2014} for details, and \citealt{Ingram2019b} for a different approach). The cross-spectra were averaged over a frequency interval of interest. Since we are interested in the thermal reprocessing component, we considered the frequency intervals where soft X-ray reverberation lags are detected (Sect. \ref{sec:softlags}). In order to exclude as much contribution as possible from the process responsible for the hard X-ray lags (which produces correlated disc and hard X-ray continuum variability at low frequencies; e.g. \citealt{Wilkinson2009,Uttley2011,BDM2015a}), we focused only on the highest sampled frequencies. Specifically, we averaged the cross-spectra over the frequencies that range between the maximum observed soft lag amplitude and $\tau\sim\tau(\nu_0)$ (the frequency intervals for the lags displayed in Fig. \ref{fig:softlags} are indicated by the blue arrows). In general, the tested frequencies range between $\sim$6 Hz and $\sim$ 90 Hz. Since $kT_{\rm in,covar}$ is expected to increase with frequency (as higher frequencies test regions of the disc closer to the BH), we verified that even limiting our analysis to the highest-frequency end of the chosen frequency intervals does not influence significantly the results reported here.

\begin{figure}
        \includegraphics[width=\columnwidth]{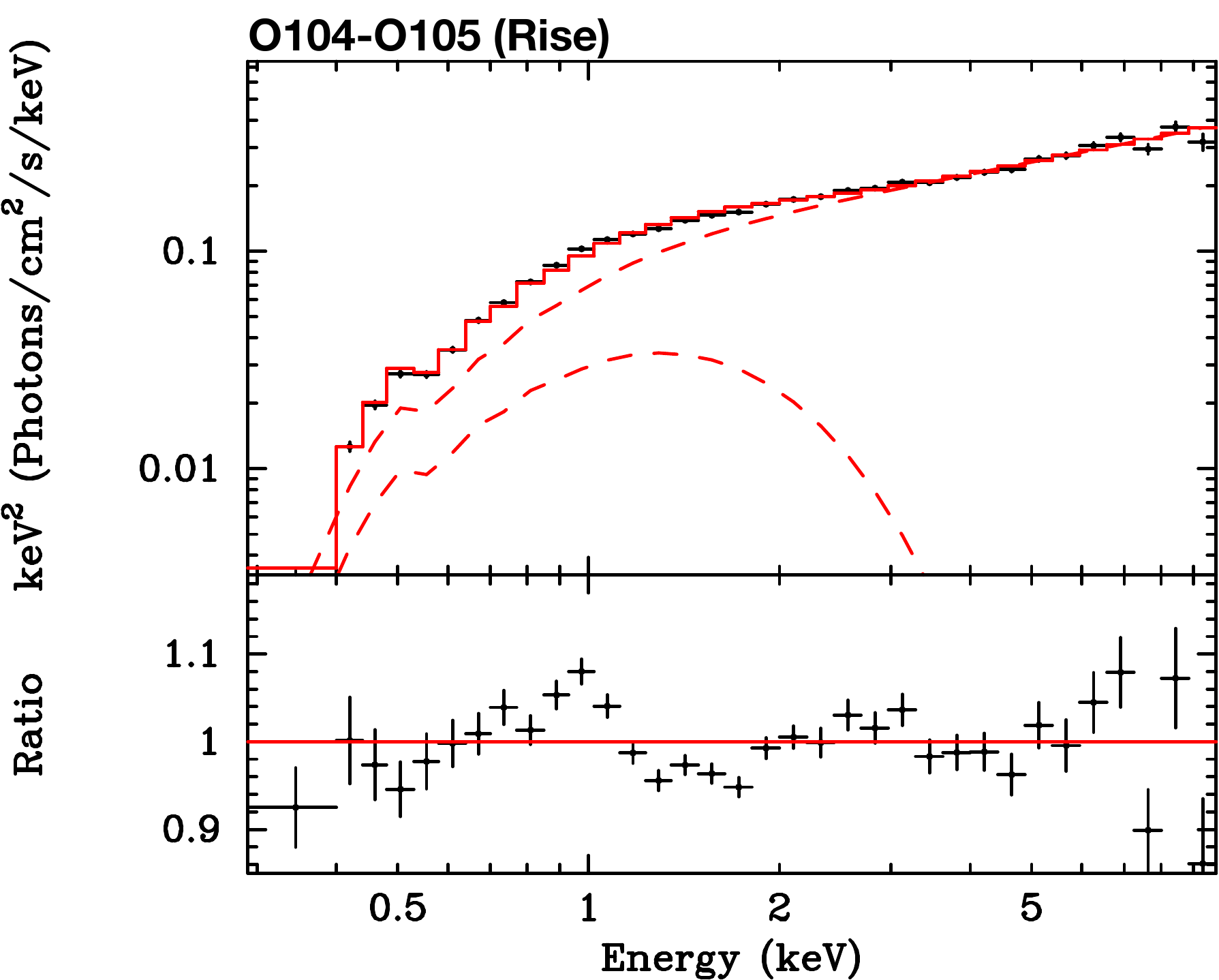}
    \caption{Unfolded best-fit \texttt{TBabs[deltadisk + nthComp]} model for the high-frequency ($7-30$ Hz) covariance spectrum of combined observations O104-O105. The residuals observed at $\sim$1--2 keV are likely due to \nicer's calibration systematics (\nicer's effective area shows features due to O, Si, and Au at 0.5, 1.8, and 2.2 keV).}
    \label{fig:covarO104105}
\end{figure}

We first fit the covariance spectra of each observation at energies $E\geq3$ keV with a Comptonisation model, absorbed by a column of cold gas (\texttt{TBabs*nthComp} in Xspec, \citealt{Zdziarski1996,Zycki1999,Wilms2000}). The seed photons temperature was fixed at 0.2 keV, the photon index, and normalisation were left free to vary. Since \nicer\ data are not sensitive to the Comptonisation high energy cut-off, the temperature of the Comptonising plasma was fixed at $kT_{\rm e}=30$ keV \citep[][]{Zdziarski2021b}. The column density was fixed at $N_{\rm H}=1.4\times10^{21}\rm{cm^{-2}}$ (\citealt{Kajava2019}; Dzie{\l}ak et al. 2021). The best-fit models were then extrapolated down to $E=0.3$ keV, revealing significant residuals at soft energies. 
In order to fit such residuals, we added a quasi-thermal contribution from a disc component. However, since we are fitting covariance spectra (which are essentially obtained from the difference between the time-dependent spectrum and the time-averaged spectrum of the source, \citealt{Wilkinson2009}) we need to account for the fact that the observed thermal emission in such spectra deviates from that of a disc blackbody \citep[][]{vanParadijs1986}. In other words, fitting covariance spectra with a disc blackbody model such as \texttt{diskbb} in Xspec \citep[][]{Makishima1986} would lead to an overestimate of the intrinsic disc temperature. Therefore, following \cite{vanParadijs1986}, within Xspec we defined the model \texttt{deltadisk}\footnote{\texttt{deltadisk}$=$\texttt{diskbb}$(kT_{\rm in,covar})-$\texttt{diskbb}$(kT_{\rm in})$\\$\approx (\partial {\tt diskbb}/\partial T_{\rm in})\vert_{T_{\rm in,covar}}\Delta T_{\rm in}$.} that represents the derivative of the \texttt{diskbb} model over $T_{\rm in}$. This corresponds to computing the variable spectrum as a small perturbation of the time-averaged one. In this model, the temperature of the disc in covariance spectra $T_{\rm in,covar}$ was left free to vary, and the seed photons temperature of \texttt{nthComp} was tied to $T_{\rm in,covar}$. 
 The fits were performed in the 0.3--10 keV band, simultaneously for groups of consecutive observations belonging to the same phase of the outburst (as reported in Table \ref{tab1} and Fig. \ref{fig:hid}), thus resulting in a total of four groups. Due to its short exposure, O168 did not allow us to constrain the parameters of the model well, so it was discarded. Results of the fits are reported in Table \ref{tab2} for each fitting group. As an example, the best-fit model to high-frequency ($7-30$ Hz) covariance spectra of combined observations O104-O105 is shown in Fig. \ref{fig:covarO104105}.

The \texttt{TBabs[deltadisk + nthComp]} model is statistically preferred to the simple Comptonisation model in the joint fits for each phase of the outburst (for the four fitting groups the $\chi^2$ improves by 128, 167, 74, and 71, respectively, for a difference of 7, 18, 12, and 8 in the number of degrees of freedom). 
These results point to the presence of linearly correlated variability between the primary source of hard X-ray photons and the disc on the short timescales corresponding to the sampled frequencies. This is expected if the excess soft emission is produced via hard X-ray irradiation of the inner disc, and the process is linear at least to first order (Sect. \ref{sec:discussion}). This is also in agreement with the interpretation of the soft X-ray lags detected in MAXI J1820+070 as due to thermal reprocessing (Sect. \ref{sec:softlags}).

\begin{figure*}
        \includegraphics[width=\columnwidth]{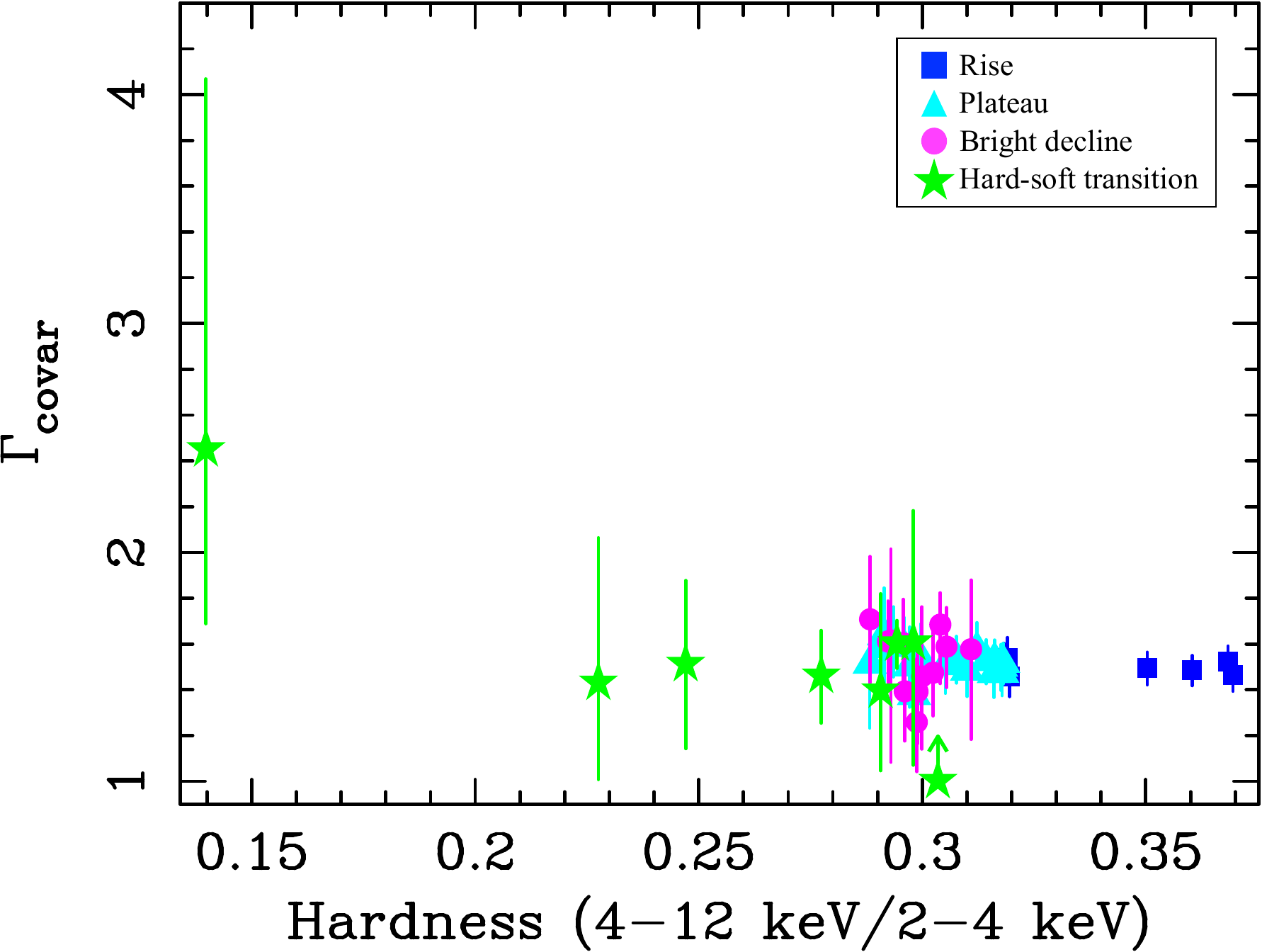}
        \hspace{0.2cm}  
        \includegraphics[width=\columnwidth]{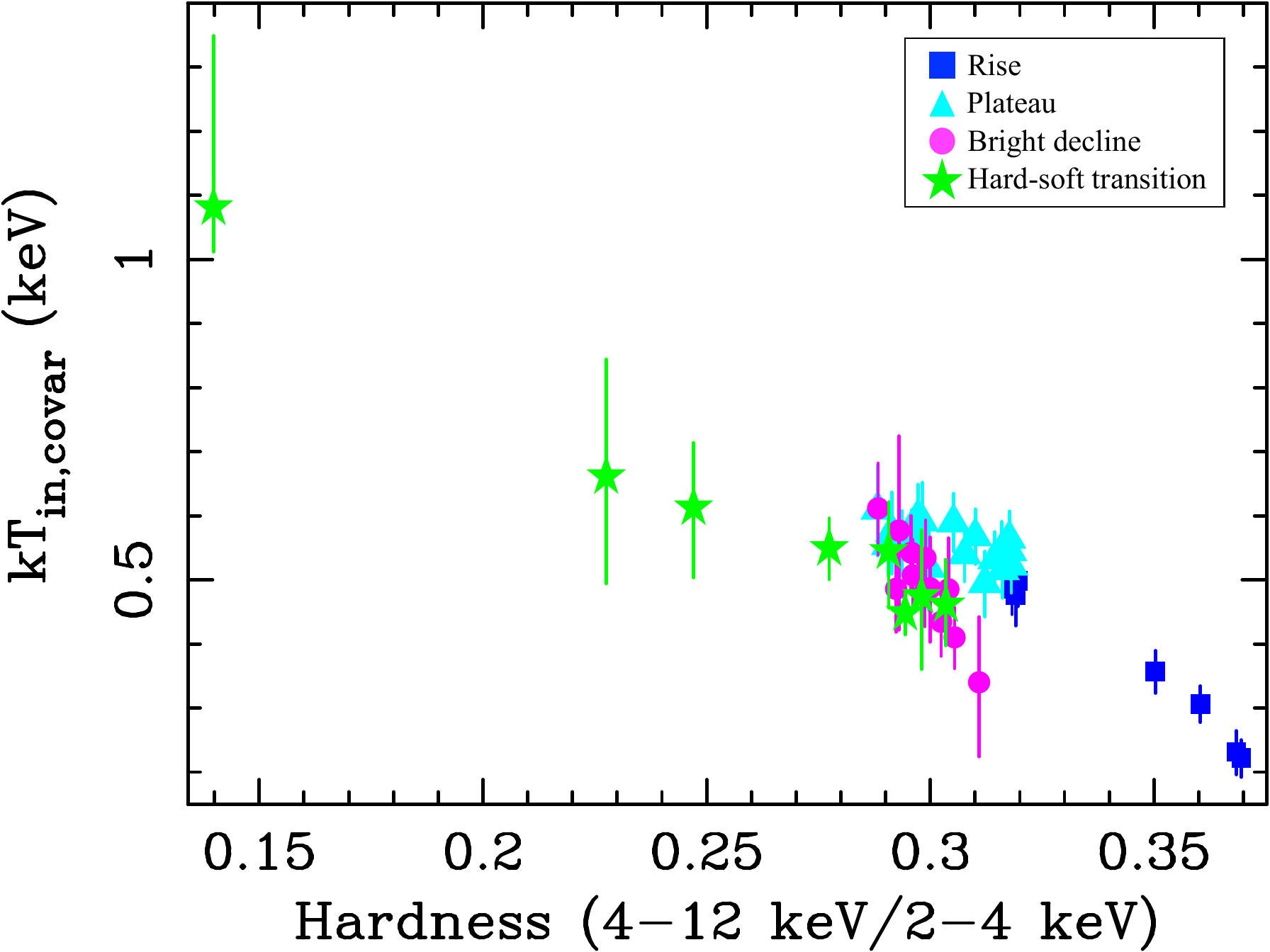}
    \caption{Best-fit values of spectral parameters from fits of high-frequency covariance spectra with the \texttt{TBabs[deltadisk + nthComp]} model. The parameters are plotted as a function of spectral hardness (i.e. the ratio between the 4--12 keV and the 2--4 keV total count rates). \emph{Left panel:} Spectral slope of the hard Comptonisation component, $\Gamma_{\rm covar}$. \emph{Right panel:} Inner disc temperature, $kT_{\rm in,covar}$.}
    \label{fig:Tin_Gamma_hr}
\end{figure*}

In Fig. \ref{fig:Tin_Gamma_hr} we report the best-fit values of the $\Gamma_{\rm in,covar}$ and $kT_{\rm in,covar}$ parameters as a function of spectral hardness. The slope of the hard X-ray Comptonisation component does not show significant changes throughout this part of the outburst (left panel of Fig. \ref{fig:Tin_Gamma_hr}). This indicates that the spectral structure of the innermost parts of the hot plasma does not vary significantly throughout the hard and hard-intermediate states: it remains quite hard, although the errors become large towards the transition and a spectral softening cannot be excluded during the last observation.

On the other hand, the temperature of the disc black body component in high-frequency covariance spectra is observed to steadily vary with hardness (right panel of Fig. \ref{fig:Tin_Gamma_hr}). We checked that letting the $N_{\rm H}$ free to vary in covariance spectra results in a slight, systematic shift of $kT_{\rm in,covar}$ towards lower values, which does not significantly influence the obtained results. 
The measured disc black body component in covariance spectra becomes hotter, by a factor of $\sim$5, as the time-averaged spectrum softens. 
Notably, during the rise, plateau, and bright decline, the characteristic frequency $\nu_{0}$ of the soft X-ray lag (and thus its intrinsic amplitude) is consistent with an increase (decrease in intrinsic lag amplitude) by a similar or slightly smaller factor (of $\sim3-5$; Fig. \ref{fig:nu0_vs_HR}, left panel). This strongly hints at a connection between the two in these phases of the outburst. 
Indeed, the steady increase in the characteristic frequency of soft X-ray lags (Fig. \ref{fig:nu0_vs_HR}) in these phases of the outburst suggests a change in the geometry of the innermost accretion flow. This might ultimately lead to a change in the irradiation of the disc, causing (according to Eq. \ref{eq:S-B_law}) the observed increase in the inner disc temperature $kT_{\rm in,covar}$ (Fig. \ref{fig:Tin_Gamma_hr}, right panel).

Therefore, we assumed that the scale height of the X-ray source is relatively small, and used the best-fit $kT_{\rm in, covar}$ values from high-frequency covariance spectra to put constraints on the inner radius of the irradiated disc via Eq. \ref{eq:S-B_law2}.
The effective temperature of the disc is influenced by the variable hard X-ray flux, but intrinsic dissipation as well irradiation from constant (or variable on longer timescales) hard X-ray emission may also contribute to it. Thus, in general Eq. \ref{eq:S-B_law2} yields a lower limit on the inner radius, unless heating from irradiation dominates in these regions ($kT_{\rm in}/f_{\rm col}=kT_{\rm eff}\sim kT_{\rm eff,irr}$).

We assumed $f_{\rm col}=1.7$ \citep[][]{Shimura1995,Kubota1998,Gierlinski1999,Davis2005}. 
In order to estimate the total irradiating luminosity $L_{\rm irr}$ we performed systematic fits of time-averaged spectra in the energy band 3--10 keV. Details of these fits are reported in Appendix \ref{sec:timeave}. The Eddington-scaled irradiating luminosity $l_{\rm irr}$ was estimated from the 0.1--1000 keV flux ($F_{\rm irr,0.1-1000 keV}$ reported in Table \ref{tab2}) of the extrapolated best-fit Comptonised incident spectrum and assuming the constraints on the distance from \cite{Atri2020}.  
For the albedo we assumed a value of $a=0.5,$ which is an average of the values observed in the hard state of other BHXRBs ($\sim0.3-0.7$ \citealt{ZdziarskiBDM2020}). We accounted for the uncertainty on this parameter in the computations of the errors (see below). 
We kept the reflection fraction fixed at $\mathcal{R}=1$ (as expected for an isotropic, static X-ray source, covering the inner disc), since time-averaged fits yield estimates very close to this value.
In order to compute the 90 per cent confidence errors on the inferred values of $R_{\rm in}$ we randomised the measured irradiating fluxes from time-averaged spectra (according to a log-normal distribution, \citealt{Uttley2005}) and the best-fit values of $kT_{\rm in,covar}$ (according to a normal distribution) within their range of uncertainties. The error was then inferred from the resulting distribution of $R_{\rm in}$ values. In the computation of the errors we also included the uncertainty on the assumed values of $f_{\rm col}$ and $a$ by considering the extreme values of these parameters within the ranges $f_{\rm col}\sim 1.3-2.2$ \citep[e.g.][]{Salvesen2021} and $a\sim0.3-0.7$ \citep[][]{ZdziarskiBDM2020}. 
Given that this computation yields a lower limit on $R_{\rm in}$, we conservatively accounted for the uncertainties on the BH mass and source distance (Sect. \ref{sec:intro}) by considering their $1\sigma$ confidence values that minimise the inner radius (Sect. \ref{sec:intro}).

\begin{figure}
        \includegraphics[width=\columnwidth]{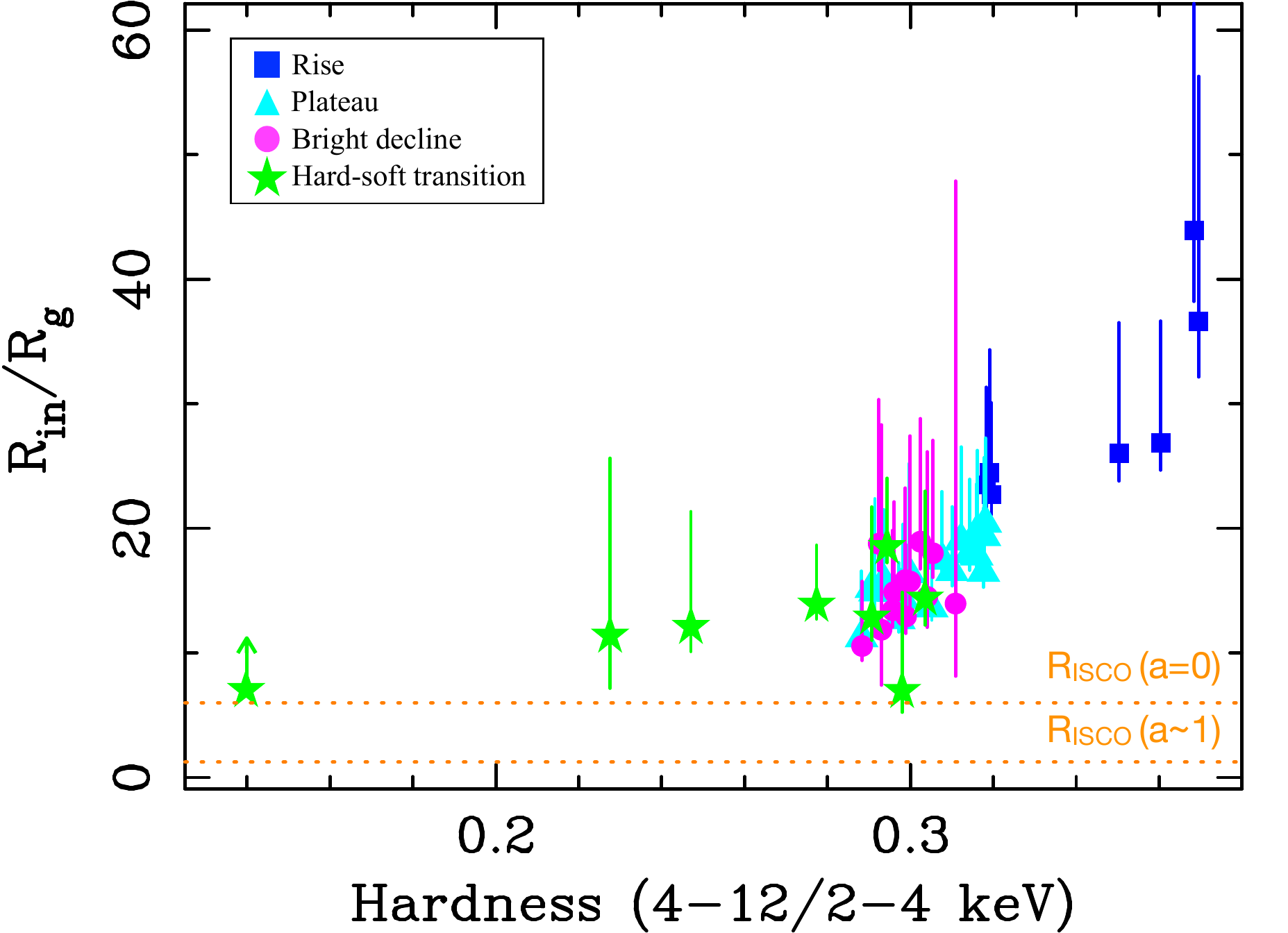}
    \caption{Lower limits on the inner disc radius as a function of spectral hardness, as obtained from constraints on the quasi-thermal component in high-frequency covariance spectra. The horizontal dotted lines mark the ISCO for a Schwarzschild and Kerr BH. Results are obtained assuming an isotropic static X-ray source irradiating the inner disc (reflection fraction $\mathcal{R}=1$) and with a small scale height compared to the disc radius.}
    \label{fig:Rin_vs_hr}
\end{figure}

Figure \ref{fig:Rin_vs_hr} reports the results as a function of spectral hardness. These are consistent with a significant disc-truncation throughout most of the analysed observations, with a net decrease in the disc inner radius (by a factor of $\sim$5--6 at most) between the first and the last observation.

\section{Discussion}
\label{sec:discussion}

Our analysis shows that the geometry of the innermost accretion flow of MAXI J1820+070 changes significantly and continuously 
throughout the hard and hard-intermediate states of its 2018 outburst.
This result was obtained via two independent X-ray spectral-timing methods applied to \nicer\ data, namely the analysis of soft X-ray reverberation lags, and of the soft quasi-thermal  component in high-frequency covariance spectra.

\subsection{The evolution of the inner flow geometry}
\label{sec:discussion1}
\begin{figure}
        \includegraphics[width=\columnwidth]{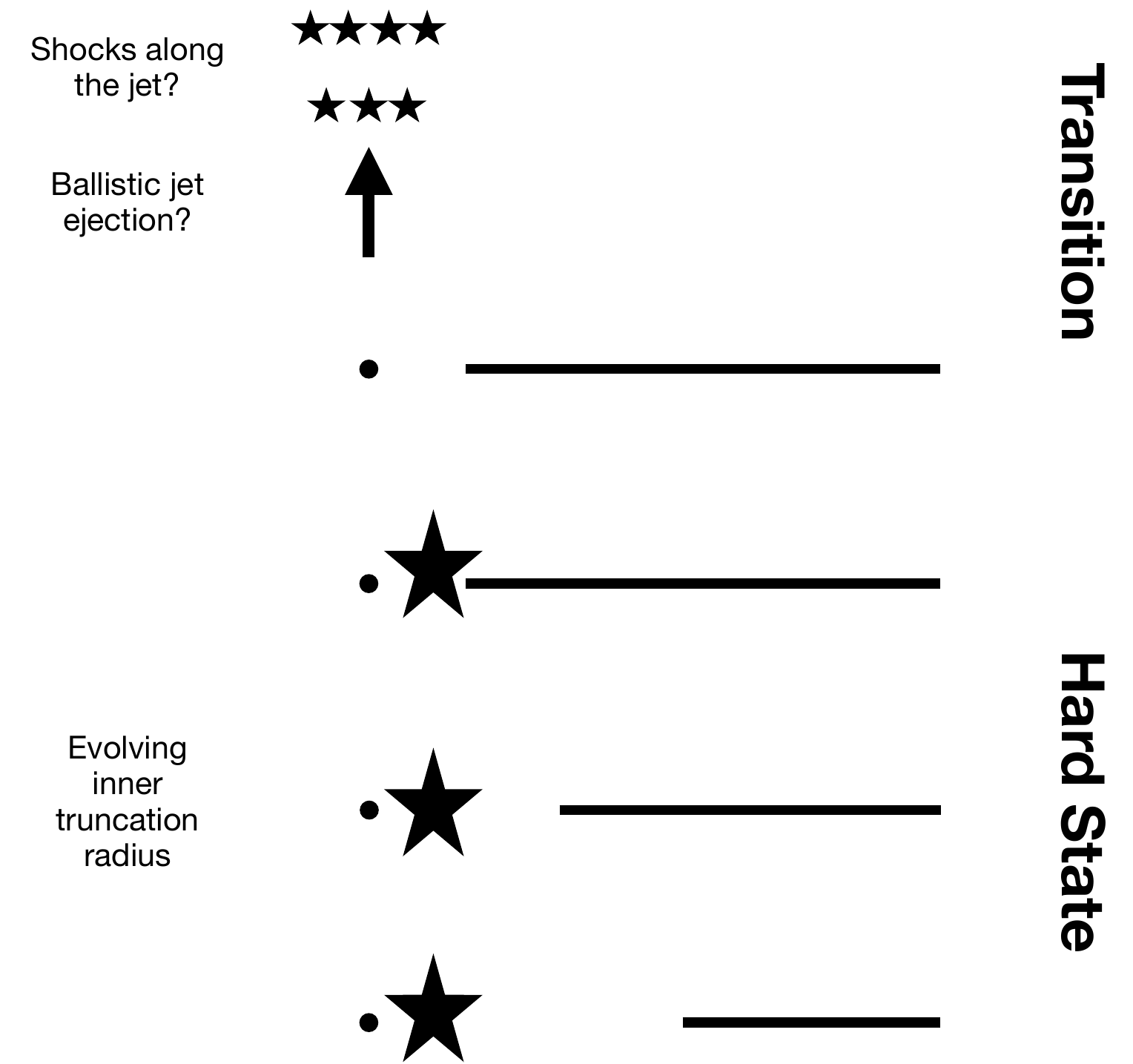}
    \caption{Sketch of the physical scenario proposed to explain the observed evolution of soft X-ray reverberation lags and of the quasi-thermal component from irradiation in high-frequency covariance spectra observed in MAXI J1820+070.}
        \label{fig:cartoon}
\end{figure}

During the rise, the plateau, and the bright decline (Fig. \ref{fig:hid}) the soft X-ray lags in MAXI J1820+070 show a steady increase (decrease) of characteristic frequency $\nu_{0}$ (intrinsic amplitude) as the source softens (Fig. \ref{fig:nu0_vs_HR}, left panel). This trend is in qualitative agreement with the evolution of the frequency of the type-C QPO in the PSDs of the source \citep[][]{Buisson2019}. Soft X-ray lags in BHXRBs are ascribed to disc thermal reverberation of incident hard X-ray photons \citep[e.g.][]{Uttley2011,BDM2015b,Kara2019}. Therefore, the behaviour observed in MAXI J1820+070 can be interpreted in terms of a reduction of relative distance between the Comptonising region and the disc as the source softens (such behaviour is schematically illustrated in Fig. \ref{fig:cartoon} in the framework of the evolving inner radius of a truncated disc; see discussion below).
However, during the hard-soft transition this trend breaks (Fig. \ref{fig:nu0_vs_HR}, right panel).
In particular, during observations at spectral hardness $\lsim 0.28$, $\nu_{0}$ suddenly decreases (by a factor of $\sim$2--3), implying an increase in the reverberation lag intrinsic amplitude. We notice that this occurs simultaneously to an intense hard (15--50 keV) X-ray flare, as detected in \emph{Swift}/BAT light curves \citep{WangY2020}, suggesting a major change in the innermost emitting regions.
Moreover, during O197, when the transition took place (as identified via X-ray and radio analysis; \citealt{Shidatsu2019,Homan2020}) there is no evidence of hard lags (the lack of a hard lag is observed at least down to $\sim 0.05$ Hz), and a long ($\sim$ 15--20 ms) reverberation lag is observed to dominate the entire frequency band (Fig. \ref{fig:softlags} bottom right panel). We suggest that such a behaviour might be the consequence of a different component dominating the X-ray emission at this time, possibly associated with the observed \citep[][]{Bright2019,Espinasse2020,Wood2021} ballistic jet ejections from the source (see Fig. \ref{fig:cartoon} and discussion below). State-of-the-art X-ray spectral-timing models may not be able to reproduce the soft X-ray lags, as well as the (apparently total) suppression of hard X-ray lags in this phase of the outburst.

At the frequencies of the soft X-ray reverberation lag covariance spectra show excess soft X-ray emission, well-modelled by a quasi-thermal component due to the disc (Fig. \ref{fig:covarO104105}, Sect. \ref{sec:covar}). 
This indicates that this component is produced via X-ray heating, and that the thermal response of the disc to variable hard X-ray irradiation preserves a certain level of coherence.  
We found that the inner temperature of the disc black body component in the high-frequency covariance spectra of MAXI J1820+070 steadily increases as the source softens (Fig. \ref{fig:Tin_Gamma_hr}, right panel). Following the method outlined in \cite{ZdziarskiBDM2020} and assuming a constant and relatively small scale height for the X-ray source, we used these measurements to obtain constraints on the inner radius of the irradiated disc region responsible for the observed reverberation lags. 
Our estimates are consistent with an evolution of $R_{\rm in}$ (Fig. \ref{fig:Rin_vs_hr}), causing changes in the irradiation of the disc, and consequently in its temperature (Fig. \ref{fig:Tin_Gamma_hr}, right panel). In most of the hard state (i.e. during the rise, plateau, and bright decline) the observed increase (decrease) in the irradiated inner disc temperature (radius) as the source softens is consistent with the scaling characterising the frequency of soft X-ray reverberation lags (a change by a factor  of $\sim3-4$ is observed in all these cases, Fig. \ref{fig:nu0_vs_HR}, Fig. \ref{fig:Tin_Gamma_hr} right panel, and Fig. \ref{fig:Rin_vs_hr}). This behaviour is qualitatively in agreement with predictions of disc-truncated models \citep[e.g.][]{Esin1997,Done2007}. 
However, the observed trend characterising the irradiated inner disc temperature and radius persists till transition, not showing any clear break as instead seen for the soft lags. 
In other words, these two independent diagnostics of geometry appear to stop probing the same regions when the source is close to transition.

Therefore, we interpret the behaviour near the transition to soft-intermediate states as due to a drastically different configuration of the innermost accretion flow than that characterising the preceding phases of the outburst.
In particular, by tracing back the motion of the discrete, relativistic ejecta from the central X-ray source detected in radio and X-ray observations of MAXI J1820+070, \cite{Bright2019} and \cite{Espinasse2020} estimate a launch date around MJD 58306, while an additional earlier ejection (MJD 58305.6) has been recently reported \citep[][]{Wood2021}. The latter coincides with the last of our observations (O197). 
This suggests a link between these discrete ejections, the long reverberation lag and the suppression of hard X-ray lags.

The scenario we propose to explain the observed phenomenology is illustrated in the simplified sketch of Fig. \ref{fig:cartoon}. 
The inner radius of the disc decreases steadily throughout most of the hard state, as implied by the increase (decrease) of lag frequency (amplitude) and as traced by the inner temperature of the irradiated inner disc in high-frequency covariance spectra. Close to transition, while the inner radius is consistent with settling near (or at) the ISCO (as traced by the steady evolution of the quasi-thermal component in high-frequency covariance spectra), the structure of the X-ray source undergoes major changes (as traced by the break in the scaling of soft X-ray lags and the suppression of hard X-ray lags). Indeed, the intrinsic reverberation lag depends on the location of the region(s) of hard X-ray energy dissipation and the reprocessing region in the disc. If the emission due to internal shocks in a ballistic jet becomes significant close to the transition, then the dissipation of hard X-rays would predominantly occur in a larger or more distant region, which would cause the irradiation of a larger area of the disc. This would qualitatively explain the generally lower observed characteristic lag frequencies (Fig. \ref{fig:nu0_vs_HR}), and in particular, the very long reverberation lag (Fig. \ref{fig:softlags} bottom rightmost panel) at state transition. 
Moreover, if the hard X-ray lags commonly observed during the outburst are produced in the inner accretion flow (e.g. as proposed in models of propagating accretion rate fluctuations; e.g. \citealt{Lyubarskii1997}), a change in the dominant process of hard X-ray emission would also explain the observed suppression of these hard lags. In particular, this is expected if the dominant hard X-ray component is largely decoupled from fluctuations generated in the accretion flow (e.g. if this component originates in shocks along the jet, \citealt{Vincentelli2018}).
We observe that the soft X-ray lags start deviating from the trend observed at higher spectral hardness about 4 days before the transition (Fig. \ref{fig:nu0_vs_HR}, right panel), namely on MJD 58302. This might indicate that the major changes in the inner accretion flow leading to the launch of relativistic ejecta start occurring a few days before the ejection. Interestingly, \cite{Buisson2021} reported the detection of a much longer ($\sim$20 s) soft X-ray lag $\sim$ 0.4 day after the transition (i.e. in soft-intermediate states). The authors interpreted it as due to the disruption and refilling of the inner disc following the jet ejection.

We again emphasise that the estimates of the disc inner radius through Eq. \ref{eq:S-B_law2} (Fig. \ref{fig:Rin_vs_hr}) are strictly valid for a X-ray source with a height above the disc small compared to the disc radius (for example a corona above the inner disc or a lamppost located at a few $R_g$ from the BH, Sect. \ref{sec:covar}). Moreover, our computations assume that the height of the source remains approximately constant. If this condition does not apply, then variations in the height of the X-ray source are expected to play a significant role too in the observed changes of disc illumination. 
Recently, \cite{Kara2019} and \cite{Buisson2019} proposed a scenario whereby the vertical extent of the X-ray source, rather than the inner edge of the accretion disc, decreases throughout the hard state, driving the observed changes in the thermal reverberation lags. In particular, this was proposed in order to justify the remarkable stability of the broad component of the Fe K line in the hard and hard-intermediate states of MAXI J1820+070. However, this scenario entails constant illumination of the inner parts of the disc, which ultimately implies that the geometry of the regions of the X-ray source closer to the disc (and responsible for the irradiation of the inner disc) does not change significantly. Our results appear to disfavour such a scenario because in this case we would expect to measure a constant inner disc temperature, while we find the temperature to increase steadily (Fig. \ref{fig:Tin_Gamma_hr}, right panel). Therefore, our results appear not in agreement with the scenario proposed by \cite{Kara2019} and \cite{Buisson2019}. On the other hand, given the above mentioned assumptions of our analysis, we cannot completely rule out the possibility that the X-ray source varies in its height (rather than in its vertical extent) during the outburst, thus contributing to the observed trend of inner disc temperature. However, this scenario seems hard to reconcile with the observed evolution of type-C QPOs in this source \citep[][]{Buisson2019}.
Therefore, we favour a solution whereby the hard state is dominated by a decrease in the inner radius. This is also supported by a recent re-analysis of four \emph{NuSTAR} observations of MAXI J1820+070 at different hard state epochs \citep[][]{Zdziarski2021b}, according to which the lack of a spectral evolution of the broad component of the Fe K line is indicative of modest relativistic effects, as expected when the inner radius does not reach the ISCO. The apparent lack of a clear relativistic component in the high-frequency covariance spectra of the source (Fig. \ref{fig:covarO104105}) further corroborates this interpretation.

\subsection{The truncation radius of the disc} 
\label{sec:discussion2}
Our analysis of high-frequency covariance spectra (Sect. \ref{sec:covar}) yields a lower limit 
on the disc inner radius for each of the analysed observations (Fig. \ref{fig:Rin_vs_hr}).
These estimates depend on the value of the reflection fraction parameter $\mathcal{R}$. In our treatment the parameter $\mathcal{R}$ is used to account for a possible reduction of hard X-ray flux reaching the disc for causes other than changes of disc radius or X-ray source height \citep[][]{ZdziarskiBDM2020}. Assuming isotropic illumination of the disc $\mathcal{R}=1$ we found truncation at $\gsim30\ R_g$ at the beginning of the monitoring, corresponding to bolometric luminosities of a few per cent of $L_{\rm E}$ \citep[][]{Zdziarski2021b}. The disc turns out to be significantly truncated ($\gsim10\ R_g$) throughout most of the bright hard and hard-intermediate states, and appears to reach close to the ISCO only at transition. 
Assuming a lower value of $\mathcal{R}$ (e.g. as found in \citealt{Zdziarski2021b}) would reduce the inferred value for $R_{\rm in}$ by a factor $\mathcal{R}^{1/2}$, thus not affecting our inference of a disc significantly truncated throughout most of the analysed phases of the outburst.
However, even accounting for the possible effects of a smaller $\mathcal{R}$, the data remain consistent with a disc at the ISCO of a low spinning BH during the transition (O197). This is in agreement with other independent results suggesting the BH in MAXI J1820+070 has spin $a<1$ \citep[][]{Buisson2019,Atri2020}.

 As discussed in Sect. \ref{sec:softlags}, the soft X-ray reverberation lags can also be used to infer an estimate of the disc truncation radius. However, this requires a more complicated modelling approach that relies on the simultaneous fit of variability, lag and time-averaged spectra \citep[e.g.][]{Mahmoud2019,Mastroserio2019}, and depends on the assumptions on the geometry of the Comptonising region. 
 This problem has been widely discussed in the literature \citep[][]{Uttley2011,BDM2017,WangJ2020}, and can be understood by considering the intrinsic soft lag amplitude implied by its characteristic frequency $\nu_0$ (Fig. \ref{fig:nu0_vs_HR}), and the distance that this amplitude maps. The corresponding Euclidean distance is $d_{\rm rev}=c\tau_{\rm rev}(1+\cos[\theta-i])^{-1}\sim c(2\nu_{0})^{-1}(1+\cos[\theta-i])^{-1}$, where $\theta$ is the angle between the line of sight to the X-ray source and the straight path linking the X-ray source to the disc reprocessing region. The case $\theta=\pi/2$ approximates a truncated disc illuminated by a central source (in this case the derived distance coincides with the inner disc radius). 
Assuming $\theta=\pi/2$ and $i=70^\circ$, our estimates of $\nu_0$ (Sect. \ref{sec:softlags} and Table \ref{tab2}) yield Euclidean distances between the X-ray source and the reprocessing region in the disc ranging between a few tens and a few hundreds of $R_{\rm g}$. These distances are a factor  of $\sim3-10$ times larger than the disc inner radius inferred from covariance spectra.
In fact the distance inferred from reverberation lags is a weighted average among all possible light paths between the X-ray source and the disc (i.e. corresponding to the mean lag of the impulse response function; e.g. \citealt{Wilkins2013}). Therefore, in case of complicated geometries (i.e. more intricate impulse response function), such as for an extended X-ray source, the link between the measured lag and the geometrical parameters of the disc and of the X-ray source may be quite complex. This has been also shown in \cite{Mahmoud2019}, where the inner truncation radius resulted significantly smaller (by a factor of $\sim4$) than the distance mapped by the observed reverberation lag. This would explain why the soft X-ray lags measured in MAXI J1820+070 apparently imply distances larger than the disc inner radius inferred from covariance spectra.
 
It is interesting to note that such large discrepancies between the distances implied by X-ray reverberation lags and the disc inner radius inferred by other independent methods is generally not observed in active galactic nuclei (AGN; e.g. \citealt{Fabian2009,BDM2013,Marinucci2014,Kara2016,Alston2020}). This might be due to a substantial difference between the geometry of the X-ray source in hard state BHXRBs as compared to that of those AGN (all type-1 non-jetted Seyferts or narrow line Seyferts) for which X-ray reverberation has been measured. In particular, dissipation of hard X-rays in these AGN likely occurs in a more compact region \cite[e.g.][]{Fabian2015}, thus reducing the number of possible light paths and thus the width of the impulse response function. Such inference is also in agreement with the fact that the characteristic timescales of variability appear systematically offset towards lower frequencies in hard state BHXRBs as compared to AGN \citep[e.g.][]{Koerding2007}.

Finally, it is important to point out that in the previous discussion we make the implicit assumption that the detected reverberation lags are dominated by the light travel distance between the X-ray source and the disc, while other timescales are negligible. In particular, the thermalisation timescale should also contribute to the observed delays. However, this quantity scales inversely with the density and directly with the temperature of the disc surface layers irradiated by the X-ray source \citep[e.g.][]{Collin2003}. For certain values of these two parameters (e.g. density $\lsim10^{18}\rm cm^{-3}$ for temperatures of $\sim$1 keV) the thermalisation timescale may be of the order of the observed lags (i.e. $\sim0.1-1$ ms). However, while the density of the disc is currently a not well-constrained parameter, recent studies suggest it to be quite high in BHXRBs ($\gsim10^{19-20}\rm cm^{-3}$, \citealt{Tomsick2018,Jiang2019,ZdziarskiBDM2020}). If so, the corresponding reprocessing time is negligible compared to the measured lag. Moreover, contrary to what observed, if the lags were dominated by the thermalisation timescale, the observed process would be highly non-linear (incoherent; e.g. \citealt{Vaughan1997}).

 \section{Conclusions}
We performed a systematic X-ray spectral-timing study of the first part of the 2018 outburst of the BHXRB system MAXI J1820+070. Our analysis covers the hard and hard-intermediate states as well as the transition to the soft-intermediate states. 
Our main results are as follows:
\begin{itemize}
\item The soft X-ray thermal reverberation lags show a steady evolution throughout most of the hard state (specifically during the rise, plateau, and bright decline). This evolution is indicative of a decrease in the relative distance between the X-ray source and the disc as the source softens.\\
\item The quasi-thermal component associated with the soft X-ray reverberation lags and observed in high-frequency covariance spectra shows a steady increase in the inner temperature. This is a consequence of changes in the hard X-ray irradiation of the inner disc and is consistent with a decrease in the disc truncation radius, in agreement with the observed evolution of soft X-ray lags. The disc is consistent with being highly truncated at the beginning of the outburst ($R_{\rm in}\gsim 30 R_{\rm g}$) and with remaining significantly truncated throughout most of the hard and hard-intermediates states ($R_{\rm in}\gsim 10 R_{\rm g}$).\\
\item Major changes occur before the transition to the soft-intermediate states. The thermal reverberation lags suddenly increase (decrease) in amplitude (frequency), breaking the trend observed during the preceding phases of the outburst. This suggests an increase in the relative distance between the X-ray source and the disc. At the same time, the hard X-ray lags intrinsic to the primary continuum get suppressed. 
On the other hand, the observed trend characterising the quasi-thermal component in high-frequency covariance spectra persists until transition, not showing any clear break. This trend is consistent with the inner edge of the disc increasingly approaching the ISCO, and possibly settling there at (or close to) transition.\\
\end{itemize}

In order to explain the observed X-ray spectral-timing phenomenology, we propose a scenario whereby the inner radius of the disc decreases steadily throughout the hard and hard-intermediate states. This scenario is an alternative to the contracting vertical extent of the corona proposed in \cite{Kara2019} and appears more in line with the detected steady increase in the irradiated disc temperature. While the inner disc maintains this trend close to the transition to soft-intermediate states, reaching (or nearing) the ISCO, the X-ray source undergoes major structural changes. In particular, we propose that in this phase the dissipation of hard X-rays predominantly occurs in the ballistic jet (possibly through internal shocks). This would qualitatively explain the behaviour of soft X-ray reverberation lags and the suppression of hard X-ray lags (provided the latter are intrinsically produced in the accretion flow). Interestingly,  
the soft X-ray lags start deviating from the trend observed during most of the hard state about 4 days before the appearance of a type-B QPO (which marks the hard-to-soft transition; \citealt{Homan2020}) and the ejection of relativistic plasma (as later detected at radio and X-ray wavelengths; \citealt{Bright2019,Espinasse2020}), possibly representing the precursor to such events. Future intensive monitoring campaigns of BHXRB systems with \nicer\ will allow us to further test this scenario.

 \begin{landscape}
\begin{table}
\centering
\caption{Log of analysed \nicer\ observations of MAXI J1820+070.}  
\label{tab1}
\resizebox{1.3\textwidth}{!}{\begin{tabular}{ccccccccccccccc} 
\hline
 \noalign{\smallskip}
Obs ID                 &    Exp  &{\tt N}$_{\rm{FPM}}$  &   Phase   & Date   &Obs ID         &    Exp &{\tt N}$_{\rm{FPM}}$  &  Phase  &  Date & Obs ID         &    Exp  &{\tt N}$_{\rm{FPM}}$  &  Phase  & Date \\
\noalign{\smallskip}
                           &          [s]      &                           &               &           &         &           [s]     &                                  &        &     &         &           [s]       &                                  &     &     \\
                           \noalign{\smallskip}
\hline
\noalign{\smallskip}
1200120101       &   6274               & 50      &    Rise   &  2018-03-12    & 1200120132  &   3773   & 50       & Plateau  & 2018-04-20   & 1200120165   &   1303   & 50        & Bright decline  & 2018-05-31\\ 
                          &                     &            &              &                           &1200120133  &   4727   & 50       & Plateau   & 2018-04-21    & 1200120166  &   1349   & 50       &  Bright decline & 2018-06-01 \\ 
1200120102       &   4513               & 50      &    Rise   &  2018-03-13    & 1200120134 &  6345    &  50       & Plateau  &  2018-04-21    &  1200120167 &   1529  & 50        & Bright decline & 2018-06-02    \\
                          &                     &            &              &                           &                     &               &            &                &                        &                       &             &             &                       &                   \\  
1200120103       &   9474               & 50      &    Rise   & 2018-03-13      &1200120135  &  4077   & 50       & Plateau    & 2018-04-23    & 1200120168  &   417    & 50        & Bright decline & 2018-06-03   \\ 
                          &                     &            &              &                          &1200120136  &   1570   & 50        & Plateau    & 2018-04-24    &                       &             &              &                         &                               \\
1200120104       &   6231               & 50      &    Rise   &  2018-03-15    &1200120137  &   6870   & 50      & Plateau    &     2018-04-25    & 1200120169   &   1761  & 50      & Bright decline  & 2018-06-04 \\
1200120105       &   3060               & 50      &    Rise   & 2018-03-16     &1200120138  &   3131   &  50   & Plateau     & 2018-04-25     & 1200120170   &   3592  & 50      & Bright decline  &   2018-06-05     \\ 
                          &                     &            &               &                         &1200120139 &   560     & 50      & Plateau     & 2018-04-27     &                        &             &         &                         &     \\
1200120106       &   4446               & 50      & Rise      & 2018-03-21    &                      &              &          &                  &                             & 1200120171   &   2373   &50      &   Bright decline  &  2018-06-06\\
1200120107$^{a}$&   9293        & 50      & Rise      & 2018-03-22    & 1200120140 &  1844  & 50     &  Plateau     & 2018-04-29       &                       &              &            &                         &               \\
                          &                     &            &               &                      &1200120141  &   904    & 50      & Plateau     & 2018-05-01      & 1200120172    &   2249 & 50       & Bright decline  & 2018-06-07    \\ 
1200120107$^{b}$&   3703        & 42      & Rise      & 2018-03-22   &                     &            &           &                    &                    &                        &              &            &                         &         \\
1200120108       &   5296          & 42      & Rise      &      2018-03-22   &1200120142  &   3195 & 50     & Plateau       & 2018-05-02     & 1200120173   &   3284    & 50       & Bright decline  & 2018-06-08 \\
                           &                     &           &               &               &                       &           &          &                     &                            & 1200120174  & 2698       & 50      & Bright decline &  2018-06-09 \\
1200120109       &   11472      & 50      & Rise      & 2018-03-24   & 1200120143  & 2075 & 50   & Plateau          & 2018-05-03      &                       &                 &           &                         &    \\
                          &                      &           &               &               &1200120144  & 2992   & 50     & Plateau        & 2018-05-04     & 1200120175  &   2156   & 50       & Bright decline   & 2018-06-10  \\
1200120110       &   19360      & 50      & Plateau  & 2018-03-24   &                      &             &         &                     &                      & 1200120176  &   5311     & 50       & Bright decline   &         2018-06-11 \\
                          &                      &           &               &                      &1200120145  &   4863   & 50    & Plateau       & 2018-05-05    &                       &               &           &                          &             \\
1200120111       &   14893              & 50      & Plateau  & 2018-03-26  &1200120146 &   4756   & 50     & Plateau        & 2018-05-06    & 1200120177  &  435        &          50  &  Bright decline   &     2018-06-12\\
                           &                     &           &               &                       &                      &               &          &                    &                         & 1200120178   &   2625    & 50    & Bright decline   & 2018-06-13 \\ 
1200120112       &   12033      & 50      & Plateau  & 2018-03-26 &1200120147   &   4609  & 50     & Plateau        & 2018-05-07   & 1200120179  & 1237       & 50    & Bright decline     & 2018-06-14 \\
1200120113       &   1980               & 50      & Plateau  & 2018-03-28  &                      &              &         &                      &                      & 1200120180  & 2246      &  50   & Bright decline   &  2018-06-15       \\ 
1200120114       &   2701               & 50      & Plateau  & 2018-03-29  &1200120148  &  2650   & 50    & Plateau         &  2018-05-08    &                      &            &           &                         &    \\
1200120115       &   3137               & 50      & Plateau  & 2018-03-30  &                     &              &         &                       &                      &  1200120182   &   1120   & 50       & Bright decline  & 2018-06-17   \\
                          &                      &           &                &                      &1200120149 &   2650   & 50    & Plateau         & 2018-05-09  & 1200120183   &      438        &  50     &   Bright decline         & 2018-06-18  \\ 
1200120116       &   9209               & 50      & Plateau  & 2018-03-31    & 1200120150 &  1652     & 50    & Plateau       & 2018-05-10  & 1200120184  &   448     & 50       & Bright decline &  2018-06-19     \\  
1200120117       &   2752               & 50      & Plateau  & 2018-04-01   &1200120151  &  1037    & 50   &  Plateau         & 2018-05-11   & 1200120185 &   578      &  50     &   Bright decline    &  2018-06-20   \\
                          &                      &           &                &               &                       &               &         &                        &                   &                      &               &            &                                 &   \\
1200120118       &   8105               & 50      & Plateau   & 2018-04-02 & 1200120152 &   2102    & 50     &  Plateau       & 2018-05-12  & 1200120186  &  916        &    50    &  Hard-soft Transition    & 2018-06-23  \\
1200120119       &   2935               & 50      & Plateau   & 2018-04-03   &  1200120153 & 356      &     50    &  Plateau      & 2018-05-13  &  1200120187  &   1395    & 50        & Hard-soft Transition  & 2018-06-24 \\ 
                           &                     &           &                &                &                       &              &         &                    &                      &                        &                &            &                    &      \\
1200120120       &   6049               & 50      & Plateau   & 2018-04-04   & 1200120155  &   326     &  50   &  Bright decline &2018-05-21  &  1200120188 &   1244   & 50        & Hard-soft Transition  & 2018-06-27   \\    
1200120121       &   2955               & 50      & Plateau   & 2018-04-05   &  1200120156  &    2886    &50     & Bright decline   & 2018-05-22  &                  &                &           &                                 &    \\
1200120122       &   3247               & 50      & Plateau   & 2018-04-06   &                       &               &           &                          &                     & 1200120189 &   11441 & 50  & Hard-soft Transition & 2018-06-28  \\ 
1200120123       &   752                & 50      & Plateau   & 2018-04-07   & 1200120157   &   3584     & 50     & Bright decline  & 2018-05-23  &                      &              &         &                                 &   \\ 
                           &                     &           &                &                &                        &               &           &                         &                        & 1200120190   &   2150  & 50    &   Hard-soft Transition &  2018-06-29 \\  
1200120124       &   177                & 50      & Plateau   & 2018-04-09   & 1200120158 &   1730      & 50   &   Bright decline   & 2018-05-24  &                       &               & &      &  \\
1200120125       &   307                & 50      & Plateau   & 2018-04-10   & 1200120159 &   1960     & 50      & Bright decline   & 2018-05-25  &   1200120194 &    4738   & 50     &   Hard-soft Transition   &2018-07-03 \\
1200120126       &   896                & 50      & Plateau   & 2018-04-11    &1200120160  &   1688     & 50      & Bright decline  &  2018-05-26 &                          &               &            &    &    \\
1200120127       &   354                & 50      & Plateau   & 2018-04-12    & 1200120161   &  2540   & 50     & Bright decline  &  2018-05-27   &   1200120195   &   691     & 50    &    Hard-soft Transition  & 2018-07-04  \\ 
                           &                     &           &                &                 & 1200120162   &   574    & 50      & Bright decline  & 2018-05-28   &                         &               &           &     &   \\ 
 1200120130      &   7493         & 50       & Plateau   & 2018-04-16   &                        &               &           &                          &                       &  1200120196   &   541    & 50    &   Hard-soft Transition  &  2018-07-05  \\  
 1200120131      &   6571         & 50       & Plateau   & 2018-04-17   & 1200120163   &   1040   & 50       & Bright decline  & 2018-05-29  &                         &              &        &                                       &           \\
                           &                    &            &                 &                      & 1200120164     &   375      & 50     & Bright decline & 2018-05-30  &  1200120197$^{c}$ &   12429    & 27    &   Hard-soft Transition  & 2018-07-06 \\  
                           \noalign{\smallskip}
                \hline
        \end{tabular}}
\tablefoot{The table reports: \nicer\ observations IDs; net on-source exposure time after screening and filtering of $<10$ s GTIs; number of FPMs used for the analysis (\texttt{N}$_{\rm FPM}$); position of the source in the HID as sketched in Fig. \ref{fig:hid}; date of the observation (yyyy:mm:dd). The adopted grouping highlights observations combined for the X-ray spectral-timing analysis.\\
\tablefoottext{a,b}{First and second part of observation 1200120107, characterised by a change in the number of active FPMs; \\$^c$ First part of observation 1200120197, including the time before and during the transition.}}
\end{table}
\end{landscape}

\begin{acknowledgements}

This work is part of the EU-funded ``BHmapping'' project.
While this work was being reviewed, \citealt{WangJ2021} was published, also showing the behaviour of reverberation lags at transition.
The authors thank G. Mastroserio for providing the version of {\tt RELTRANS} used in this paper, M. Mendez for helpful discussions on fits of variability spectra, G.K. Jaisawal for help with \nicer\ data reduction, and the anonymous referee for their useful comments.
BDM acknowledges support from the European Union's Horizon 2020 research and innovation programme under the Marie Sk{\l}odowska-Curie grant agreement No. 798726 and Ram\'on y Cajal Fellowship RYC2018-025950-I. AAZ and MD acknowledge support from the Polish National
Science Centre under the grants 2015/18/A/ST9/00746 and
2019/35/B/ST9/03944. GP acknowledges funding from the European Research Council (ERC) under the European Union’s Horizon 2020 research and innovation programme (grant agreement No 865637). TMB acknowledges financial contribution from the agreement ASI-INAF n.2017-14-H.0 and PRIN-INAF 2019 n.15. 
We acknowledge support from the International Space Science Institute (Bern). 
\end{acknowledgements}

\bibliographystyle{aa}
\bibliography{biblio}

\begin{appendix} 

\section{Stationarity}
\label{sec:PSD}

\begin{figure}
        \includegraphics[width=0.86\columnwidth]{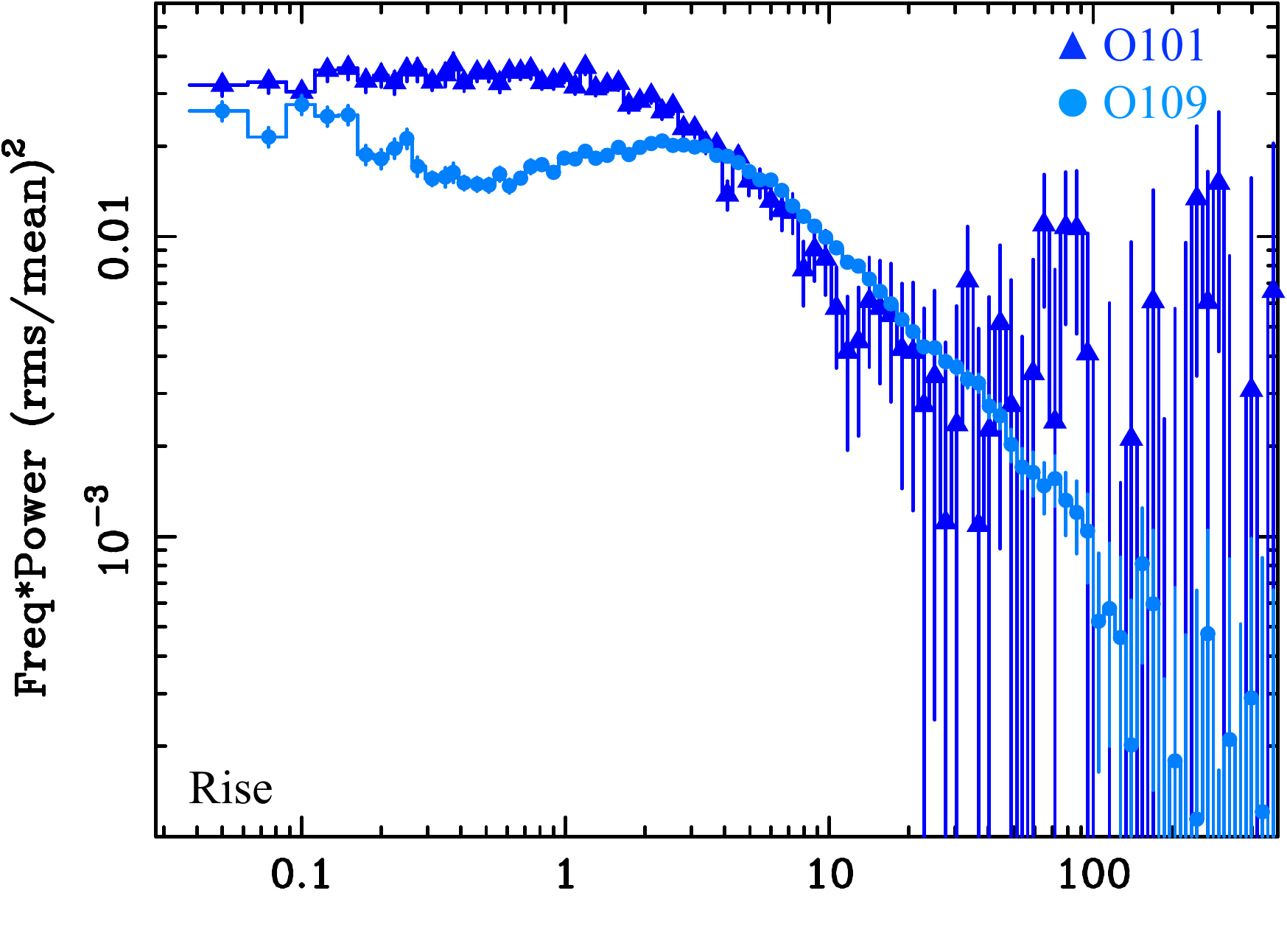}\\
        \includegraphics[width=0.86\columnwidth]{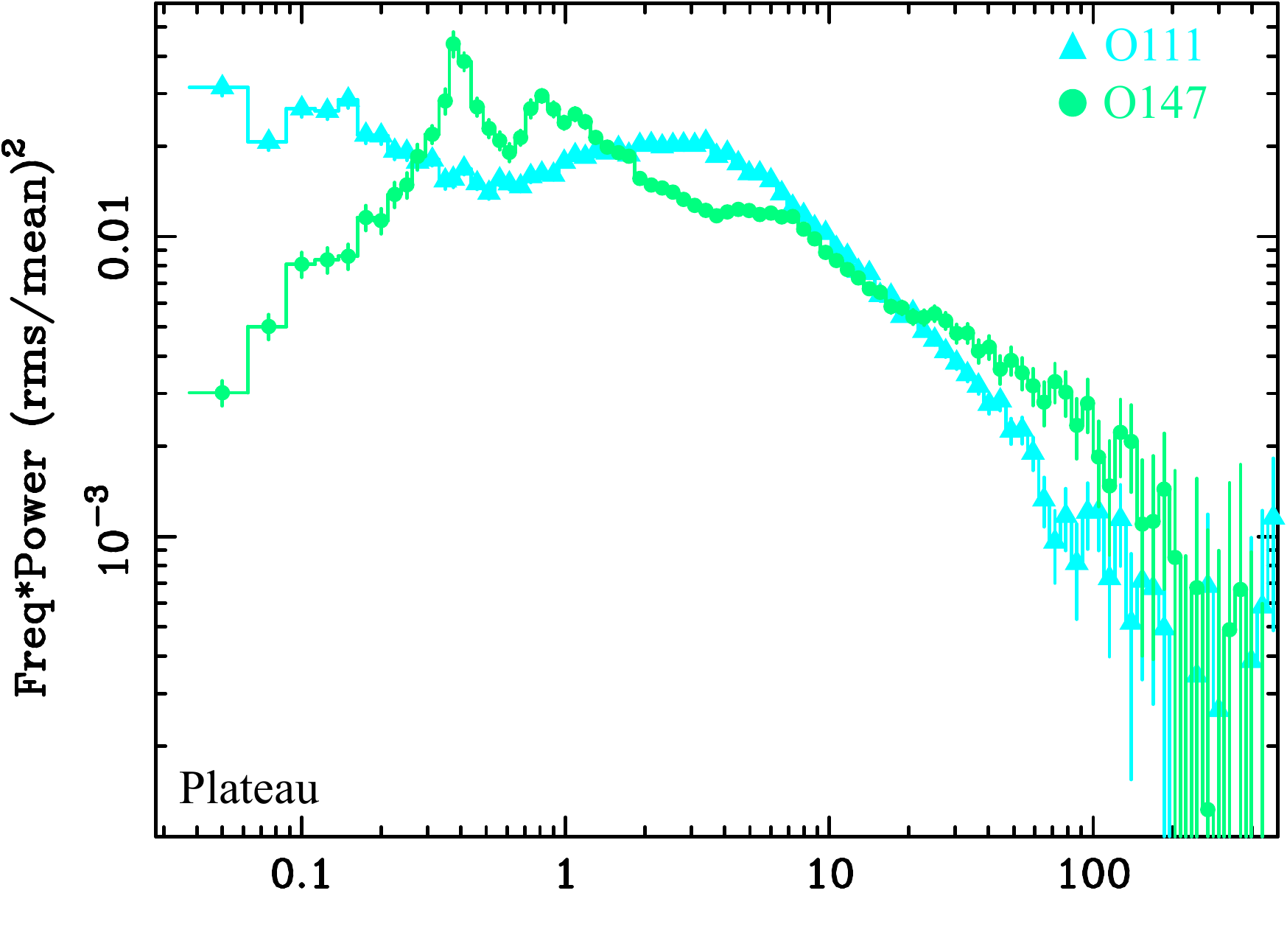}\\
        \includegraphics[width=0.86\columnwidth]{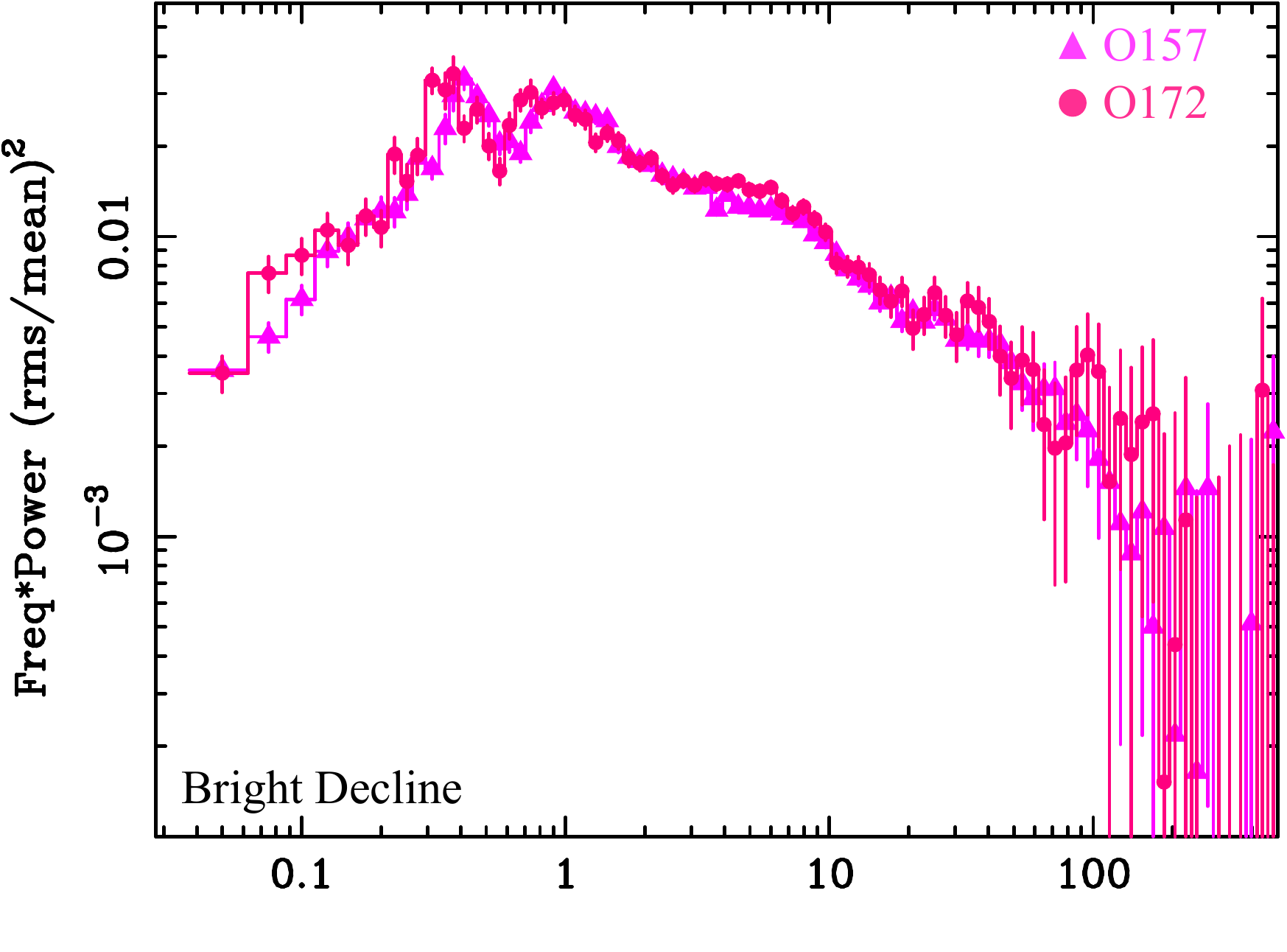}\\
        \includegraphics[width=0.86\columnwidth]{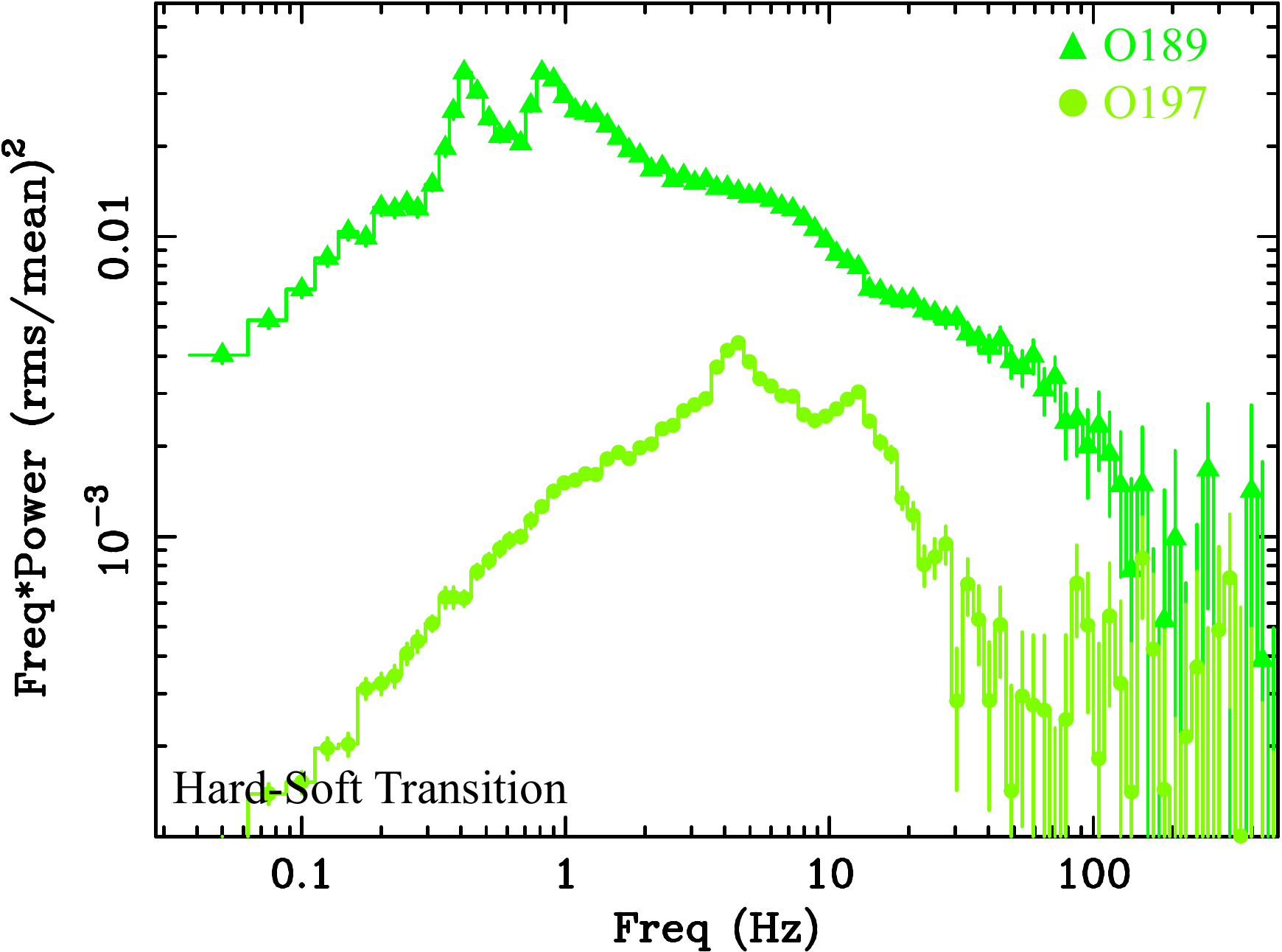}
    \caption{Selection of PSDs of MAXI J1820+070 during different phases (colour-coded according to Fig. \ref{fig:hid} and reported in the labels) of the first part of the 2018 outburst (two examples are shown for each phase). The PSDs are extracted in the energy band 2--11 keV and reveal a strong and complex evolution of the X-ray timing properties of the source.}
    \label{fig:psd}
\end{figure}

The condition of stationarity was verified by measuring the PSD in the energy band 2--11 keV. We extracted PSDs from light curve segments of 40 s, and measured an average PSD for each observation, or part of observation characterised by a constant number of FPMs (the assumption here is that the source is stationary during the time spanned by each dataset). The light curves have a time bin of 0.4 ms (Sect. \ref{sec:data}), and thus each PSD covers the frequency interval from 0.025 Hz up to a Nyquist frequency of 1250 Hz. This allowed us to obtain a good sampling of both the broadband X-ray variability components and the Poisson noise contribution. The latter was fit at high frequencies ($\nu>300$ Hz) with a constant model and subtracted out. The PSDs were normalised using fractional rms-squared units \citep[][]{Belloni1990,Miyamoto1992}.
Only observations visually showing compatible PSD shape (i.e. characteristic frequencies and normalisation) were combined for our X-ray spectral-timing analysis (i.e. an average cross-spectrum was computed from which the lag and covariance spectrum was estimated, Sect. \ref{sec:spec-tim}). Examples of PSDs showing the non-stationary behaviour of the source are shown in Fig. \ref{fig:psd}.

\section{Observed soft X-ray lag amplitude and its evolution as a function of spectral hardness}
\label{sec:lagampl}

\begin{figure}
        \includegraphics[width=\columnwidth]{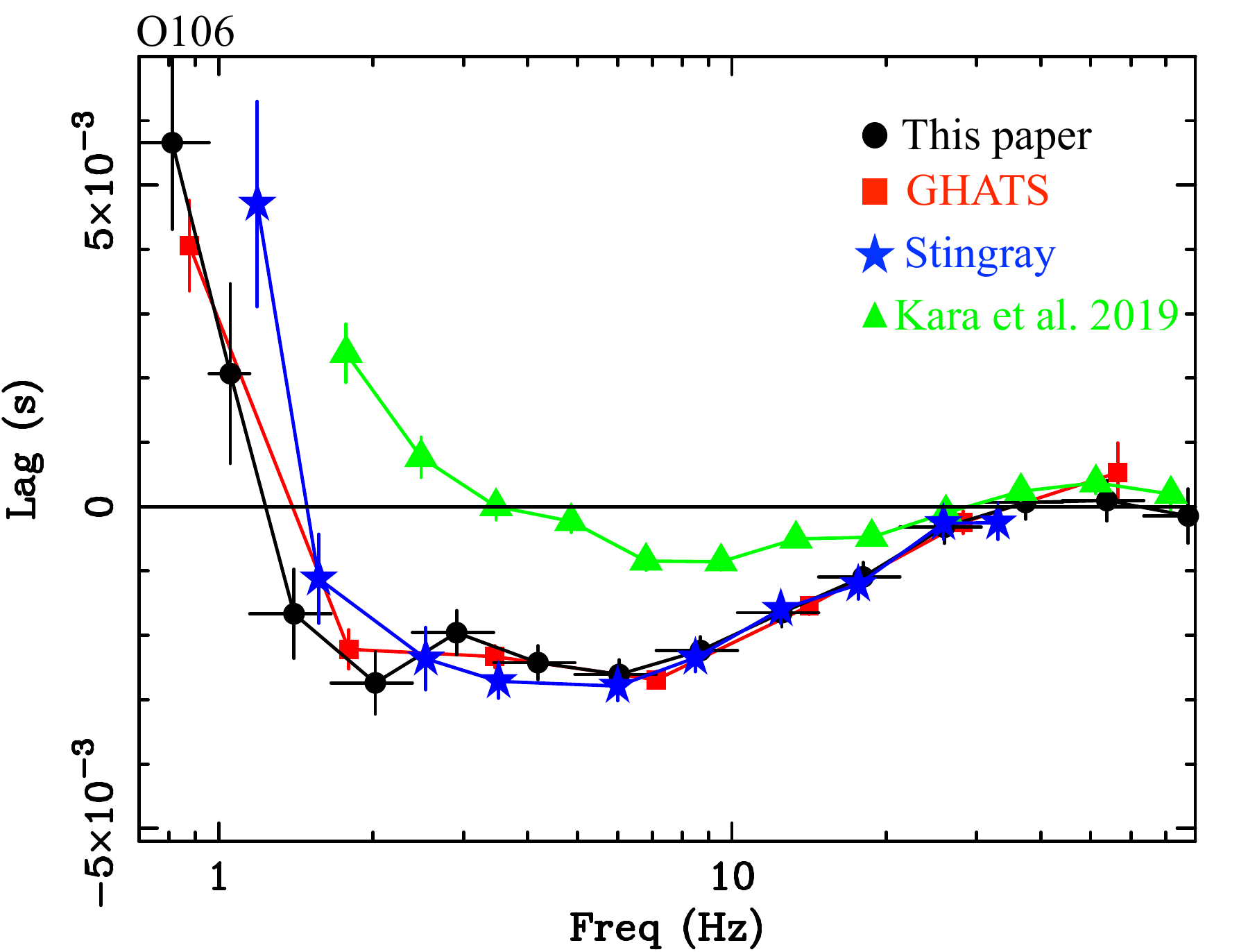}            
    \caption{Lag-frequency spectrum of observation O106 independently computed with the code used in this paper (implemented in IDL and used e.g. in \citealt{BDM2015b,BDM2017}), with the General High-Energy Aperiodic Timing Software (GHATS; used e.g. in \citealt{Belloni2005}), and with Stingray (developed by \citealt{Huppenkothen2019a,Huppenkothen2019b} and publicly available). The scatter in the data is due to slight differences in the extraction and binning. These lags are compared to those reported in \cite{Kara2019} for the same observation.}
    \label{fig:lags_codes}
\end{figure}

The computations of frequency-dependent lags reported in Sect. \ref{sec:softlags} were double-checked using three different codes for spectral-timing analysis. Two of the codes were independently implemented by us in the past \citep[e.g.][]{Belloni2005,BDM2015b,BDM2017,Belloni2020}. In addition, we processed the data using the public software for spectral-timing analysis Stingray\footnote{\url{https://stingray.science}} \citep{Huppenkothen2019a,Huppenkothen2019b}. 
In all the cases the observed amplitude of the soft X-ray lags results a factor of $\sim3$ longer than reported in \cite{Kara2019}. 
The existence of such a discrepancy was verified using the same energy bands (i.e. 0.5--1 keV vs. 1--10 keV) as in \cite{Kara2019}. As an example, in Fig. \ref{fig:lags_codes} we contrast the lag-frequency spectrum of observation O106 obtained from the three different codes and the spectrum reported in \cite{Kara2019}. We notice that, while the interpretation given in \cite{Kara2019} of a contracting corona in the hard state does not depend on the observed amplitude of the soft X-ray lags, the longer lags might not be easily reconciled with their conclusion of a disc extending close to the ISCO.

\begin{figure*}
        \includegraphics[width=\columnwidth]{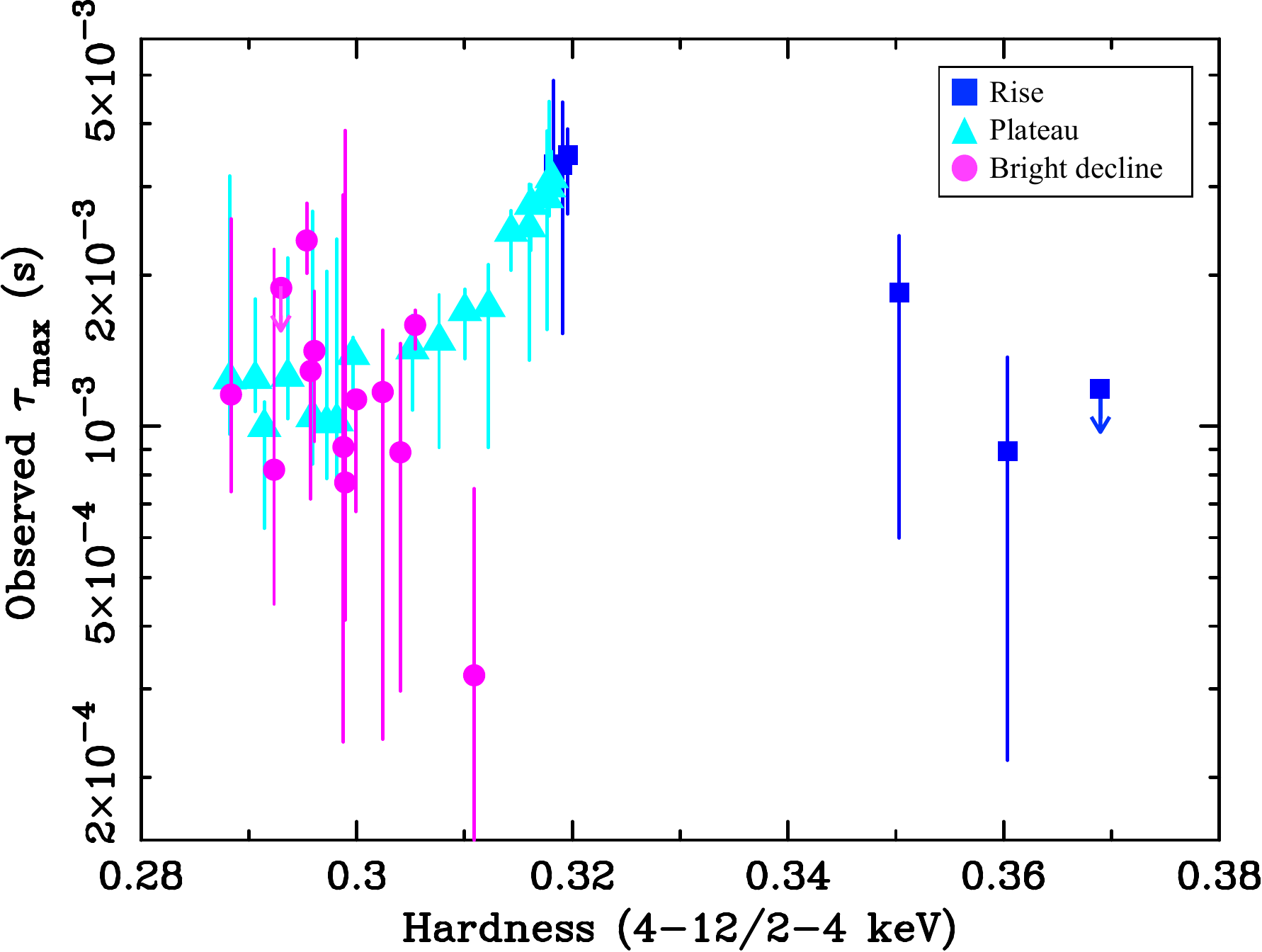}           
        \hspace{0.2cm}  
        \includegraphics[width=\columnwidth]{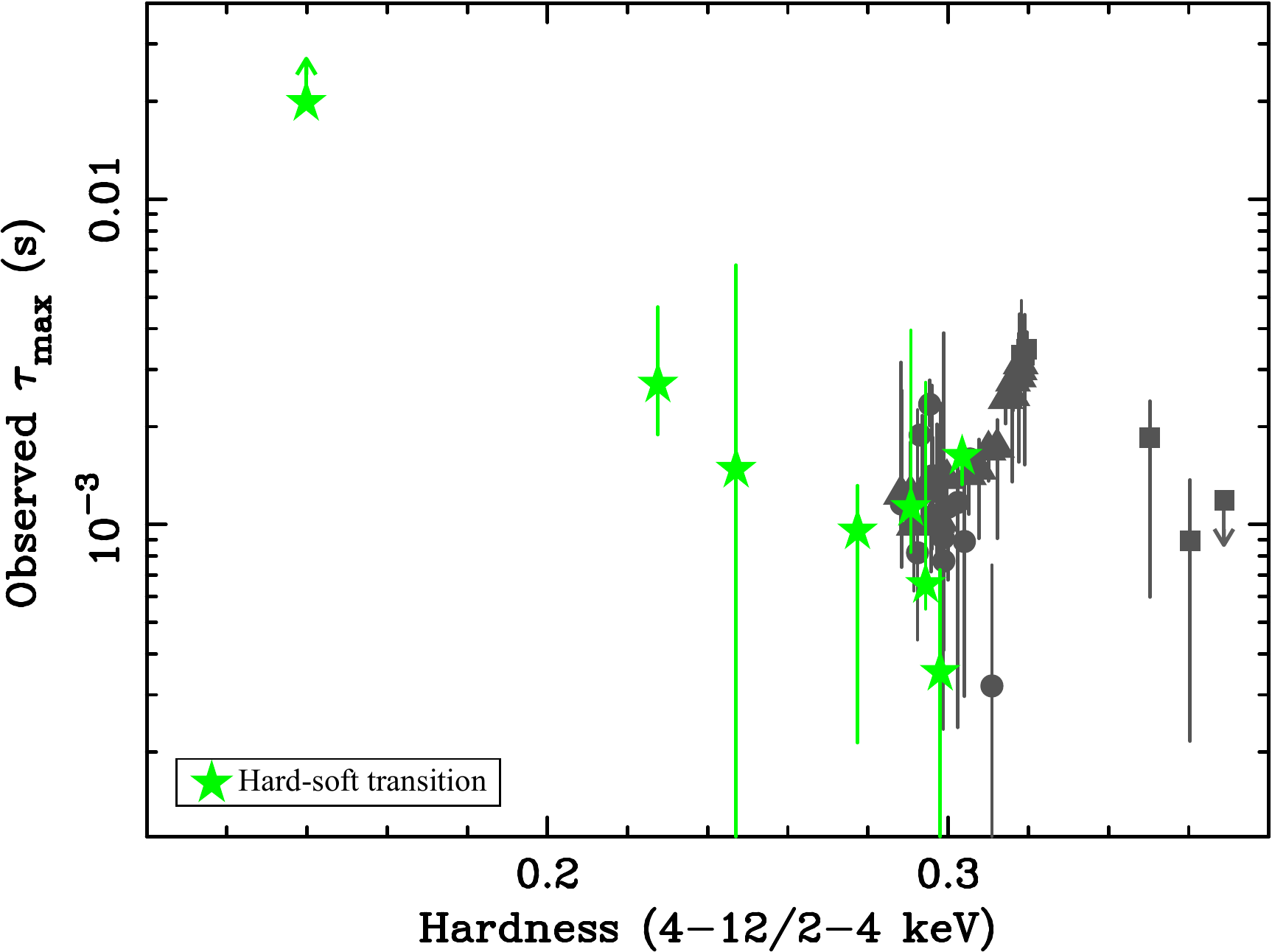}
    \caption{Evolution of observed maximum soft lag amplitude (as inferred from the best-fit models) in lag-frequency spectra as a function of spectral hardness. For comparison with Fig. \ref{fig:nu0_vs_HR}, the \emph{left panel} reports results from the rise, plateau, and bright decline phases, while the \emph{right panel} reports results from the hard-soft transition phase as compared to the other phases (grey symbols).}
    \label{fig:obslag_vs_HR}
\end{figure*}

For further comparison with \cite{Kara2019}, in Fig. \ref{fig:obslag_vs_HR} we also report the evolution of the maximum observed soft X-ray lag amplitude (from the best-fit {\tt RELTRANS} models) as a function of spectral hardness. This amplitude does not show the same steady trend as inferred from the lag characteristic frequency $\nu_{0}$ (Fig. \ref{fig:nu0_vs_HR} and Sect. \ref{sec:softlags}). The rise is characterised by an increase in the observed maximum amplitude of the lag. Then, during the Plateau this amplitude decreases (as previously reported in \citealt{Kara2019} for selected observations). During the last part of the Plateau, the entire Bright decline and the beginning of the hard-soft transition the observed lag amplitude does not have a well-defined trend, and it is consistent with being constant. Finally, it increases during the last observations of the hard-soft transition phase (Fig. \ref{fig:obslag_vs_HR}, right panel). We argue that this behaviour and its apparent inconsistency with the evolution of $\nu_{0}$ is a consequence of the concurrent complex evolution of the hard X-ray lags, which are likely to have a strong influence on the observed maximum amplitude of the soft lag (Fig. \ref{fig:hardsoftlags}). It is worth noting that, contrary to what affirmed in \cite{Kara2019}, the observed amplitude of the longest soft lags detected in MAXI J1820+070 (mostly at the beginning of the outburst, Fig. \ref{fig:obslag_vs_HR}), are consistent with that of the soft lag detected in GX 339-4 during its bright hard state \citep[][]{BDM2015b,BDM2017}.

\section{Independence of $\nu_0$ on the hard X-ray lags of the continuum}
\label{sec:nuzero}

\begin{figure}
        \includegraphics[width=\columnwidth]{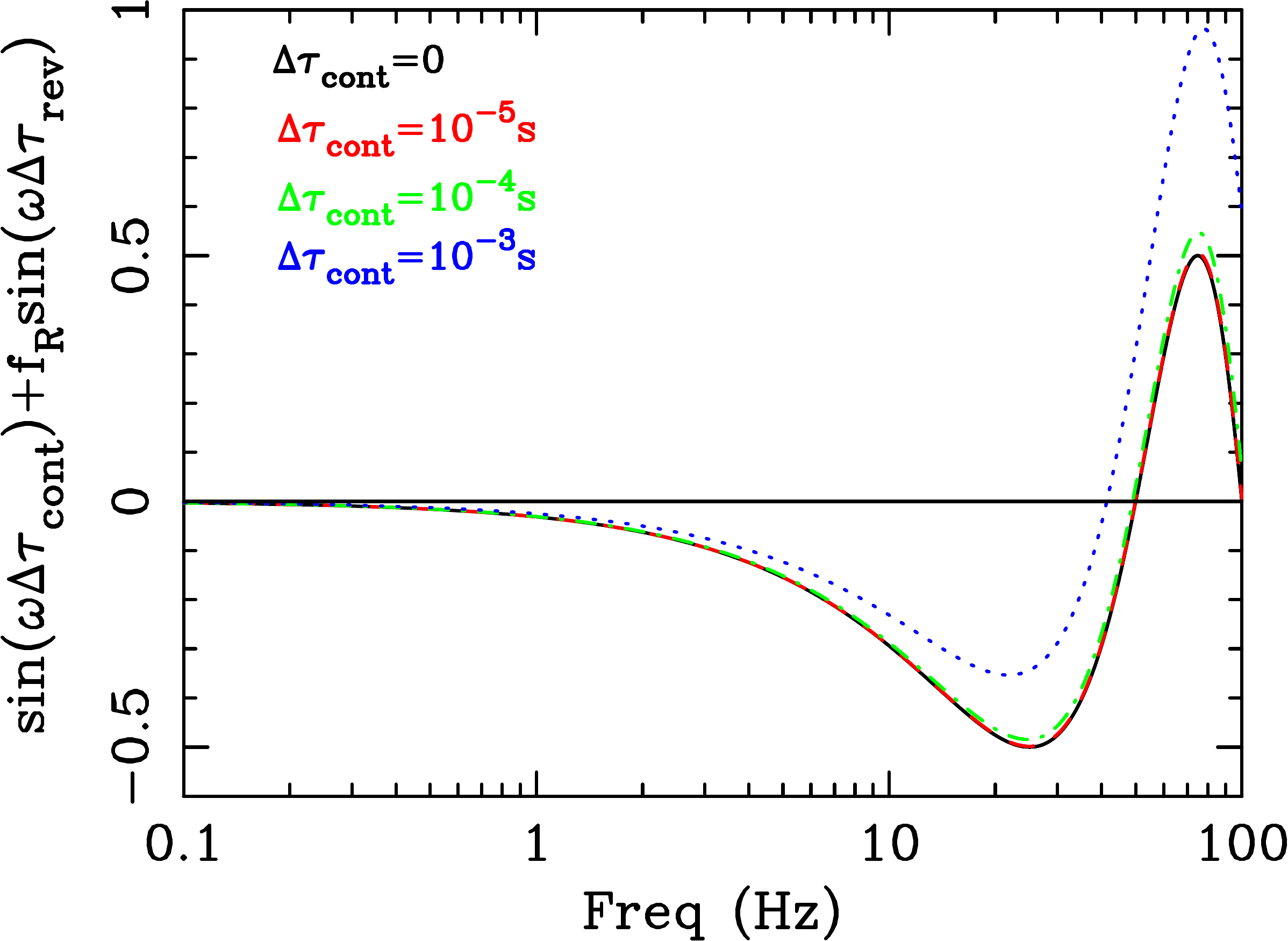}               
    \caption{Solutions of Eq. \ref{eq:phi2} for three different values of $\Delta \tau_{cont}$ (and fixed $\Delta \tau_{rev}=-0.01$s and $f_R=0.5$). The first high-frequency zero-crossing point is the solution that corresponds to $\nu_0$.}
    \label{fig:zerocross}
\end{figure}

In Sect. \ref{sec:softlags} we used the first frequency at which the reverberation lag drops to zero, $\nu_0$, as an unbiased diagnostic of the intrinsic reverberation lag amplitude. Here we show that the hard X-ray lags intrinsic to the X-ray continuum do not affect this frequency. For simplicity, we assume that one band, usually called channel-of-interest (CI, as in \citealt{Uttley2014}) contains both reprocessed and continuum photons, while the other band, usually called the reference band (REF), contains only continuum photons. This assumption is particularly appropriate for the energy bands used in our analysis, where the soft band is chosen to have contribution from the thermal component but includes contribution from the Comptonisation continuum as well, while the hard band is dominated by the primary Comptonised emission and has negligible contribution from the thermal/reflection component. It can be easily shown that in this case, the observed phase lag is:
\begin{eqnarray}\label{eq:phi}
\phi_{obs}=\arctan \frac{\sin(\Delta \tau_{cont}\omega)+f_{R}\sin(\Delta \tau_{rev}\omega)}{\cos(\Delta \tau_{cont}\omega)+f_{R}\cos(\Delta \tau_{rev}\omega)}
,\end{eqnarray}
where $\omega=2\pi\nu$, $\Delta \tau_{rev}\omega$ is the phase shift due to reverberation between reprocessed photons in the CI and continuum photons in the REF band; $\Delta \tau_{cont}\omega$ is the phase shift of continuum photons in the CI with respect to continuum photons in the REF band; and $f_{R}$ is the ratio between reprocessed and continuum photons in the CI. According to Eq. \ref{eq:phi}, the observed phase lag is zero, $\phi_{obs}=0$, when
\begin{eqnarray}\label{eq:phi2}
\sin(\Delta \tau_{cont}\omega)+f_{R}\sin(\Delta \tau_{rev}\omega)=0
.\end{eqnarray}
The hard X-ray lags intrinsic to the continuum show a steadily decreasing trend with frequency (Fig. \ref{fig:hardsoftlags}). This behaviour is commonly observed in BHXRBs \citep[e.g.][]{Nowak1999,Pottschmidt2000,Uttley2011,BDM2015b}, and predicted by models \citep[e.g.][]{Arevalo2006}. Given the tendency of hard X-ray lags to get suppressed at high frequencies, it is reasonable to assume that their contribution is negligible ($\Delta \tau_{cont}\sim0$) around $\nu_0$. In this case Eq. \ref{eq:phi2} shows that the observed phase lag is zero when $\Delta \tau_{rev}\omega=n\pi$, with $n=1,2..$, so that the first zero-crossing frequency can be used to infer the intrinsic amplitude of the reverberation lag \citep[see also][]{Mizumoto2018}. However, we verified that $\nu_0$ is not significantly biased by the presence of hard lags even when the assumption that $\Delta \tau_{cont}\sim0$ around $\nu_0$ does not strictly hold. To this aim we fixed the value of the reverberation lag at $\Delta \tau_{rev}=-0.01$s (so that the first zero-crossing frequency in the absence of continuum hard lags is $\nu_0=50$ Hz) and solved Eq. \ref{eq:phi2} for three different values of $\Delta \tau_{cont}$ (namely $\Delta \tau_{cont}=10^{-3}$s, $10^{-4}$s, and $10^{-5}$s; the highest value of $10^{-3}$s is chosen so as to resemble the value typically measured at the frequencies where the reverberation lag starts to dominate, Fig. \ref{fig:hardsoftlags}). The results are shown in Fig. \ref{fig:zerocross} (where the zero-crossing points are solutions of Eq. \ref{eq:phi2}; here $f_R$ was fixed at 0.5). It is clear that for $\Delta \tau_{cont}\lsim10^{-3}$s the shift induced on $\nu_0$ by the presence of hard X-ray lags is negligible. The maximum shift (about 8 Hz) of $\nu_0$ compared to the $\Delta \tau_{cont}=0$ case is obtained for $\Delta \tau_{cont}=10^{-3}$s; however, this shift is smaller than the estimated uncertainties on $\nu_0$ (Table \ref{tab2}).

Interestingly, observations of MAXI J1820+070 show that the amplitude of the hard lags at a given fixed (high) frequency increases as the source softens (Fig. \ref{fig:hardsoftlags}). This behaviour is generally observed in BHXRBs \citep[e.g.][]{Kylafis2018}. However, according to Fig. \ref{fig:zerocross}, the longer the hard lags, the larger would be the shift induced on $\nu_0$ towards lower frequencies. This implies that, if driven by the observed changes in the hard lags, we would expect to observe a shift of $\nu_0$ towards lower frequencies as the source softens. This is at odds with one of our main results (Fig. \ref{fig:nu0_vs_HR}). Thus we conclude that the hard lags in MAXI J1820+070 do not affect our estimates of $\nu_0$ and do not influence its observed increasing trend with decreasing spectral hardness.

\section{Fits of lag-frequency spectra}
\label{sec:lagfits}

Fits of lag-frequency spectra were performed using the {\tt RELTRANS} spectral-timing model (\citealt{Ingram2019c}) in the version presented in \citeauthor{Mastroserio2019} (\citeyear{Mastroserio2019,Mastroserio2020}), which also allows for the fit of hard lags. We let the X-ray source height, $h$, and the disc inner radius, $R_{in}$ free to vary.  
We also let the parameters regulating the spectral pivoting of the inverse-Compton spectrum \citep{Mastroserio2018} free to vary in order to account for contributions from the hard lags in the fitted frequency band.
We let the BH mass vary within its estimated range of uncertainties (Sect. \ref{sec:intro}). 
We fixed the BH spin at its maximum value (in order to allow for the largest range of values for the disc inner radius, and thus the highest values of $\nu_{0}$), the column density at $\rm{N_{\rm H}\sim1.4\times 10^{21} cm^{-2}}$ (\citealt{Kajava2019}; Dzie{\l}ak et al. 2021), the boosting factor and the normalisation at 1. We also assumed $\Gamma=1.6$, $\rm{log(\xi/erg\ cm\ s^{-1})}\sim 3.3$ and $\rm{E_{cut}=30}$ keV. The latter parameters were fixed in order to speed-up the fits. Nonetheless, letting them free to vary does not affect our estimates of $\nu_{0}$ as this parameter, for a given value of BH mass, depends only on the height of the X-ray source and on the inner radius of the disc.
The errors on $\nu_{0}$ were obtained considering the 90 per cent confidence interval of the $h$ and $R_{\rm in}$ parameters within Xspec.

\section{Estimating the total irradiating flux and the thermally re-emitted fraction}
\label{sec:timeave}
We used \texttt{relxillCp} (v1.4.3; \citealt{Dauser2016} and references therein) to model the broadband continuum in time-averaged spectra. The spectrum is absorbed by a column of cold gas (modelled with \texttt{TBabs}) with column density fixed to $\rm{N_{\rm H}\sim1.4\times 10^{21} cm^{-2}}$ (\citealt{Kajava2019}; Dzie{\l}ak et al. 2021). The emissivity index was fixed at 3, the spin parameter was fixed at its maximum value $a=0.998$, and the high energy cut off, as done for the fit of covariance spectra, was fixed at $kT_{\rm e}=30$ keV \citep[][]{Zdziarski2021b}. The spectral index of the Comptonised incident spectrum, the ionisation parameter, the reflection fraction, and the disc inner radius were left free to vary. Finally, the iron abundance and inclination angle were determined from the simultaneous fit of observations O103, O137, O170, and O197 (chosen so as to sample different phases of the outburst). We obtained a best-fit value of $i=71.1^{\circ+0.6}_{-0.7}$ for the inclination and an upper limit of $A_{\rm Fe}<0.51$ for the iron abundance. Thus we fixed these parameters to $i=71^\circ$ and $A_{\rm Fe}=0.5$ in the individual fits of the single observations.
The fits of the single observations yield high values of reflection fraction consistent with $\mathcal{R}=1$. The resulting 0.1--1000 keV irradiating fluxes are reported in Table \ref{tab2}. 

While time-averaged spectra yield the total irradiating flux (which contributes to the observed inner disc temperature), high-frequency covariance spectra yield the true thermal response of the inner disc to hard X-ray irradiation, with negligible contribution from disc internal dissipation.
This follows from the fact that while disc emission due to internal dissipation may be variable on the fastest timescales, this emission is more easily suppressed near the region where it is produced in a geometrically thin optically thick accretion flow \citep[e.g.][]{Churazov2001,Mushtukov2018}, thus not contributing significantly to the overall high-frequency variability. Moreover, in order to contribute to covariance spectra, the variable intrinsic disc emission should be linearly correlated with the observed primary hard X-ray continuum, most likely through Compton up-scattering \citep[e.g.][]{Nowak1996} or downward propagation of mass accretion rate perturbations \citep[e.g.][]{Lyubarskii1997, Kotov2001,Bollimpalli2020}.
However, both processes would produce hard X-ray lags, whereas in our case the lag-frequency spectrum is dominated by soft X-ray lags at high frequencies (Fig. \ref{fig:softlags} and Sect. \ref{sec:softlags}). Therefore, we expect any contribution from disc internal dissipation producing variable soft X-ray emission and a coherent hard X-ray response to be negligible in high-frequency covariance spectra.

In order to measure the variable thermally reprocessed flux emitted by the disc we can fit high-frequency covariance spectra with the model \texttt{thcomp(deltadisk)}, where the use of the convolution model \texttt{thcomp} \citep[][]{Zdziarski2020} allows us to account for the contribution from both the up-scattered and direct disc photons. For example, applying this model to the high-frequency covariance spectrum of O103, we find a thermally reprocessed-to-total variable flux fraction of about 10 per cent. 
Such a fraction is significantly lower than expected if all of the observed flux in the variable primary continuum was reprocessed in the disc (either reflected or thermalised, and assuming an albedo of $\sim$ 0.3--0.7). This is an interesting result that does not have an immediate explanation, thus deserving future investigations. Nonetheless, we argue that a more complex physical situation than that of a single power law emitting-region may determine the observed high-frequency covariance spectra. In this event, the fraction of Comptonised photons that effectively irradiates the disc should be lower, with negligible contribution from other (possibly inner) regions of the accretion flow (e.g. as in the scenarios proposed in \citealt{Zdziarski2021a} and Dzie{\l}ak et al. 2021).

\longtab[1]{
\begin{longtable}{cccccc}
\caption{\label{tab2} Results of the X-ray spectral-timing analysis of MAXI J1820+070.}\\ 
\hline\hline
\noalign{\smallskip}
       Observations    &   $\Gamma_{\rm covar}$          & $kT_{\rm in,covar}$               &         $F_{{\rm irr,0.1-1000\ keV}}$               &  $R_{\rm in}$                      &  $\nu_{0}$           \\ 
       \noalign{\smallskip}
                             &                           &  [keV]                          &      [$10^{-8} {\rm erg\ s^{-1}\ cm^{-2}}$]        &  [$R_{\rm g}$]      &      [Hz]                      \\
\noalign{\smallskip}
 \hline\hline      
\endfirsthead
\caption{continued.}\\
\hline\hline
\noalign{\smallskip}
       Observations    &   $\Gamma_{\rm covar}$          & $kT_{\rm in,covar}$            &         $F_{{\rm irr,0.1-1000\ keV}}$               &  $R_{\rm in}$                      &  $\nu_{0}$           \\ 
       \noalign{\smallskip}
                             &                           &  [keV]                            &      [$10^{-8} {\rm erg\ s^{-1}\ cm^{-2}}$]        &  [$R_{\rm g}$]      &      [Hz]                      \\
\noalign{\smallskip}
 \hline\hline
\endhead
\hline
\endfoot
\noalign{\smallskip}
 \multicolumn{6}{c}{{\bf Rise}}\\
 \noalign{\smallskip}
 \hline
 \noalign{\smallskip}
 O101   &         $1.46\pm0.07$ &   $0.22\pm0.03$                & $ 1.08^{+0.02}_{-0.01}$    & $37^{+20}_{ -4}$  & $\tablefootmark{a}15^{+15}_{-14}$            \\
                &                                            &                                                                  &                                                   &                                      \\
O102    &        $1.52\pm0.07$ &   $0.23^{+0.03}_{-0.04}$                 & $ 1.84^{+0.03}_{-0.02}$    & $44^{+29}_{ -6}$ &  $\tablefootmark{a}15^{+15}_{-14}$        \\
                &                                              &                                                                               &                                                      &                                \\
O103    &       $1.49^{+0.06}_{-0.07}$  &   $0.31\pm0.03$                   & $ 2.11^{+0.03}_{-0.02}$     & $27^{+10}_{ -2}$  &$15^{+2}_{ -6}$     \\
                &                                            &                                                                                         &                                          &                                   &             \\
O104--O105      &       $1.49\pm0.07$     &    $0.36\pm0.03$                       &$ 3.66^{+0.04}_{-0.03}$      & $26^{+10}_{ -2}$  &  $29^{+15}_{-10}$     \\
                        &       &                                   &                                                                                &                              \\
O106--O107      &       $1.46\pm^{+0.08}_{-0.09}$ &    $0.50\pm0.04$        & $10.57\pm0.03           $    & $23^{+7}_{ -2}$  &   $42^{+17}_{-6}$           \\
                        &       &                                   &                                          &                             &                                     \\
O107--O108      &       $1.54\pm{0.09}$ &   $0.48\pm0.05$       & $10.19\pm0.04      $   & $18^{+12}_{ -2}$ & $24^{+10}_{-2}$          \\
                        &       &                                   &                                                            &                                                \\
O109            &       $1.52^{+0.08}_{-0.09}$ &   $0.49\pm0.04$                  & $10.29\pm0.04    $     & $23^{+8}_{ -2}$ & $83^{+13}_{-24}$        \\
\noalign{\smallskip}
                \hline
                \noalign{\smallskip}
$\chi^2_{\rm covar}/d.o.f$      &       \multicolumn{2}{c}{202/210}  & & &          \\
\noalign{\smallskip}
\hline\hline
\noalign{\smallskip}
 \multicolumn{6}{c}{{\bf Plateau}}\\
 \noalign{\smallskip}
 \hline
 \noalign{\smallskip}
O110    &       $1.52\pm{0.09}$ &    $0.52\pm{0.04}$       & $10.29\pm0.04  $ & $20^{+7}_{-1}$  & $39^{+2}_{-5}$         \\
                        &                                            &                                                                    &           &                                    \\
O111             &     $1.47^{+0.09}_{-0.10}$    &$0.54\pm0.04$              & $10.93\pm0.04  $  & $19^{+6}_{-1}$   & $91^{+14}_{-32}$       \\
                        &                                           &                                                                      &         &                             \\
O112--O115      &$1.50\pm{0.10}$        &  $0.56\pm0.04$                  & $ 9.12\pm0.02 $  &$16^{+5}_{-1}$  &  $62^{+33}_{-18}$         \\
                        &                                           &                                                                        &           &                         \\
O116--O117      &$1.53\pm{0.09}$        & $0.52\pm0.04$                    & $ 9.11\pm0.03 $ & $20^{+7}_{-1}$  &  $39^{+10}_{-5}$         \\
                        &                                           &                                                            &               &                        \\
O118--O119      &$1.47^{+0.10}_{-0.11}$ & $0.55\pm0.04$                      & $ 9.44\pm0.03  $ & $18^{+6}_{-1}$   & $56^{+30}_{-16}$                    \\
                        &                                           &                                                                    &               &                        &        \\
O120--O123      &$1.52^{+0.09}_{-0.10}$ & $0.53\pm0.04$        & $ 8.63\pm0.02 $  & $18^{+6}_{-1}$    &  $32^{+9}_{-4}$                      \\
                        &                                           &                                                            &               &                \\
O124--O127      &$1.57^{+0.12}_{-0.13}$ & $0.49\pm{0.05}$        & $ 7.18\pm0.04 $  & $19^{+8}_{-2}$  &  $26^{+27}_{-7}$                       \\
                        &                                           &                                                            &               &                                \\
O130--O131      &$1.48^{+0.10}_{-0.11}$ & $0.57\pm0.04$                   & $ 9.47\pm0.03 $ & $17^{+5}_{ -1}$     & $35^{+14}_{-5}$                                     \\
                        &                                           &                                                            &               &                        \\
O132--O134      & $1.53\pm{0.10}$          & $0.54\pm0.04$                  &  $ 8.59\pm0.02 $  & $17^{+5}_{ -1}$     &  $42^{+23}_{-9}$            \\
                        &                                           &                                                                    &               &                 \\
O135--O139      &$1.51\pm{0.12}$            & $0.59\pm0.05$                 & $ 7.45\pm0.02 $ & $14^{+4}_{ -1}$  & $42^{+11}_{-9}$                     \\
                        &                                           &                                                                    &               &                 \\
O140--O141      &$1.53^{+0.15}_{-0.17}$ & $0.52^{+0.06}_{-0.07}$       &  $ 6.45^{+0.04}_{-0.03}$ & $16^{+9}_{ -2}$  &  $47^{+18}_{-10}$                    \\
                        &                                           &                                                            &               &                \\
O142    &$1.38^{+0.20}_{-0.21}$ & $0.59\pm{0.06}$                &  $ 8.30\pm0.04 $ &   $14^{+6}_{ -1}$  & $146^{+58}_{-51}$          \\
                        &                                           &                                                            &               &                \\
O143--O144      &$1.50^{+0.17}_{-0.18}$ & $0.60\pm{0.05}$         & $ 6.78\pm0.03 $ & $13^{+5}_{ -1}$ & $133^{+53}_{-38}$        \\       
                        &                                           &                                                            &               &                \\
O145--O146      &$1.60\pm{0.13}$        & $0.56\pm0.05$                   & $ 7.46\pm0.02$ & $15^{+5}_{ -1}$  &  $146^{+22}_{-68}$       \\
                        &                                           &                                                                    &               &         \\
O147    &$1.61\pm{0.15}$        & $0.56\pm0.05$                      & $ 7.97\pm0.03 $ &  $16^{+6}_{ -1}$ & $146^{+39}_{-42}$       \\
                        &                                           &                                                            &               &                \\
O148    &$1.64\pm{0.21}$            & $0.57^{+0.06}_{-0.07}$       & $ 8.04\pm0.03 $ & $15^{+7}_{ -2}$ & $39^{+38}_{-13}$       \\
                        &                                           &                                                            &               &          \\
O149--O151      &$1.58\pm{0.15}$        & $0.55\pm0.05$                  & $ 7.22\pm0.02 $ & $15^{+5}_{ -1}$ & $121^{+32}_{-16}$       \\
                        &                                           &                                                            &               &               \\
O152--O153      &$1.51^{+0.26}_{-0.28}$ & $0.61\pm{0.07}$      &  $ 5.79\pm0.02 $ &   $11^{+5}_{ -1}$ & $146^{+39}_{-75}$    \\
\noalign{\smallskip}
\hline
\noalign{\smallskip}
$\chi^2_{\rm covar}/d.o.f$      &       \multicolumn{2}{c}{635/540} & & &          \\    
\noalign{\smallskip}    
\hline
\noalign{\smallskip}
\multicolumn{6}{c}{{\bf Bright Decline}}\\
\noalign{\smallskip}
 \hline
 \noalign{\smallskip}
O155--O156      & $1.71\pm{0.30}$        &$0.61\pm{0.07}$       & $ 5.18^{+0.04}_{-0.02}$ & $11^{+5}_{-1}$ & $110^{+17}_{-51}$   \\
                        &                                           &                                                                 &                  &                                                         \\
O157    & $1.61^{+0.18}_{-0.19}$         &$0.49^{+0.06}_{-0.07}$                    & $ 6.52\pm0.03 $ & $19^{+12}_{ -2}$    & $133^{+114}_{-62}$                         \\
                        &                                           &                                                      &            &                   &                                                      \\
O158--O162     & $1.60^{+0.19}_{-0.20}$  &$0.54\pm{0.06}$          & $ 5.18\pm0.02 $  & $13^{+6}_{-1}$ & $47^{+107}_{-6}$     \\
                        &                                           &                                                      &           &                    &                                                      \\
O163--O164     & $1.62^{+0.40}_{-0.53}$ &$0.58^{+0.15}_{-0.16}$         & $ 5.18^{+0.03}_{-0.06}$ & $12^{+16}_{-4}$   & $56^{+15}_{-8}$            \\
                        &                                           &                                                      &         &                      &                                                      \\
O165--O167     & $1.39\pm{0.22}$         &$0.51\pm0.06$                     & $ 4.86\pm0.02 $ & $15^{+7}_{ -2}$ & $56^{+129}_{-19}$      \\
                        &                                           &                                                      &         &                      &                                                      \\
O168                &  -                                        &                       -                                   & -      &  - &  $67^{+34}_{-23}$    \\
                        &                                           &                                                      &         &                      &                                                      \\
O169--O170     & $1.39\pm{0.23}$            &$0.53\pm{0.06}$      & $ 4.52\pm0.02 $ & $13^{+6}_{ -1}$ &  $<128$         \\                
                        &                                           &                                                      &             &                 &                                                       \\
O171                 & $1.46^{+0.30}_{-0.32}$       &$0.49\pm{0.08}$        & $ 4.62\pm0.03 $ & $16^{+12}_{ -3}$ & $51^{+248}_{-11}$       \\
                        &                                           &                                                      &         &                      &                                                      \\
O172                & $1.26^{+0.28}_{-0.22}$       &$0.48\pm{0.05}$        & $ 4.50\pm0.03 $ & $16^{+7}_{ -2}$  &  $100^{+15}_{-63}$      \\
                        &                                           &                                                      &          &                     &                                                      \\
O173--O174     & $1.47^{+0.18}_{-0.19}$  &$0.43\pm0.05$                       & $ 4.25\pm0.02 $ &  $19^{+10}_{ -2}$ & $83^{+22}_{-24}$       \\
                        &                                           &                                                      &         &                      &                                                      \\
O175--O176     & $1.68^{+0.14}_{-0.26}$  &$0.49^{+0.08}_{-0.09}$             & $ 3.87\pm0.02 $  & $15^{+12}_{-2}$  & $133^{+53}_{-54}$                          \\
                        &                                           &                                                      &          &                     &                                                      \\
O177--O180     & $1.59^{+0.17}_{-0.18}$  &$0.41\pm0.05$                      & $ 3.05\pm0.02 $ &  $18^{+9}_{ -2}$ & $42^{+11}_{-9}$         \\
                        &                                           &                                                      &          &                     &                                                      \\
O182--O185    & $1.58^{+0.30}_{-0.39}$    &$0.34^{+0.10}_{-0.12}$     & $ 0.87^{+0.21}_{-0.22}$   &$14^{+34}_{ -6}$ &  $32^{+40}_{-9}$                   \\
\noalign{\smallskip}
\hline
\noalign{\smallskip}
$\chi^2_{\rm covar}/d.o.f$      &       \multicolumn{2}{c}{295/360}     & & &   \\                
\noalign{\smallskip}
\hline\hline
\noalign{\smallskip}
\multicolumn{6}{c}{{\bf Hard-soft transition}}\\
\noalign{\smallskip}
\hline
\noalign{\smallskip}
 O186--O187     & $>1.00$           &$0.46^{+0.07}_{-0.06}$        & $ 3.09\pm0.03 $ &$14^{+9}_{-2}$ & $62^{+17}_{-13}$       \\
                        &                                           &                                                      &           &                    &                                                      \\
O188            & $1.61^{+0.58}_{-0.53}$            &$0.47^{+0.10}_{-0.11}$         & $ 0.81\pm0.01$ & $7^{+8}_{-2}$ & $47^{+252}_{-19}$        \\
                        &                                           &                                                      &         &                      &                                                      \\
O189            & $1.60^{+0.10}_{-0.11}$        &$0.45\pm0.03$                       & $ 4.55\pm0.02$ & $18^{+6}_{ -1}$ & $161^{+43}_{-82}$    \\
                        &                                           &                                                      &         &                      &                                                      \\
O190            & $1.40^{+0.42}_{-0.35}$    &$0.54^{+0.08}_{-0.09}$        & $ 4.77\pm0.02 $ & $13^{+9}_{ -2}$ & $146^{+58}_{-87}$       \\
                        &                                           &                                                      &          &                     &                                                      \\
O194            & $1.46^{+0.20}_{-0.21}$        &$0.55\pm0.05$          & $ 5.77\pm0.01 $ & $14^{+5}_{ -1}$ &  $56^{+30}_{-16}$      \\
                        &                                           &                                                            &       &                        &                                                \\
O195            &  $1.51\pm0.37$   &$0.61^{+0.10}_{-0.11}$              & $ 6.79^{+0.02}_{-0.01}$ & $12^{+9}_{ -2}$  & $>18$      \\
                        &                                           &                                                            &        &                      &                                                 \\
O196            &  $1.43^{+0.63}_{-0.42}$    &$0.66^{+0.18}_{-0.17}$       & $ 8.17\pm0.04$  & $11^{+14}_{ -4}$  & $78^{+53}_{-33}$          \\
                        &                                           &                                                            &            &                   &                                                \\
O197            &  $2.45^{+1.62}_{-0.76}$    &$1.08^{+0.27}_{-0.07}$            & $41.86^{+0.35}_{-0.40}$  & $\gsim7$  & $71^{+105}_{-19}$           \\
\noalign{\smallskip}
\hline
\noalign{\smallskip}
$\chi^2_{\rm covar}/d.o.f$      &       \multicolumn{2}{c}{234/240} & & &          \\
\noalign{\smallskip}            
\hline\hline
\end{longtable}
\tablefoot{The table reports: the single/combined observations (the groups correspond to those reported in Table \ref{tab1}); the best-fit values of the $\Gamma_{covar}$ and $kT_{\rm in,covar}$ parameters as obtained from the fits of high-frequency covariance spectra with the \texttt{TBabs[deltadisk + nthComp]} model; 
the 0.1--1000 keV irradiating flux as estimated from fits of time-averaged spectra; the estimated lower limits on the inner disc radius $R_{\rm in}$ as inferred from Eq. \ref{eq:S-B_law2} (Sect. \ref{sec:covar} and Fig. \ref{fig:Rin_vs_hr}); the best-fit values of $\nu_{0}$ from the modelling of the lag-frequency spectrum (Sect. \ref{sec:softlags} and Fig. \ref{fig:nu0_vs_HR}). The $\chi^2_{\rm covar}/d.o.f.$ refers to the simultaneous fits of the high-frequency covariance spectra of all the single/combined observations belonging to each phase of the outburst.\\
\tablefoottext{a}{Observations O101 and O102 were combined for the fit of lag-frequency spectra only (Sect. \ref{sec:softlags}).}}
}

\end{appendix}

\end{document}